\documentclass[aps,prb,twocolumn,eqsecnum,showpacs]{revtex4-1}
\usepackage{amssymb}
\usepackage{graphics}
\usepackage{dcolumn}
\usepackage{pifont}
\usepackage{graphicx}
\usepackage{amsmath}
\usepackage{verbatim}
\usepackage{epstopdf}
\usepackage{color}
\usepackage{dsfont} 

\newcommand{\bom}{\mathbf}

\newcommand{\bg}{\boldsymbol}
\newcommand{\nn}{\nonumber}
\newcommand{\beq}{\begin{equation}}
\newcommand{\eeq}{\end{equation}}
\newcommand{\bea}{\begin{eqnarray}}
\newcommand{\eea}{\end{eqnarray}}
\newcommand{\bse}{\begin{subequations}}
\newcommand{\ese}{\end{subequations}}
\newcommand{\bwt}{\begin{widetext}}
\newcommand{\ewt}{\end{widetext}}

\newcommand{\bk}{{\bf k}}
\newcommand{\bq}{{\bf q}}
\newcommand{\bp}{{\bf p}}

\newcommand{\bpp}{{\bf p}'}
\newcommand{\bl}{{\bf l}}
\newcommand{\blp}{{\bf l}'}

\newcommand{\bv}{{\bf v}}

\newcommand{\bsu}{\begin{subequations}}
\newcommand{\esu}{\end{subequations}}
\newcommand{\bs}{\hat{\boldsymbol{\sigma}}}
\newcommand{\ez}{{\bf e}_z}

\newcommand{\up}{{\bf e}_{\bf{p}}}
\newcommand{\upp}{{\bf e}_{\bf{p}'}}
\newcommand{\la}{\langle}
\newcommand{\ra}{\rangle}
\newcommand{\paren}[1]{\left( #1 \right)}
\newcommand{\bra}[1]{\left[ #1 \right]}

\begin{document}
\title{Theory of a chiral Fermi liquid: general formalism}

\author{Ali Ashrafi$^{1}$, Emmanuel I. Rashba$^{2}$,  and Dmitrii L. Maslov$^{1}$}

\begin{abstract}
We extend the Fermi-liquid (FL) theory to include spin-orbit (SO) splitting of the energy bands, focusing on the Rashba SO coupling as an example. We construct the phenomenological Landau interaction function for such a system using the symmetry arguments and verify this construction by an explicit perturbative calculation.  The Landau function is used to obtain the effective mass, compressibility, and stability conditions of the FL.  It is shown that although the charge-sector properties, such as the effective mass and compressibility, are determined solely by well-defined quasiparticles, the spin-sector properties, such as the spin susceptibility, contain a contribution from damped states in between the spin-split Fermi surfaces, and thus cannot be fully described by the FL theory, except for the case of weak SO coupling.  We derive some specific properties of a chiral FL and show, in particular, that for contact interaction spin-splitting of the Fermi velocities of Rashba subbands occurs because of the Kohn anomaly, also modified by SO coupling.
\end{abstract}
\date{\today}
\affiliation{$^1$Department of Physics, University of Florida, P. O. Box 118440,
Gainesville, FL 32611-8440\\
$^2$Department of Physics, Harvard University, Cambridge, Massachusetts 02138, USA}
\maketitle
\section{\label{sec:Int}INTRODUCTION}
A large number of many-body systems are characterized by non-trivial correlations between spin and orbital degrees of freedom.  To name just a few, these are two-dimensional (2D) electron and hole gases in semiconductor heterostructures with broken inversion symmetry,~\cite{zutic:2004,winkler:book} non-centrosymmetric normal metals~\cite{samokhin:2009} and superconductors,~\cite{sigrist:1991, *mineev_sigrist}surface/edge states of three-dimensional (3D)/2D topological insulators,~\cite{hasan:2010, *hasan:2011, *qi:2011, *alicea:2012,  *moore:2012}
conducting states at oxide interfaces,~\cite{cen:2009} atomic Bose~\cite{lin:2009, *cold_Rashba_Bose:2011} and Fermi~\cite{wang:2012, *cheuk:2012}  gases in simulated non-Abelian magnetic fields, etc.
Spin-orbit (SO) coupling inherent to all these systems locks electron spins and momenta into patterns characterized by chirality.
\cite{[{Pseudospins of Dirac fermions in graphene are also correlated to their momenta, and thus graphene can be viewed as a \lq\lq pseudochiral\rq\rq\/ system, see} ]polini:2007}
Electron-chiral materials are endowed with unique properties that are interesting from both the fundamental and applied points-of-view. 
 
 Recently, there has been a surge of interest in the interplay between chirality and the electron-electron ({\em ee}) interaction, which is necessarily present in all these system. The current tendency -- driven in part by the need to develop semiconductor-based spintronic devices--is to enhance the strength of the SO interaction relative to other energy scales, namely, the Fermi energy, and  the potential energy of Coulomb repulsion. In semiconductor heterostructures, this can be achieved by reducing the carrier number density, even to the point when only the lowest of the spin-split Rashba subbands is occupied. \cite{quay:2010, *mourik:2012, *das:2012} In other systems, such as the Au-intercalated graphene-Ni  interface, \cite{varykhalov:2008, *marchenko:2012}, Bi/Ag(111)\cite{ast:2007}  and Bi$_x$Pb$_{1-x}$/Ag(111)\cite{meier:2009} surface alloys, and  bismuth tellurohalides, \cite{bitei:2011, *eremeev:2012, *crepaldi:2012} SO coupling is enhanced due to the presence of heavy, e.g., Au and Bi, atoms.  
If the three energy scales (SO, Fermi, and Coulomb) become comparable to each other, one ventures into a regime where new phases of matter become possible.  For example, a strong enough Coulomb repulsion can force all electrons to occupy only one Rashba subband  \cite{chesi:2007b, *juri:2008}, or to form a spin-nematic phase with spin-split Fermi surface (FS) but zero net magnetization, \cite{wu:2004, *wu:2007, *chubukov:2009}  or else form a number liquid-crystal and crystalline phases, if only the lowest Rashba subband is occupied.~\cite{berg:2012} An important task is, therefore, to classify and analyze new phases of electronic matter that become possible due to the interplay of the {\em ee} and SO interactions.  

Analysis of possible new phases usually starts with the Fermi-liquid theory (FL). The well-established theory of a non-chiral [or SU(2)-symmetric (SU2S)] FL~\cite{agd:1963, *nozieres:1966, *lifshitz:1980} allows one to classify possible Pomeranchuk instabilities that break rotational but not translational symmetry of the Fermi surface, and to derive stability conditions in terms of the Landau parameters. 
With the exception of quantum wires and surface states of 3D topological insulators at the Dirac point, all other examples of electron-chiral system are believed to behave as FLs with well-defined quasiparticles near the FS (or multiple FSs in case of spin-split states), as long as the {\em ee} is too weak to break some symmetries. This conclusion is supported by a large number of studies in which various FL quantities--the effective mass,\cite
{saraga:2005,agarwal:2011,aasen:2012,yu:2013} quasi-particle renormalization factor,\cite{saraga:2005,yu:2013} quasi-particle lifetime,\cite{saraga:2005,yu:2013} spin and charge susceptibilities,\cite{saraga:2005,pletyukhov:2006, pletyukhov:2007,barlas:2007,chesi:2007b,chesi:2007c,chesi:thesis,chesi:2009,zak:2010,culcer:2011,zak:2012} spin-Hall conductivity,\cite{shekhter:2005,vignale:2006} charge- and spin-drag resistivities,\cite{tse:2007,vignale:2002} profiles of Friedel oscillations,\cite{badalyan:2010,zak:2012} plasmon spectra and Drude weights,\cite{agarwal:2011} etc.--were calculated within some version of the perturbation theory in the {\em ee} interaction. Yet, it is obvious that the conventional FL theory, based on the assumption of SU(2) invariance of electron spins, is not applicable to electron-chiral systems.  Indeed,  the Landau function of an SU2S FL  \beq
{\hat f}(\bp,\bpp)= f^{\mathrm{s}}(\bp,\bpp)\mathds{1}\mathds{1}^{\prime}+f^{\mathrm{a}}(\bp,\bpp)\bs\cdot\bs'.
\label{LFSU2}
\eeq
contains only two invariants: unity and the scalar product of electron spins. 
(Here and it what follows, the product of two matrices is to be understood for projections onto the spin basis as $f_{\alpha,\beta;\gamma,\delta}=f^{\mathrm{s}}\delta_{\alpha\gamma}\delta_{\beta\delta}+f^{\mathrm{a}}\bs_{\alpha\gamma}\cdot\bs_{\beta\delta}$. \cite{agd:1963, *nozieres:1966, *lifshitz:1980})
Clearly, broken SU(2) symmetry of a chiral FL allows for a much richer variety of invariants, which cannot be separated into the charge and spin parts. Although the SO interaction is a relativistic effect, a relativistic generalization of the FL theory, \cite{baym:1976, *song:2001} developed in the context of dense nuclear matter, is also not applicable to SO-coupled electron systems because the latter are never fully Lorentz-invariant. Another line of work focuses on FLs with SO-coupling arising because of the dipole-dipole interaction of fermion spins. \cite{betouras:2009, *fregoso:2009, *fregoso:2009a, *li:2012} While this type of theories is relevant to atomic gases,  the dipole-dipole SO coupling--as opposed to simulated non-Abelian fields -- is very much different from both the Rashba- and Dresselhaus-type couplings, \cite{winkler:book} which exist already at the single-particle level. Actually, the closest analogy to a chiral FL is a ferromagnetic or partially spin-polarized FL, e.g., He$^3$ in a magnetic field, which have a long history of studies, starting from  Refs.~\onlinecite{abrikosov:1958,silin:1958, *silin:1959,kondratenko:1965,platzman:1967,ma:1968,leggett:1970,dzyalosh:1976} and culminating in a series of detailed papers by Meyerovich et al.~\cite{bashkin:1981, *meyerovich:1983,meyerovich:1992a,
 meyerovich:1994, *meyerovich:1994b} (for a detailed discussion of more recent studies, see Refs.~\onlinecite{mineev:2004, *mineev:2005, *mineev:2011}.)
The difference between partially-polarized and chiral FLs is in that fermions in the latter experience an effective (and momentum-dependent) rather than real magnetic field.  In this paper, we will 
employ this analogy, while also emphasizing the differences between the two cases.  Also, a chiral FL with two occupied spin-split bands shares some properties with any multi-component FL;\cite{ashcroft:1981, *bedell:1991} yet its spin sector is very special and, as will we will show later, cannot be treated simply by assigning extra indices to FL parameters.
References~\onlinecite{raikh:1999} and \onlinecite{shekhter:2005} considered an SU2S FL perturbed by a weak SO coupling; in particular,  Ref.~\onlinecite{shekhter:2005} studied FL renormalization of spin-chiral resonances.  To the best of our knowledge,  a FL theory for a generic and not necessarily weak SO interaction was addressed only in Ref.~\onlinecite{fujita:1987}; however, subtleties of the spin sector of such a FL were not considered in that paper. 

In this paper,  we explore a possibility of constructing a general theory of chiral Fermi liquids (FL). For concreteness, we will focus mostly on linear Rashba SO coupling, relevant for  a number of 2D electron heterostructures; however, we will also discuss briefly the cubic coupling relevant, e.g., for 2D hole gases in III-V semiconductor heterostructures\cite{winkler:2000, *winkler:2002} and the surface state of SrTiO$_3$. \cite{nakamura:2012,zhong:2013} 
We argue that whether a particular quantity of a chiral FL can be described within the FL formalism depends on whether this quantity is spin-independent, e.g., the effective mass or charge susceptibility,  {\em or} spin-dependent, e.g., the spin susceptibility. For spin-independent properties, which are local in the SU(2) spin space, the original Landau's FL theory can be extended to include the effects of SO with rather straightforward modifications similar to any two-band FL theory.~\cite{ashcroft:1981,bedell:1991}  Spin-dependent properties, on the other hand,  are controlled by particles with momenta in between the two spin-split FSs.
If SO splitting is not small compared the Fermi energy, these particles are away from their respective FSs and thus strongly damped. In general, therefore, one cannot describe the spin sector of a chiral FL  in terms of non-interacting quasiparticles. \footnote{Although orbital and spin degrees of freedoms cannot be, strictly speaking, separated in the presence of SO coupling, we will still loosely refer to \lq\lq charge-\rq\rq\ and \lq\lq spin-sector\rq\rq\/ quantities, in the sense specified above.}
On a technical level, this means that one has to consider off-diagonal components of the density matrix, and also that some familiar relation of the SU2S theory, which are local in the momentum space, are replaced by non-local integral equations. For example, we show that the renormalized $g$ factor of chiral quasiparticles satisfies an integral equation, where the integral goes over the entire interval of the momentum in between spin-split subbands. We show, however, it is still possible to account for first-order effects in SO while keeping the interaction arbitrary.
As it was pointed out by C. Herring already in 1966,\cite{herring} the same difficulty arises also in the theory of ferromagnetic and partially spin-polarized FLs.  Later research\cite{meyerovich:1992a,meyerovich:1994, *meyerovich:1994b} confirmed his arguments;  see, in particular, a detailed discussion of the importance of the off-diagonal components in Ref.~\onlinecite{meyerovich:1992a}.
(Quite foresightfully, Herring also anticipated a similar  problem for a system with SO coupling in the absence of inversion symmetry, which is the subject of this paper).

An important topic of current interest are the collective modes of a chiral FL. \cite{shekhter:2005,raghu:2010,ashrafi:2012,zhang:2012a} We will discuss this topic in a separate forthcoming publication. \cite{ashrafi:partII}

The rest of the paper is organized as follows. In Sec.~\ref{sec:LF}, we construct the phenomenological Landau function  of a chiral FL. Having been equipped with the Landau function, we proceed to calculate the effective masses of chiral quasiparticles in Sec. \ref{sec:m}, and discuss Overhauser-type\cite{overhauser:1971} spin-splitting of effective masses in the zero magnetic field, which is similar to effective-mass splitting in a partially spin-polarized FL.~\cite{suhas:2005,wei:2011} Thermodynamic properties of a chiral FL are discussed in Sec.~\ref{sec:Thermo}. In  particular, the compressibility of a chiral FL is derived in Sec.~\ref{sec:Comp}, Pomeranchuk stability conditions are specified in Sec.~\ref{sec:Pom}, and the spin susceptibility is discussed in Sec.~\ref{sec:Spin}. In particular,
the FL formalism for the spin susceptibility is presented in Sec.~\ref{sec:chi_FL}, where the out-of-plane spin susceptibility is shown explicitly to contain a contribution from the states in between the spin-split FSs. In Sec.~\ref{sec:TR}, we discuss the physical implications of this result, and also demonstrate that the physical origin of mutual cancellation between induced magnetic moments of occupied Rashba subbands is time-reversal symmetry (at fixed magnetic field). 
In Sec.~\ref{sec:Zero}, we establish the relation between the phenomenological Landau function and microscopic interaction vertices. In Sec.~\ref{sec:examples}, we apply the formalism developed in the previous sections to a number of specific cases. In Sec.~\ref{sec:Pert.LF},   we evaluate explicitly the Landau function to second order in a  finite-range {\em ee} interaction, and confirm the phenomenological form of the Landau function obtained in Sec.~\ref{sec:LF}. The microscopic Landau function from Sec.~\ref{sec:Pert.LF} is used to calculate mass renormalization in Sec.~\ref{sec:results}, where we show that spin-splitting of the masses occurs (to second order in both {\em ee} and SO interactions) as a non-analytic effect, 
resulting from the Kohn anomaly modified by the SO interaction. Simple limiting cases for the spin susceptibility are considered in Sec.~\ref{app:B}. Our conclusions are given in Sec.~\ref{sec:Con}. Some technical details of the calculations are delegated to Appendices A-D.


\section{\label{sec:LF}Landau Function}
We consider a 2D electron system in the presence of linear Rashba SO coupling, described by 
the Hamiltonian \cite{bychkov:1984}
\beq \hat{H}=\hat{H}_{f}+\hat{H}_{\mathrm{int}}=\frac{p^2}{2m}\mathds{1}+\alpha\left(\bs\times\bom{p}\right)\cdot{\bf e}_z+\hat{H}_{\mathrm{int}},
\label{H}
\eeq
where $m$ is the effective electron mass, $\bs$ are the Pauli matrices, ${\bf e}_z$ is the unit normal vector, and $\hat{H}_{\mathrm{int}}$ 
entails
a non-relativistic, density-density {\em ee} interaction.
(Here and in the rest of the paper, $\alpha$ is chosen to be positive, while the Planck and Boltzmann constants 
are set to unity. Also, the index $f$ will denote properties of an interaction-free system.) The space group of Hamiltonian (\ref{H}) is $C_{\infty v}$, and the Rashba SO term is the only combination of the spin and momentum that is  invariant under the symmetry operations of this group.

Eigenvectors and eigenenergies of $\hat{H}_f$ are given by
\beq
\vert s,\bom{p}\rangle=\frac{1}{\sqrt{2}}\left(
\begin{array}{ccc}
1 \\ -ise^{i\theta_{\bp}}
\end{array} 
\right)
\label{basis}
\eeq
and 
\beq
\epsilon^{\bp,f}_{s}=p^2/2m+s\alpha p,\label{e_free}\eeq correspondingly, where $\theta_{\bp}$ is the azimuth of the momentum $\bom{p}$, and $s=\pm 1$ denotes the chirality, i.e. the winding direction of spins around the FS. 

In the rest of the paper, we assume that the chemical potential, $\mu$, is positive and intersects both Rashba subbands. The (bare) Fermi momenta and Fermi velocities of the individual subbands are given by
\bse
\bea
p^f_{\pm}&=&\sqrt{(m\alpha)^2+2m\mu}\mp m\alpha=m\paren{v_0\mp\alpha}\label{ppm}\\
v^f_{F \pm}&=&v_0\equiv  \sqrt{\alpha^2+2\mu/m}.\label{v0}
\eea
\ese

The central object of the FL theory is the Landau function describing the interaction between quasiparticles with momenta
$\bp$ and $\bpp$, and spins $\bs$ and $\bs'$. Since SO coupling 
reduces $SU(2)$ spin symmetry down to $U(1)$, the Landau function will contain more invariants compared to the SU2S case [Eq.~(\ref{LFSU2})]. The task of finding all invariants is simplified by noting that the Rashba coupling is equivalent to the effect of a non-Abelian magnetic field  $\boldsymbol B_R(\bp)=(2\alpha/g\mu_B)\bp\times{\bf e}_z$, where $g$ is the electron's $g$-factor and $\mu_B$ is the Bohr magneton. The most general form of the Landau function must include 
all invariants formed
out of
 the six objects--
$\mathds{1}$, $\bs$ and $\boldsymbol{B}_R(\bp)$ for each of the two quasiparticles--that are invariant under $C_{\infty v}$, time reversal, and permutations of quasiparticles.
The Landau function in the real magnetic field  ${\bf B}$ contains extra \lq\lq Zeeman\rq\rq\/ terms, $\bs'\cdot {\bf B}$ and $\bs\cdot {\bf B}$, as well as their products. \cite{abrikosov:1958} 
In addition to these couplings, the Rashba field can be coupled to spins in more ways because, in contrast to the real-field case,  the cross product ${\bf B}_R(\bp)\times {\bf B}_R(\bpp)\propto \bp\times\bpp$ is non-zero and acts on spins as another effective magnetic field along the $z$ axis.

Exploring all possible invariants, we arrive at the following Landau function
\bwt
\begin{eqnarray}\label{LF}
\hat{f}&=& f^{\mathrm{s}}\mathds{1}\mathds{1}^{\prime}+f^{\mathrm{a}\|}(\hat{\sigma}_x^{{}}\hat{\sigma}^{\prime}_x+\hat{\sigma}_y^{{}}\hat{\sigma}^{\prime}_y)
+f^{\mathrm{a}\perp}\hat{\sigma}_z^{{}}\hat{\sigma}^{\prime}_z
+ \frac 1 2 g^{\mathrm{ph}}\left[\mathds{1}\bs'\times({\boldsymbol{p^{\prime}}}
-{\boldsymbol{p}})\cdot{\bf e}_z-\mathds{1}^{\prime} \bs\times({\boldsymbol{p^{\prime}}}
-{\boldsymbol{p}})\cdot{\bf e}_z\right]\notag\\
&+& \frac 1 2 g^{\mathrm{pp}}\left[\mathds{1}\bs'\times({\boldsymbol{p^{\prime}}}+{\boldsymbol{p}})\cdot{\bf e}_z+\mathds{1}^{\prime} \bs\times({\boldsymbol{p^{\prime}}}+
{\boldsymbol{p}})\cdot{\bf e}_z\right]+h^{(1)}(\bs\times\up\cdot{\bf e}_z)
(\bs'\times\upp\cdot{\bf e}_z)+h^{(2)} (\bs\times\upp\cdot{\bf e}_z)
(\bs'\times\up\cdot{\bf e}_z)
\notag\\
&+&\frac 1 2 h \left[(\bs\times\up\cdot{\bf e}_z)(\bs'\times\up\cdot{\bf e}_z)
+(\bs\times\upp\cdot{\bf e}_z)(\bs'\times\upp\cdot{\bf e}_z)\right],\notag\\
\label{LFR}
\end{eqnarray} 
\ewt
where ${\bf e}_{\bk}\equiv {\bf k}/k$ for any $\bk$. 
The scalar functions $f^{\mathrm{s}}$
through $h$ depend on $p^2$, $p^{\prime 2}$, and $\bp\cdot\bpp$; for brevity, we do not display these dependences in Eq.~(\ref{LFR}).
 Superscripts \lq\lq ph\rq\rq\/ and \lq\lq pp\rq\rq\/ in the \lq\lq $g$\rq\rq\/ terms indicate that, in the perturbation theory, these terms come from the interaction in the particle-hole and particle-particle (Cooper) channels, respectively. The electron momenta enter $\hat f$ via the Rashba field, ${\bf B}_R(\bp)$. In the $h^{(1)}$, $h^{(2)}$, and $h$ terms, which contain bilinear combinations of $\bp$ and $\bpp$, we took the product $pp'$ out of the invariants and absorbed it into the scalar functions. In the $g^{\mathrm{ph}}$ and $g^{\mathrm{pp}}$ terms, which contain the $\bp\mp \bpp$ combinations, we kept the full vectors $\bp$ and $\bpp$. In Sec.~\ref{sec:Zero}, we show how all the terms in Eq.~(\ref{LF}) are generated by the perturbation theory for the interaction vertex.

As is to be expected, SO coupling breaks spin-rotational invariance of the exchange, $\bs\cdot\bs'$, term  of an SU2S FL.  Anisotropy in the exchange part of ${\hat f}$
comes from the combination $\left[\bs\cdot(\boldsymbol{B}_R\times\boldsymbol{B^{\prime}}_R)\right]
\left[\bs'\cdot(\boldsymbol{B}_R\times\boldsymbol{B^{\prime}}_R)\right]
\propto \hat{\sigma}_z^{{}}\hat{\sigma}^{\prime}_z$, which affects only the  $\hat{\sigma}_{z}^{{}}\hat{\sigma}_{z}'$ part of the exchange interaction.
[Here, ${\bf B}_R\equiv {\bf B}_R(\bp)$ and ${\bf B}_R'\equiv {\bf B}_R(\bpp)$.] 
Consequently, anisotropic exchange interaction [the second and third terms in the first line of Eq.~(\ref{LF})] contain different couplings ($f^{\mathrm{a}||}$ and $f^{\mathrm{a}\perp}$) parameterizing the interaction between the in-plane and out-of-plane spins, correspondingly. Recent perturbative calculations \cite{chesi:thesis,zak:2010,zak:2012} show
that anisotropy is of the Ising type, i.e, that $f^{\mathrm{a}\perp}>f^{\mathrm{a}||}$. 
In addition to breaking the rotational symmetry of the exchange interaction, SO coupling generates effective Zeeman terms [with coupling constants $g^{\mathrm{ph}}$, $g^{\mathrm{pp}}$, $h^{(1)}$, and $h^{(2)}$], which depend explicitly on $\bp$ and $\bpp$.
It is worth pointing out that the \lq\lq $h$\rq\rq\/ terms can be written in two equivalent forms.
For example,  the $h^{(1)}$ term can be written down either as $(\bs\times\bom{p}\cdot {\bf e}_z)(\bg{\hat{\sigma}'}\times\bom{p'}\cdot {\bf e}_z)$ or as $(\bs\cdot\bom{p})
(\bg{\hat{\sigma}'}\cdot\bom{p'})$,
which results in equivalent Landau functions upon a redefinition of $f^{\mathrm{a}\|}$.

Spins in Eq.~(\ref{LF}) are not yet represented in any particular basis. However, in order to project electrons' momenta
on the FSs, we need to specify the basis. 
Since the {\em ee} interaction commutes with spins, the spin structure of quasiparticles' states is still governed by the Rashba term,
as long as the interaction is below the threshold value for 
a Pomeranchuk instability.  This argument is nothing more than the usual assumption that symmetries of the system do not change if the interaction is switched on adiabatically.
Microscopic calculations in Sec.~\ref{sec:Zero} 
indeed show that the spin structure of quasiparticles is the same as of non-interacting electrons.
Therefore, we take the chiral basis of Eq.~(\ref{basis})
as the eigenbasis for quasiparticles of a Rashba FL.

\section{\label{sec:m}Effective Mass}
The Landau's derivation of the FL effective mass is restricted to Galilean-invariant systems, and thus cannot be applied to our case because  the SO term breaks Galilean invariance. A generalization of the Landau's derivation for a relativistic FL  \cite{baym:1976,song:2001} is also not applicable here because our Hamiltonian [Eq.~(\ref{H})] is not fully Lorentz invariant either. Therefore, we need to devise an argument which involves neither Galilean nor Lorentz boosts. 
To this end, we notice that the position operator commutes with $\hat{H}_{\mathrm{int}}$ because $\hat{H}_{\mathrm{int}}$ depends only on the positions of electrons but not on their velocities.  Therefore, the velocity operator is the same as in the absence of the {\em ee}  interaction:
\beq \bom{\hat{v}}_j
=-i[\bom{\hat{x}}_j,\hat{H}]=-i[\bom{\hat{x}}_j,\hat{H}_{f}]=\frac{\bom{p}_j}m \mathds{1}+\alpha ({\bf e}_z\times\bs)
\label{v}\eeq
Summing Eq.~(\ref{v}) over all particles, we obtain 
\beq \mathrm {Tr} \sum_{j}{\bom{v}_j}=\frac 1 m \mathrm{Tr}\sum_{j}[ \bom{p}_j\mathds{1}+ m\alpha ({\bf e}_z\times\bs) ].
\label{eq:3}\eeq
 By definition, the total flux of particles is  equal to that of quasiparticles.
The same is true for the total momentum and total spin. Hence the sums over particles can be converted into sums over quasiparticles. Introducing a $2\times 2$ matrix of occupation numbers of quasiparticles  $\hat{n}^{\bom{p}}$, we have
\beq\mathrm {Tr}\int \left[\bom{p}\mathds{1}+m\alpha({\bf e}_z\times\bs)\right] \hat{n}^{\bom{p}}(dp)=m \mathrm{Tr}\int \partial_\bom{p}\hat{\varepsilon}^{\bom{p}} \hat{n}^{\bom{p}}(dp),
\label{flux}
\eeq
where $(dp)\equiv d^2p/(2\pi)^2$,
$\hat{\varepsilon}^{\bom{p}}$ is the energy functional for quasiparticles,
and $\partial_\bom{p}\hat{\varepsilon}^{\bom{p}}$ 
is their velocity (which are also $2\times 2$) matrices).
Since both the Hamiltonian and $\hat n^{\bom{p}}$ are diagonal in the chiral basis, i.e., $(\varepsilon^\bp)_{ss'}=\delta_{ss'}\varepsilon_s^\bp$ and $(\hat n^{\bp})_{ss'}=\delta_{ss'} n^\bp_s$, projecting Eq.~(\ref{flux}) onto this basis, we obtain 
\beq \sum_s\int [\bom{p}+m\alpha({\bf e}_z\times\bs_{ss})] n_{s}^{\bom{p}}(dp)=m \sum_s\int \partial_\bom{p}\hat{\varepsilon}_s^{\bom{p}} n_{s}^{\bom{p}}(dp),
\label{flux1}
\eeq
Now we apply arbitrary independent variations to
the diagonal elements of the occupation number, keeping off-diagonal elements to be zero: $
n_s^{\bom{p}}=n^{0\bom{p}}_s+\delta n_s^{\bom{p}}$, where $n^{0\bom{p}}_s=\Theta(p_s-p)$ at $T=0$ and $p_s$ is the Fermi momentum of fermions with chirality $s$. (Here, superscript $0$ refers to the unperturbed FL  with no variations in the occupation number.) A variation of the diagonal element of the quasiparticle energy 
is related to $\delta n_s^{\bom{p}}$ via 
\beq
\delta\varepsilon_s^{\bom{p}}=\sum_{s'}\int f_{ss'}(\bom{p},\bom{p'})\delta n_{s'}^{\bom{p'}}(dp'),
\label{deltaeps}
\eeq
 where the diagonal part of the Landau function is given by
 \beq
 f_{ss'}(\bp,\bpp)\equiv f_{s,s';s,s'}(\bp,\bpp)
 \eeq
 and $f_{s_1,s_2;s_3,s_4}(\bp,\bpp)$ is obtained by projecting Eq.~(\ref{LF}) onto the chiral basis.
 (It is understood that the unprimed Pauli matrices come with with $s_1s_3$ indices the primed ones with $s_2s_4$ indices).
Varying both sides of Eq.~(\ref{flux1}), we obtain
\bea && \sum_s \int \left[\bom{p}+m\alpha({\bf e}_z\times\bs_{ss})\right] \delta n_{s}^{\bom{p}}(dp)\notag\\
 && =m \sum_s\int \partial_\bom{p}\varepsilon^{0\bom{p}}_s \delta n_{s}^{\bom{p}}(dp)+m \sum_s\int \partial_\bom{p}\delta\varepsilon_s^{\bom{p}} n_{s}^{0\bom{p}}(dp).\notag\\\label{6}
\eea
Integration by parts in the first term of the right-hand side of Eq.~(\ref{6}) gives, upon relabeling $s\longleftrightarrow s'$, $\bp\longleftrightarrow \bpp$ 
and using symmetry $f_{ss'}(\bp,\bpp)=f_{s's}(\bpp,\bp)$, 
\bwt
\bea && \sum_s\int  \left[\bom{p}+m\alpha({\bf e}_z\times\bs_{ss})\right] \delta n_{s}^{\bom{p}}(dp)
=m\sum_s (\partial_\bom{p}\varepsilon^{0\bp}_s)\delta n_s^{\bp} -m\sum_{s,s'}\int f_{ss'}(\bom{p},\bom{p'})(\partial_\bom{p'}n^{0\bom{p'}}_{s'})\delta n^{\bp}_s(dp)(dp').\label{7}
\eea
\ewt
Since variations are arbitrary, the integrands themselves must be equal. Using $\partial_\bom{p'}n^{0\bom{p'}}_{s'}=-\delta(p_{s'}-p')
\bf{e}_{\bpp}
$ 
and $\partial_\bom{p}\varepsilon^{0\bom{p}}_s=
\bp_s/m^{*}_s$ as a definition of the quasiparticle effective mass (with $\bp_s\equiv 
p_s{\bf e}_\bp$), projecting Eq.~(\ref{7}) onto  
${\bf e}_\bp$ and setting $\bp=\bp_s$, we obtain
\bea
&& p_s\left[1+\frac{m\alpha}{p_s}({\bf e}_z\times\bs_{ss})\cdot{\bf e}_\bp\right]\notag\\
&& =p_s \frac m {m^*_s}+m\sum_{s'}\int f_{ss'}(
\bp_s,
\bom{p'})\delta(p_{s'}-p'){\bf e}_\bp\cdot{\bf e}_{\bpp} 
(dp'),\notag\\
\label{mstar3}
\eea
 To simplify the final form of $m^{*}_s$, we define 
\beq 
F_{ss'}(\bom{p},\bom{p'})= 
\nu_{s}
f_{ss'}(\bom{p},\bom{p'})\label{F}
\eeq
with $\nu_s=m^{*}_s/2\pi$ being the density of states of the subband $s$.
All the scalar functions in the $\hat{f}$ matrix can be re-defined in the same fashion, e.g.,
\bea F^{\mathrm s}(\bom{p},\bom{p'})= 
\nu_sf^{\mathrm{s}}(\bom{p},\bom{p'}),
\eea
and similarly for the $F^{\mathrm{a}\|}$, $F^{\mathrm{a}\perp}$, $G^{\mathrm{ph}}$, $G^{\mathrm{pp}}$, $H^{(1)}$, $H^{(2)}$, and $H$ functions. 
With these definitions, Eq.~(\ref{mstar3}) can be written as 
\beq \frac{m^*_s}{m}\left[1+\frac{m\alpha}{p_s}({\bf e}_z\times\bs_{ss})\cdot{\bf e}_\bp\right]=
1+\sum_{s'} \frac{p_{s'}}{p_s} 
F
^{,1}_{ss'},
\label{mstar}
\eeq
where
\beq 
F
^{,\ell}_{ss'}\equiv \int 
F
_{ss'}(\bp_s,\bp_{s'}) \cos(\ell\theta_{\bp\bp'})\frac{d\theta_{\bp\bpp}}{2\pi}
\label{harmonics}\eeq
and $\theta_{\bp,\bpp}$ is the angle between $\bp$ and $\bpp$. 
Using explicit forms of the Pauli matrices in the chiral basis  (\ref{basis})
\begin{widetext}
\beq \hat{\sigma}_x =\left(\begin{array}{ccc}\sin\theta_{\bp} & i\cos\theta_{\bp} \\-i\cos\theta_{\bp} & -\sin\theta_{\bp} \
\end{array} \right),
\hat{\sigma}_y =\left(\begin{array}{ccc}-\cos\theta_{\bp} & i\sin\theta_{\bp} \\-i\sin\theta_{\bp} & \cos\theta_{\bp} \
\end{array} \right),
\hat{\sigma}_z =\left(\begin{array}{ccc}0&1 \\1&0 \
\end{array} \right),
\label{chiralpauli}
\eeq
we find
$({\bf e}_z\times\bs_{ss'})\cdot{\bf e}_\bp=s\delta_{ss'}$. (This result simply follows form the fact the Hamiltonian is diagonal in the chiral basis.)
Projecting Eq.~(\ref{LF}) onto the chiral basis, we obtain
\bea
F
_{s,s'}{(\bom{p}_s,\bom{p}_{s'})}&=&F^{\mathrm{s}}(\bom{p}_s,\bom{p}_{s'})+F^{\mathrm{a}\|}
(\bom{p}_s,\bom{p}_{s'})ss'\cos\theta_{\bp\bpp}
+G^{\mathrm{ph}}(\bom{p}_s,\bom{p}_{s'})\left[sp_s+s'p_{s'}-(sp_{s'}+s'p_{s})\cos\theta_{\bp\bpp}\right]/2\notag\\
&+& G^{\mathrm{pp}}(\bom{p}_s,\bom{p}_{s'})\left[sp_s+s'p_{s'}+(sp_{s'}+s'p_{s})\cos\theta_{\bp\bpp}\right]/2
+H^{(1)}(\bom{p}_s,\bom{p}_{s'})ss'\cos(2\theta_{\bp\bpp})
\notag\\&+&H^{(2)}(\bom{p}_s,\bom{p}_{s'})ss'\cos^2\theta_{\bp\bpp}
+H(\bom{p}_s,\bom{p}_{s'})ss'\cos\theta_{\bp\bpp}.\label{12}
\eea
\end{widetext}
Notice that the $f^{\mathrm{a}\perp}$ component of the Landau function in Eq.~(\ref{LF}) does not enter Eq.~(\ref{12}), because  $\hat{\sigma}_z $ is off-diagonal in the chiral basis and hence $f^{\mathrm{a}\perp}$ drops out from the diagonal part of the Landau function. 
We see that, in contrast to 
the case of a SU2S FL,
where the effective mass contains only the $\ell=1$ harmonic of $F^{\mathrm{s}}$,  
the $\ell=1$ harmonic of $F_{ss'}$ in Eq.~(\ref{mstar}) contains also both lower and higher harmonics of partial components of $F_{ss'}$. Namely, $F_{ss'}^{,1}$ contains
the $\ell=0$ and $\ell=2$ harmonics of $F^{\mathrm{a}\|}$, $G^{\mathrm{ph}}$, $G^{\mathrm{pp}}$, and $H$, and the $\ell=3$ harmonics of $H^{(1)}$ and $H^{(2)}$. 
The final result for the effective masses can be cast into the following form:
\beq 
\frac{m^*_s}{m_s}=
1+\sum_{s'} \frac{p_{s'}}{p_s} F^{,1}_{ss'},
\label{mstar1}
\eeq
where 
\beq
m_s\equiv 
\frac{m}{1+s\frac{m\alpha}{p_s}}.
\label{ms}
\eeq

Notice that the Luttinger theorem\cite{luttinger:1961b} guarantees only that the total area of the FS, $\pi(p_+^2+p_-^2)$,  is not renormalized by the interaction but says nothing about the partial areas. Therefore, $p_{+}$ and $p_{-}$ in Eqs. (\ref{mstar1}) and (\ref{ms}) must be considered as renormalized Fermi momenta.
The mass $m_s$ in Eq.~(\ref{ms}) has the same form as the effective mass of free Rashba fermions $m^{f}_s=p_s^f/\partial_p\epsilon_s^{f}=
m/(1+sm\alpha/p_s^f)$, except for the Fermi momenta of an interaction-free system, $p_s^f$, are now replaced by the renormalized ones, $p_s$. From Eqs.~(\ref{mstar1}) and (\ref{ms}), we see that a chiral FL liquid is different from an SU2S
 one is that the effective mass (and, as will see later, other FL quantities) is not determined only by the Landau parameters: one also needs to know the {\em renormalized} Fermi momenta. 
 In this respect, a chiral FL is similar to other two-component FLs. \cite{meyerovich:1983,bedell:1991}
 
 Due to hidden symmetry of the Rashba Hamiltonian--conservation of the square of the velocity\cite{rashba:2005}-- the Fermi 
 velocities of free Rashba fermions with opposite chiralities are the same, cf. Eq.~(\ref{v0}). (However, the masses are different because of the difference in the Fermi momenta). It is reasonable to expect that the {\em ee} interaction breaks hidden symmetry and leads to splitting of the subbands' velocities. However, no such splitting occurs to first order in SO coupling, \cite{saraga:2005} even if the {\em ee} interaction is accounted for exactly.\cite{chesi-theorem} In the remainder of this section, we present a microscopic calculation of the effective masses and show that the velocities are indeed split even  by a weak {\em ee} interaction but only at larger values of SO coupling.

We adopt a simple model of a screened Coulomb potential in the high-density limit ($r_s\ll 1$). The leading-order result for mass renormalization, $\left(m^*/m-1\right) \sim r_s\ln r_s$, is obtained already for a static screened Coulomb potential, 
\beq U^{\mathrm{TF}}_q=\frac{2\pi e^2}{q+p_{\mathrm{TF}}},\label{TF}
\eeq where $p_{\mathrm{TF}}=\sqrt{2}r_s  p_F$ is the inverse Thomas-Fermi screening radius. SO coupling does not affect the form of $U^{\mathrm{TF}}_q$ because the total density of states remains the same as long as both Rashba subbands are occupied.   The self-energy of an electron in the subband $s$ is given by 
\bea \Sigma_s(\bp,\omega)=&-&\frac12\sum_{s^{\prime}}\int (dq)\int\frac{d\Omega}{2\pi}
U^{\mathrm{TF}}_q
\\&\times&[1+ss^{\prime} \cos(\theta_{\bom{p+q}}-\theta_{\bom{p}})]g_{s^{\prime} }(\bom{p+q},\omega+\Omega),\notag
\label{selfenergy}
\eea
where 
\beq 
g_{s}(\bp,\omega)=\frac{1}{i\omega-\epsilon_s^{\bp,f}+\mu}
\label{free_g}
\eeq
is the free (Matsubara) Green's function in the chiral basis with $\epsilon_s^{\bp,f}$ given by Eq.~(\ref{e_free}).
At small $r_s$, typical momenta transfers are small: $q\sim p_{\mathrm{TF}}\ll p_F$; therefore, the difference between $\theta_{\bp+\bq}$ and $\theta_{\bp}$ can be neglected and intersubband transitions (with $s'=-s$) drop out.  Simplifying Eq.~(\ref{selfenergy}) in the way specified above and subtracting the self-energy evaluated at the FS, we obtain
\bea
\Delta\Sigma_s &\equiv&\Sigma_s{(\bp,\omega)}-\Sigma_s{(\bp_s,0)}
\\&=&-\sum_{q,\Omega } {U^{\mathrm{TF}}_q}\left[g_{s}(\bom{p+q},\omega+\Omega)-g_{s}(\bom{p}_s+\bom{q},\Omega)\right].\notag
\eea
Integration over $\Omega$ gives (for $\varepsilon^{\bp,f}_s\equiv \epsilon^{\bp,f}-\mu>0$)
 \bea
\Delta\Sigma_s &=&\int_{-\frac{\varepsilon_{s}^{\bp,f}}{v_0 q}<\cos\theta_{\bp} < 0} \frac{d^2q}{(2\pi)^2} U^{\mathrm{TF}}_q
\label{sigma_a}
\\
&=&
\int_{\frac{\varepsilon_{s}^{\bp,f}}{v_0}}^{p_s}\frac{dq q}{2\pi^2} U^{\mathrm{TF}}_q
\bra{\cos^{-1}\paren{-\frac{\varepsilon_{s}^{\bp,f}}{v_0 q}}-\frac{\pi}{2}},\notag\eea
where $v_0$ is the (common) Fermi velocity of Rashba subbands given by Eq.~(\ref{v0}).
Expanding Eq.~(\ref{sigma_a}) to linear order in $\varepsilon_{s}^{\bp,f}$ and solving the resulting integral over $q$ to logarithmic accuracy, we find
\beq
\Delta\Sigma_s=\frac{\varepsilon_{s}^{\bp,f}}{\pi v_0}e^2\ln\frac {p_s}{p_{\mathrm{TF}}},
\label{mass-splitting}
\eeq
where the subband Fermi momentum, $p^f_s$, was chosen as the upper cutoff. Since the electron $Z$-factor is not affected to this order of the perturbation theory, \cite{saraga:2005} Eq.~(\ref{mass-splitting}) immediately gives renormalized Fermi velocities as 
\beq
\frac{v_{\pm}}{v_0}=1+\frac{e^2}{\pi v_0}\ln \frac{p_{\pm}^f}{p_{\mathrm{TF}}}.
\label{mstara}
\eeq
This result is the same as for a 2D electron gas without SO coupling, except for that the Fermi momentum, entering the logarithmic factor,
is specific for a given subband. 
The mechanism of such splitting is similar to Overhauser-type splitting of effective masses in a partially spin-polarized metal.\cite{overhauser:1971,suhas:2005,wei:2011} Equation (\ref{mstara}) is already enough to produce spin-splitting of the Fermi velocities 
\beq \frac{v_{+}-v_{-}}{v_0}
=\frac{e^2}{\pi v_0}\ln \frac{p_{+}^f}{p_{-}^f}.
\label{vsplit}\eeq
One needs to keep in mind, however, that Eq.~(\ref{vsplit}) was obtained as a difference of two formulas (\ref{mstara}), each of which contains a large logarithmic factor, and is thus valid if not only each of the logarithms but also their difference is large. This requires that $p^f_+ \ll p^f_-$. Such a case is realized for strong SO interaction, when $m\alpha^2/2\gg \mu >0$. In this case, $p^f_{+}\approx \mu/\alpha\ll p^f_{-}\approx 2m\alpha$, $v_0\approx \alpha$, and thus
\beq
\frac{v_+-v_-}{v_0}\approx -\frac{e^2}{\pi \alpha}\ln\frac{m\alpha^2}{\mu}.
\eeq
 In Appendix \ref{app:A}, we compute the self-energy beyond the logarithmic accuracy 
and show that Eq.~(\ref{vsplit}) is reproduced as a leading term.
Also, we present there a simple way to generate an analytic expansion of the self-energy in $\alpha$, which confirms previous
results for velocity splitting. \cite{saraga:2005,aasen:2012}  
We return to the issue of mass renormalization in Sec.~\ref{sec:results}, where we show 
that a non-analytic in SO result  for velocity splitting comes from the Kohn anomaly modified by the SO interaction.
 
\section{\label{sec:Thermo}Thermodynamic properties}
In principle, Eq.~(\ref{LF}) enables one to compute all thermodynamic properties of a chiral FL. In this section, we illustrate how this program can be carried out in the charge sector by calculating the isothermal compressibility (Sec.~\ref{sec:Comp} and deriving Pomeranchuk stability conditions (Sec.~\ref{sec:Pom}).

\subsection{\label{sec:Comp}Compressibility}
The $T=0$  compressibility of a FL is given by
\beq \kappa = \frac 1 {N^2} \left( \frac{\partial N}{\partial \mu} \right)_{\mathcal{N}}, \eeq
where $N=\sum_s \int (dp)n_s^{\bp}= \sum_sN_s$ is the number density
 and $\mathcal{N}$  is the total particle number.

The compressibility of a chiral FL is obtained by a simple generalization of the original Landau's argument. 
\cite{agd:1963}
If the exact dispersions of Rashba subbands are $\epsilon_{\pm}(p)=\varepsilon^{\bp}_{\pm}+\mu$, the  chemical potential is defined as
$\mu=\epsilon_{+}(p_+)=\epsilon_{-}(p_-)$. The variation of the chemical potential consists of two parts: the first one is due to a change in the Fermi momenta in each of the two subbands and the second one is due to a variation of the quasiparticles' dispersions. 
Since the variations of the chemical potential are the same for both subbands, we have
 \beq
 \delta\mu=\frac{\partial \epsilon_{+}(p)}{\partial p}\Big\vert_{p_+}\delta p_{+}+\delta\varepsilon^{\bp}_{+}=\frac{\partial \epsilon_{-}(p)}{\partial p}\Big\vert_{p_-}\delta p_{-}+\delta\varepsilon^{\bp}_{-},
 \label{deltamu}
 \eeq
 where $\delta\varepsilon^{\bp}_{\pm}$ are given by Eq.~(\ref{deltaeps})
 and $\delta p_{s}=(2\pi/p_s)\delta N_s$.
Assuming that the variations of $n^{\bp}_{\pm}$ in the $p$-space are localized near the corresponding Fermi momenta, we integrate $n^{\bp}_{\pm}$ over the magnitude of the momenta to obtain
 $\varepsilon^{\bp}_{s}=\sum_{s'} F^{,0}_{ss'}\delta N_{s'}/\nu_{s}$.  
 Combining Eq.~(\ref{deltamu}) with the constraint $\delta N=\delta N_{+}+\delta N_{-}$, we solve 
 the resulting system for $\delta N_{\pm}$ in terms of $\delta\mu$ and $\delta N$ to obtain
 \beq
 \kappa=\frac{1}{N^2} \frac{\nu+\nu_{+}\left(F_{--}^{,0}-F_{-+}^{,0}\right)+\nu_{-}\left(F_{++}^{,0}-F_{+-}^{,0}\right)}
 {\left(1+F_{++}^{,0}\right)\left(1+F^{,0}_{--}\right)-F_{+-}^{,0}F_{-+}^{,0}}
 \label{comp}
 \eeq
 with $\nu=\sum_s\nu_s$. The result for the 
 SU2S 
 case is recovered in the limit when the 
 intra- and intersubband components of $F^{,0}_{ss'}$ become the same. Indeed, subsitituting  $F^{,0}_{++}=F^{,0}_{--}=F^{,0}_{+-}=F^{,0}_{-+}=F^{\mathrm{s},0}/2$ into Eq.~(\ref{comp}), we obtain the familiar result~\cite{agd:1963, *nozieres:1966, *lifshitz:1980} $\kappa=\nu/\left[(1+F^{\mathrm{s},0})N^2\right]$.
 
 Notice that Eq.~(\ref{comp}) is different from the compressibility of a two-component FL 
 \cite{ashcroft:1981} because of the additional assumption used in Ref.~\onlinecite{ashcroft:1981}, namely, that not only the total number of particles but also the numbers of particles of a given type remains constant under compression. While this assumption
 is justified in the context of a degenerate electron-proton plasma considered in Ref.~\onlinecite{ashcroft:1981}, it is not applicable to our case
 when particles of opposite chiralities can be exchanged between the Rashba subbands, and thus the number of particles with given chirality is not conserved.

\subsection{\label{sec:Pom} Stability conditions}
To obtain stability condition in the charge sector, we must require that the Gibbs free energy, $\Omega$, be minimal with respect to arbitrary deformations of the FSs which do not affect their spin structure:\cite{pomeranchuk:1958}
\beq \bar{p}_s(\theta)-p_s=\sum_{n=-\infty}^{n=\infty}\Delta_{s,n}e^{in\theta}. \eeq
Corresponding variations in the occupation numbers can be written as
\beq \delta \hat{n}_{s}^{\bom{p}}=\Theta(\bar{p}_s(\theta)-p)-\Theta(p_s -p), \eeq
A change in the free energy
\beq
\delta\Omega= \mathrm{Tr} \int \hat{\varepsilon}^{\bp} \delta \hat{n}^{\bom{p}} (dp)+ \frac 1 2 \mathrm{Tr} \mathrm{Tr'}\int \hat{f}(\bom{p},\bom{p'}) \delta \hat{n}^{\bom{p}} \delta \hat{n}^{\bom{p'}}(dp)(dp')
\eeq
must be positive definite with respect to such variations. Using $\hat{\varepsilon}_{ss'}=\delta_{ss'}\frac{p_s}{m^*_s}(p-p_s)$,
the equation for $d\Omega$ is simplified to
\beq \delta\Omega=\sum_{s,n}\frac{p_s^2}{4\pi{m^*_s}}\Delta_{s,n}^2+\sum_{s,s',n} \frac{p_s p_{s'}}{4\pi{m^*_s}}\hat{F}^{,n}_{ss'}\Delta_{s,n}\Delta_{s',-n}.\eeq
We thus arrive at the following stability conditions in the charge sector
\begin{subequations}
\bea
&&1+F_{++}^{,n}>0,\label{pom_a}\\
&&1+F_{--}^{,n}>0,\label{pom_b}\\
&&(1+F_{++}^{,n})(1+F_{--}^{,n})>F_{+-}^{,n}F_{-+}^{,n}.\label{pom_c}
\eea
\end{subequations}
for any $n\geq 0$.
Conditions (\ref{pom_a}) and (\ref{pom_b}) are the same as for the single-component case, while condition (\ref{pom_c}) indicates that the two-component FL is stable only if the inter-band interaction is sufficiently weak.\cite{ashcroft:1981} For $n=0$, condition (\ref{pom_c}) coincides with the condition that the denominator of Eq.~(\ref{comp}) is positive.

Note that the divergence of $\kappa$ in gated systems does not manifest a thermodynamic instability as the total energy of the system remains positive due to a compensating effect of the classical charging energy of a parallel-plate capacitor. Instead, this divergence indicates the onset of the \lq\lq negative compressibility\rq\rq\/ regime, in which the measured capacitance is larger than the geometric one. Electron and, especially, hole FLs in semiconductor heterostructures are typically studied in this regime. \cite{[{An extensive list of references on both theoretical and experimental studies of negative compressibility can be found in, e.g.,~}] skinner:2013}
\subsection{\label{sec:Spin}Spin susceptibility}
\subsubsection{Fermi-liquid formalism}
\label{sec:chi_FL}
As mentioned in Sec.~\ref{sec:Int}, the subtleties of a non-SU2S FL are all in the spin sector. 
To illustrate this point more clearly, we proceed with evaluating the spin susceptibility. 
In order to do this within the framework of FL theory, one needs to properly find the change in the occupation numbers of quasiparticles due to an external magnetic field. 

Suppose that an 
external magnetic field is in the ${\bf e}_z$ direction and of magnitude much smaller than the  effective  SO field 
\beq g\mu_B H \ll \alpha p_F, \eeq
where $g$ is effective $g$-factor. 
(We neglect here the diamagnetic response of electrons. In principe, the spin part of the total susceptibility can be measured, e.g., via the Knight shift or neutron scattering.)
In the absence of an external field,
quasiparticles occupy chiral states $|s,\bp\rangle$ (with $s=\pm 1$) filled up to the Fermi momenta $p_{s}$.
 The presence of a magnetic field affects the spin structure of quasiparticle states. Suppose that the states in the presence of the field
are $|\bp,h\rangle$ (with $h=\pm 1$) filled up to the Fermi momenta $\tilde{p}_{h}$.
(To distinguish between the quantities in the absence and in the presence of the field, we denote the latter with a tilde over a corresponding symbol.) The energy functional of quasiparticles in the absence of the field is diagonal in the $|s,\bp\rangle$ basis with eigenvalues $\varepsilon_{\pm}^\bp$:
\beq \varepsilon^{\bp}_{ss'}=\langle s,\bp|\hat{\varepsilon}^\bp|s',\bp\rangle=\delta_{ss'}\varepsilon_s^\bp.
\eeq
The occupation number
in the absence of the field is also diagonal
\beq n^\bp_{ss'}=\langle s,\bp|\hat{n}^\bp|s',\bp\rangle=\delta_{ss'}n_s^\bp,
\eeq
where $n_{s}^\bp=\Theta(p_s-p)$. In the presence of the field, the $|s,\bp\rangle$ basis is not an eigenstate of the Hamiltonian. Therefore, the Zeeman part of the energy functional is not diagonal in this basis:
\footnote{In the SU2S case, the choice of the spin quantization axis is arbitrary, and the Zeeman energy of a quasiparticle can always be written in the diagonal form as $\sigma_z H$. In the chiral case, the choice of the spin quantization axis is unique, and the Zeeman energy cannot always be reduced to the diagonal form.}
\beq {\tilde \varepsilon}^\bp_{ss'}=\varepsilon_{ss'}^\bp+\delta\varepsilon_{ss'}^\bp,\eeq
where \beq\delta\varepsilon_{ss'}^\bp= \frac 1 2 g^*(\bp)\mu_BH\sigma^z_{ss'},\label{delta_eps}\eeq
$g^*(\bp)$ is the renormalized $g$-factor which depends on the electron momentum,  
and $\sigma^z$ in the chiral basis is given by the last formula in Eq.~(\ref{chiralpauli}). (At this point, our analysis differs from that of Ref.~\onlinecite{fujita:1987}, where the occupation number was assumed to be diagonal. On the other hand, our approach is similar to that of Ref.~\onlinecite{meyerovich:1992a}, which also emphasizes the necessity of including the off-diagonal components of $\hat n$ for the case of a partially spin-polarized FL.)    Both the energy functional of quasiparticles in the presence of the field and their occupation number are diagonal in the $|h,\bp\rangle$ basis
\bea {\tilde \varepsilon}^{\bp}_{hh'}&=&\langle h,\bp|\hat{\varepsilon}^\bp|h',\bp\rangle=\delta_{hh'}{\tilde
 \varepsilon}^\bp_h\notag\\
{\tilde n}^\bp_{hh'}&=&\langle h,\bp|\hat{n}^\bp|h',\bp\rangle=\delta_{hh'}{\tilde n}^\bp_h
\eea
There exists a unitary matrix ${\hat U}$ that diagonalizes ${\tilde \varepsilon}^\bp_{ss'}$ 
\beq
{\tilde \varepsilon}^\bp_{hh'}= U^{\dagger}_{hs}{\tilde \varepsilon}^\bp_{ss'}U_{s'h'}.
\eeq
or, equivalently,
\beq
 {\tilde \varepsilon}^\bp_{ss'}= U^{\dagger}_{sh}{\tilde \varepsilon}^\bp_{hh'}U_{h's'}.
\eeq
To first order in $H$,
\beq {\hat U}=\mathds{1}+H{\hat M}+{\mathcal O}(H^2)
,\eeq
and ${\hat M}$ is an anti-Hermitian matrix parameterized as
\bea
{\hat M}=\left(\begin{array}{ccc}ia& \beta \\-\beta^*&ib \
\end{array}\right)\label{M}
\eea
with real $a$ and $b$. The matrix ${\hat M}$ is determined from the condition that the linear-in-$H$ part of
${\tilde \varepsilon}^\bp_{ss'}$ is given by Eq.~(\ref{delta_eps}). For a diagonal ${\tilde\varepsilon}_{hh'}^{\bp}$, the diagonal elements of 
$U^{\dagger}_{sh}{\tilde \varepsilon}^\bp_{hh'}U_{h's'}$ are equal to zero, while Eq.~(\ref{delta_eps}) contains only off-diagonal elements. This fixes $\beta$ in Eq.~(\ref{M}) to be real and equal to 
\beq \beta=\frac {g^*(\bp)\mu_B}{2}\frac{1}{\varepsilon_+^\bp-\varepsilon^\bp_-}. \eeq

The diagonal components of ${\hat M}$ remain undefined to first order in $H$ but, using $\hat{U}=e^{H\hat{M}}$, they are found to be zero to all higher orders as well. %
Consequently, $\hat M$ can be written as
\beq {\hat M}=\beta\left(\begin{array}{ccc}0 & 1 \\-1 &0 \
\end{array} \right).\eeq
Since the same matrix ${\hat U}$ diagonalizes also the occupation number, we have
\beq {\tilde n}^\bp_{ss'}= U^{\dagger}_{sh}{\tilde n}^{\bp}_{hh'}U_{h's'}.\eeq
To find a linear-in-$H$ correction to ${\tilde n}^\bp_{ss'}$, it suffices to approximate ${\tilde n}^{\bp}_{hh'}$ as $\mathrm{diag}\paren{n^\bp_+,n^\bp_-}$ with field-independent $n^{\bp}_{\pm}$. Then 
\bea
\delta {\tilde n}^\bp_{ss'}&=&\beta\left[\hat M^\dagger\mathrm{diag}\paren{n^\bp_+,n^\bp_-} +\mathrm{diag}\paren{n^\bp_+,n^\bp_-}\hat M\right]\notag\\
&=&\beta H(n^\bp_+-n^\bp_-)\hat{\sigma}^z_{ss'}.
\eea
It is at this point when the main difference between the SU2S and chiral FLs occurs: for the former, the change in the occupation number is localized near the FS; for the latter, it is proportional to a {\em difference} of the occupation numbers in the absence of the field, and is thus finite for {\em all} momenta in between the FSs of chiral subbands.

With this remark in mind, we still proceed with a derivation of the equation for the renormalized $g$-factor. As in the SU2S case, we decomposing the energy variation into the Zeeman part and the part due to a variation in the occupation numbers:
\bea 
\delta{\tilde \varepsilon}^\bp_{ss'}&=&\frac 1 2 g^*(\bp)\mu_B H\sigma^z_{ss'}\notag\\
&=&\frac 1 2g\mu_BH\sigma^z_{ss'}+\sum_{tt'}\int (dp') f_{st,s't'}(\bp,\bpp)\delta {\tilde n}^{\bpp}_{t't}.\notag\\
\label{delta_n}\eea
Since $\delta {\tilde n}^{\bpp}_{t't}$ is proportional to $\sigma^z_{t't}$, the sum over $t$ and $t'$ selects the only component of the Landau function in Eq.~(\ref{LF}) that contains $\sigma^z$, i.e., $f^{\mathrm{a}\perp}$. The renormalized $g$-factor remains isotropic in the momentum space until a Pomeranchuk instability is reached: $g^*(\bp)=g^*(p)$. With these simplifications, Eq.~(\ref{delta_n}) reduces to
\bea \frac{g^*(p)}{g}&=&1+4\int (dp') f^{\mathrm{a}\perp}(\bp,\bpp)(n^{\bpp}_+-n^{\bpp}_-)\frac{\beta}{\mu_Bg}\notag\\
&=&1-2\int_{p_+}^{p_-} \frac {dp'p'}{2\pi}\frac{f^{\mathrm{a}\perp,0}(p,p')}{\varepsilon^{\bpp}_{+}-\varepsilon_-^{\bpp}}
\frac{g^*(p')}g,\label{g}\eea
where $f^{\mathrm{a}\perp,0}(p,p')$ is the $\ell=0$ angular harmonic of $f^{\mathrm{a}\perp}$
(we remind that $\varepsilon^{\bp}_{\pm}$ depend only on the magnitude of $\bp$). 
In contrast to the SU2S case, the equation for $g^*(p)$ remains an integral one, even if the external momentum
is projected onto one of the FSs. (The occurrence of integral rather than algebraic equations is also typical for the theory of a partially spin-polarized FL; see, in particular, Ref.~\onlinecite{meyerovich:1992a}.)

The out-of-plane spin susceptibility 
is then found as
\bwt
\bea \chi_{zz}&=&
\frac {g\mu_B } {2H} \sum_{ss'}\int (dp)\sigma^z_{ss'}\delta {\tilde n}_{s's}
=\frac {g\mu_B^2} {2}\int_{p_+}^{p_-} \frac {dpp}{2\pi}\frac{g^*(p)}{\varepsilon^\bp_+-\varepsilon_-^\bp}\notag\\
&=&\frac{g^2\mu_B^2}{2}\left[\int _{p_+}^{p_-}\frac{dp p}{2\pi}\frac{1}{\varepsilon^\bp_{+}-\varepsilon^\bp_{-}}
-2\int^{p_-}_{p_+} \frac{dpp}{2\pi} \int^{p_-}_{p_+} \frac{dp' p'}{2\pi}\frac{f^{\mathrm{a}\perp,0}(p,p')}{\left(\varepsilon^{\bp}_+-\varepsilon_-^{\bp}\right)\left(\varepsilon^{\bpp}_+-\varepsilon_-^{\bpp}\right)}\frac{g^*(p')}{g}\right].
\label{chiz}\eea
\ewt
In the last line, we used Eq.~(\ref{g}) for $g^*(p)$.  

Following the same steps, the in-plane spin susceptibility ($\chi_{xx}=\chi_{yy}$) can be shown to contain contributions both from each of the FSs and from the interval in between the two.  However, since there is always a finite contribution from the damped states in between the two FSs, the problem remains the same as for $\chi_{zz}$. In addition,
the in-plane spin susceptibility contains angular harmonics of all Landau parameters except for $f^{\mathrm{a}\perp}$. The final formula for $\chi_{xx}$ is not sufficiently instructive to be presented here.
\subsubsection{Physical interpretation}
\label{sec:TR}
Equation (\ref{chiz}) (and a similar formula for $\chi_{xx}$)
show that the spin susceptibility of a chiral FL has a different physical origin compared to the Pauli susceptibility of a SU2S FL. Indeed, while the latter is determined by the states right on the FS, the former comes from a finite interval of states in between the two spin-split FSs. Moreover, this difference is present already for non-interacting electrons, when the second term in Eq.~(\ref{chiz}) is absent. Physically, this difference comes about because, in the SU2S case, the induced magnetic moments of almost all spin-up and spin-down states cancel each other, except for the states with momenta in between the Zeeman-split FSs, where the spin-down states are empty but the spin-up states are occupied. In a weak magnetic field, the width of this region is proportional to the field, and this is why  only the states at the FS matter in the $H\to 0$ limit. In the chiral case, SO splitting is present even at zero magnetic field, and the whole interval $p_+<p<p_-$ contributes to magnetization. This property becomes especially clear in the Kubo-formula approach, which gives for the spin susceptibility in the non-interacting case \cite{zak:2010}
\bse
\bea
\chi^f_{zz}&=&\frac{g^2\mu_B^2}{8\pi\alpha}\int^\infty_0 dp\left[n_{-}(p)-n_{+}(p)\right]\label{chizzf}
\\
\chi^{f}_{xx}&=&\chi^f_{yy}=\frac 12\chi^f_{zz} - \frac {g^2\mu_B^2}{8}\int \frac{d^2p}{(2\pi)^2}\left(\frac{\partial n_+}{\partial\epsilon^{\bp,f}_{+}}+\frac{\partial n_-}{\partial\epsilon^{\bp,f}_{-}}\right),\notag\\
\label{chixxf}
\eea
\ese
where $n_{\pm}$ are the occupation numbers of the  Rashba subbands. The difference between Eqs.~(\ref{chizzf}) and (\ref{chixxf}) can be traced back to the difference between the corresponding Pauli matrices in the chiral basis [Eq.~(\ref{chiralpauli})]. Indeed, since $\hat\sigma_z$ is off-diagonal, $\chi_{zz}^f$ comes only from inter-subband transitions, while $\hat\sigma_x$ and $\hat\sigma_y$ contain both diagonal and off-diagonal parts, and hence $\chi_{xx}^f$ and $\chi_{yy}^f$ come from both intra- and inter-subband transitions. At $T=0$, the integrand in $\chi_{zz}^f$  is non-zero only in the interval $p_+^f\leq p\leq p_{-}^f$. The second term in Eq.~(\ref{chixxf}) is the FS contribution.
 Note that, despite the difference in the intermediate results, the finite results for the in- and out-of-plane components at $T=0$ are the same:  $\chi_{xx}^f=\chi_{yy}^f=\chi_{zz}^f=g^2\mu_B^2/4\pi$, i.e., not only the non-interacting spin susceptibility  is fully isotropic but it also coincides with the spin susceptibility in the absence of SO coupling.~\cite{zak:2010} This is a special property of a linear-in-momentum SO interaction.
(In Appendix~\ref{app:Bfree}, we show how Eq.~(\ref{chizzf}) can be obtained in a thermodynamic approach.) 

To apply Eq.~(\ref{chiz}) to the FL-case, one needs to know (renormalized) dispersions, $\varepsilon^{\bp}_{\pm}$,
and the $f^{\mathrm{a},\perp}$ component of the Landau function in the entire interval of momenta in between the FSs. If SO coupling is not weak,  one thus needs to know the properties of chiral quasiparticles far away from their respective FSs, which is outside the scope of the FL theory. This problem is not merely technical but fundamental because quasiparticles decay away from their FSs, and one thus cannot formulate the FL theory as a theory of well-defined quasiparticles. This does not mean that a chiral electron system is a non-FL; on the contrary, all microscopic calculations (cited in Sec.~\ref{sec:Int}) as well as experimental evidence point at the FL-nature of chiral electron systems. However, they are FLs with  the spin sector that cannot be described in the framework of the FL theory.  This conclusion is not restricted to the case of a FL with the Rashba SO interaction but is also true for any  
non-SU2S FL, e.g., ferromagnetic and partially spin-polarized FLs, \cite{herring,meyerovich:1992a, meyerovich:1994, *meyerovich:1994b} Nevertheless, we show in 
Sec.~\ref{app:B} that
Eq.~(\ref{chiz}) reproduces correctly both limiting cases of a weak {\em ee} interaction and a weak SO coupling. 

We should also point out that the states with momenta in the interval $p<p_{+}$, where both Rashba subbands are occupied, do not contribute to the spin susceptibility, which implies that the induced magnetic moments of the occupied Rashba subbands cancel each other.  This cancellation is not accidental but follows from a non-trivial time-reversal symmetry at a fixed magnetic field.  Indeed, a contribution  to the total spin susceptibility from the occupied electron states of both chiralities and with given momentum $\bp$ can be written as
\beq
\chi_{ii}(\bp)=\lim_{H_i\to 0}\frac{1}{H_i}\frac{ g\mu_B}{2}\sum_s\langle s,\bp;{\bf H}|\hat\sigma_i|s,\bp;{\bf H}\rangle ,
\label{magn}
\eeq
where $|s,\bp;{\bf H}\rangle$ are the basis states in the presence of the field.
To linear order in the out-of-plane magnetic field, the energies of states do not change, while
the basis vector in Eq.~(\ref{basis}) changes to
\bea
|s,\bp, {\bf H}\rangle=\frac{1}{\sqrt{2}}\left(\begin{array}{ccc}
1\\
-is\left[1-s\frac{g\mu_B H_z}{2\alpha p}\right]e^{i\theta_\bp}
\end{array}
\right).
\label{vecz}\eea
If the field is in the plane, e.g., along the $x$-axis, the energies acquire a linear-in-$H$ correction $\epsilon^{\bp,f}_s=p^2/2m+s\alpha p+s\sin\theta_\bp g\mu_BH_x/2$, while the basis vector becomes
\bea
|s,\bp, {\bf H}\rangle=\frac{1}{\sqrt{2}}\left(\begin{array}{ccc}
1\\
-is\left[1+i\cos\theta_{\bp}\frac{g\mu_B H_x}{2\alpha p}\right]e^{i\theta_\bp}
\end{array}
\right).\notag\\
\label{vecx}
\eea
(To find the induced magnetization, one does not need to take into account the corresponding changes in the normalization coefficients.) For both orientations of the field, the states $|s,\bp;{\bf H}\rangle$ and $|s,-\bp, -{\bf H}\rangle$ are the components of the Kramers doublet which correspond to the same energy. Indeed, applying the time-reversal operator, $\hat{\cal  K}=\hat\sigma_y \hat C$, where $\hat C$ is complex conjugation operator, to spinor (\ref{vecz}), we obtain
\bea
\hat{\cal  K}|s,\bp,{\bf H}\rangle&=&\left(\begin{array}{ccc}
1\\
is\left[1+s\frac{g\mu_B H_z}{2\alpha p}\right]e^{i\theta_\bp}
\end{array}
\right)
=|-s,\bp;{\bf H}\rangle\notag\\
&&=\left(\begin{array}{ccc}
1\\
-is\left[1+s\frac{g\mu_B H_z}{2\alpha p}\right]e^{i\theta_{-\bp}}\end{array}
\right)=|s,-\bp;-{\bf H}\rangle,\notag\\
\label{TR}
\eea
up to an unessential overall phase factor, and similarly for spinor (\ref{vecx}). The second line in Eq.~(\ref{TR}) expresses
full time-reversal symmetry,  which involves not only acting by the operator $\hat{\mathcal K}$ but also reversing the direction of the magnetic field. However, the first line of the same equation says that the spinors also satisfy another symmetry: namely, the operator $\hat{\mathcal K}$ reverses the chirality of the state ($s\to -s$) at fixed ${\bf H}$.  
The spin susceptibility should be even under time-reversal. Applying $\hat{\cal K}$ (at fixed ${\bf H}$) to Eq.~(\ref{magn}), and taking into account that $\hat{\cal K}\hat\sigma_i\hat{\cal K}^{-1}=-\hat\sigma_i$, we obtain
\bea
\chi_{ii}(\bp)&=&\lim_{H_i\to 0}\frac{g\mu_B}{H_i}\sum_s \langle s,\bp;{\bf H}|\hat{\cal K}^{-1}\hat{\cal K}\hat\sigma_i \hat{\cal K}^{-1}\hat{\cal K}|s,\bp;{\bf H}\rangle\notag\\
&=&-\lim_{H\to 0}\frac{g\mu_B}{H_i}\sum_s  \langle -s,\bp;{\bf H}|\hat\sigma_i |-s,\bp,{\bf H}\rangle=-\chi_{ii}(\bp),\notag\\
\label{TR1}
\eea
which proves that $\chi_{ii}(\bp)=0$. This is the physical reason for the cancellation of the contributions from the occupied states with opposite chiralities. 

For non-interacting electrons, one can certainly check this result explicitly. Indeed, the induced polarization carried by a state with chirality $s$ and momentum $\bp$ is given by
\bse
\bea
&&\langle s,\bp,H_z\ez|\hat\sigma_z|s,\bp,H_z\ez\rangle=s\frac{g\mu_B H_z}{\alpha p}\\
&&\langle s,\bp,H_x{\bf e}_x|\hat\sigma_x|s,\bp,H_x{\bf e}_x\rangle-\langle s,\bp,{\bf 0}|\hat\sigma_x|s,\bp,{\bf 0}\rangle\notag\\
&&=s\frac{g\mu_B}{2}\sin(2\theta_{\bp})H_x,
\eea
\ese
for the out- and in-plane field, correspondingly.
Summing over $s$, we get zero in both cases. We argue, however, that since property (\ref{TR1}) is guaranteed by time-reversal symmetry, it remains valid also in the presence of the {\em ee} interaction, as long as this interaction does not lead
to spontaneous breaking of this symmetry via, e.g., ferromagnetic instability.

\section{\label{sec:Zero}Correspondence between the microscopic and phenomenological theories}
The purpose of this section is to establish the connection between the Landau function and interaction vertices of the microscopic theory.
As in the theory of a SU2S FL, the relation between the Landau function and microscopic 
interaction vertices is established by deriving the equations of motion for zero-sound modes. 
Without loss in generality, we ignore the effect of impurities 
and complications arising from the electric charge of the electrons.

To study the collective modes within the framework of a phenomenological FL theory, 
we start with the (collisionless) quantum Boltzmann equation for the non-equilibrium part of the occupation number:
\beq
 \partial_t\delta \hat{n}^\bp
 +i[\hat{n},\hat{\varepsilon}^{\bp}]_{-}+\bom{v}\cdot\bg{\nabla}_{\bom{r}}\delta \hat{n}^\bp
-\frac{1}{2}
\left[
\bg{\nabla}_{\bom{r}}\delta \hat{\varepsilon}^\bp ,\partial_{\bom{p}} \hat{n}^{0\bp}
\right]_{+}=0,
\label{1}
\eeq
where 
$\left[\hat{A},\hat{B}\right]_{\pm}$ denotes (anti)commutator of $\hat{A}$ and $\hat{B}$.  (For brevity, the dependences of $\delta{\hat n}^\bp$ and $\hat{\varepsilon}^\bp$ on $\bom{r}$, and $t$ are not displayed.) We will be interested in zero-sound modes in the charge sector which correspond to variations in the diagonal elements of $\delta{\hat n}^\bp$. In the chiral basis,
\beq \label{z1}
\delta n^\bp_{s}({\bf r},t)\equiv \delta
n_{ss}^\bp({\bf r},t)=
\delta(\varepsilon_s^\bp-\mu)a_s(\theta_\bp)e^{i(\bom{q}\cdot\bom{r}- 
\Omega t)},
\eeq
where $a_s(\theta_\bp)$ describes the angular dependence of $\delta n^\bp$.
Hence, Eq.~(\ref{1}) reduces to
 \beq
 \partial_t\delta {n}^\bp_s
+\bom{v}_s\cdot\bg{\nabla}_{\bom{r}}\delta{n}^\bp_s
+\delta(\varepsilon^\bp_s-\mu)\bom{v}_s\cdot
\bg{\nabla}_{\bom{r}}\delta \varepsilon^\bp_s
= 0,
\label{bec}
\eeq
where $\bv_s=\partial_\bp \varepsilon^\bp_s$ and
\beq
\bg{\nabla}_{\bf r}\delta\varepsilon_s^\bp=\sum_{s'}\int 
f_{ss'}(\bom{p},\bom{p'})\bg{\nabla}_{\bom{r}}\delta n_{s'}^{\bpp}(dp' ).
\eeq
Using Eq. (\ref{z1}), we obtain instead of Eq.~(\ref{bec})
\beq A_s=\sum_{s'}\int F_{ss'} \Psi_{s'
} A_{s'}\frac{d\theta'}{2\pi}, \label{z2}\eeq
where
\beq A_s= a_s\Psi_{s
}^{-1}\eeq
and 
\beq \Psi_{s
}= \frac{\bom{v}_s\cdot\bom{q}}{\Omega-\bom{v}_s\cdot\bom{q}}.\label{z3}\eeq

\begin{figure}
\includegraphics[scale=0.202]{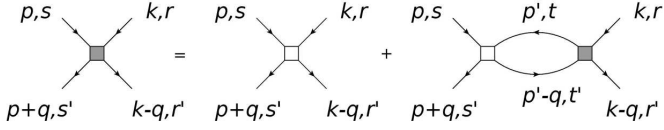}
\caption{\footnotesize The Dyson equation for the scattering vertex $\Gamma_{s,r;s',r'}(P,K;Q)$.
The first term on the right side represents the regular vertex, $\Gamma^\Omega_{s,r;s',r'}(P,K)$} 
\label{dyson}
\end{figure}

Next, we derive Eq.~(\ref{z2}) from the microscopic theory. Using the Dyson equation for the interaction vertex-- see Fig.~\ref{dyson}--we arrive at
\bea
&\Gamma&_{s,r;s',r'}(P,K;Q)={\Gamma}^{\Omega}_{s,r;s',r'}(P,K) \\
&+&\sum_{t,t'=\pm 1}\int_{P'}{\Gamma}^{\Omega}_{s,t;s',t'}(P,P')\Phi_{tt'}(P';Q)\Gamma_{t',r;t,r'}(P',K;Q),\notag
\label{d1}
\eea
where 
the $(2+1)$ momenta are defined as $P=(\bom{p},\omega)$, $P'=(\bp',\omega')$, etc., 
and $\int_P$ is a shorthand notation for $(2\pi)^{-3}\int d\omega \int d^2p\dots$. Furthermore,
 $\Gamma_{s,r';s,r'}(P,K;Q)$ is an exact vertex which contains the poles corresponding to the collective modes, ${\Gamma}^{\Omega}{s,r';s,r'}(P,K)$ is a regular vertex, 
obtained from $\Gamma$ in the limit of  $q/\Omega \rightarrow 0$ and $\Omega \rightarrow 0$, \cite{agd:1963} and $\Phi_{ss'}$ is the particle-hole correlator at fixed direction of the center-of-mass momentum of the particle-hole pair.
The off-diagonal components of $\Phi_{ss'}$ are gapped by SO splitting, while the 
diagonal ones contain singular parts given by
\beq\Phi_{ss}(P,Q)=
(2\pi iZ_s^2/v_s)\delta(\omega)\delta(p-p_s)
\Psi_{s},\label{z4}
\eeq
where $Z_s$ is the $Z$-factor of the subband $s$. 
The Landau function is related to the vertex $\Gamma^{\Omega}_{s,r;s,r}$.
Since Eq.~(\ref{d1}) must hold for any $K$, the vertex can be written as a product of two independent contributions:
\beq\Gamma_{s,r;s,r}(P,K;Q)=\eta
_s(P;Q)\eta_
r
(K;Q)\label{z5}.
\eeq
Near the poles of $\Gamma_{s,r;s,r}$, we have  
\beq
\eta_{s}=\sum_{t}{Z_t^2\nu_t}\int {\Gamma}^{\Omega}_{s,t;s,t}(\theta,\theta')
\Psi_{t}
\eta_{t}\frac{d\theta'}{2\pi},
\label{z6}
\eeq
Comparing the kernels in Eqs.~(\ref{z6}) and (\ref{z2}) and recalling definition (\ref{F}), we identify
\beq 
{f}_{ss'}=\frac{\nu_{s'}}{\nu_s}Z_{s'}^2\Gamma^{\Omega}_{s,s';s,s'}. \label{z7}\eeq
We see that, except for the ratio of the densities of states, the relation between the vertex and Landau function is the same as in the SU2S theory.
\section{Specific examples}
\label{sec:examples}
\subsection{\label{sec:Pert.LF} Landau function from the perturbation theory}
Having established a relation between the Landau function and microscopic interaction vertex, we
now show how the various components of the phenomenological Landau function in Eq.~(\ref{LF}) are reproduced by the perturbation theory for the interaction vertex. In this section, we consider a second order perturbation theory in a (spin-independent) finite-range interaction with Fourier-transform $U_q$ which, in general, depends on the magnitude of the momentum transfer $q$. The only constraint imposed on $U_q$ is that it is finite for any $q$, including $q=0$.

According to Eq.~(\ref{z7}), the Landau function is proportional the (antisymmetrized) interaction vertex $\Gamma^\Omega$. In the spin basis, 
\beq \Gamma^{\Omega}_{\alpha,\beta;\gamma,\delta}(\bp,\bpp)=\lim_{\Omega\to 0}\lim_{\frac q{\Omega}\rightarrow 0}\Gamma_{\alpha,\beta;\gamma,\delta}(P,P';Q)\left\vert_{\omega=\omega'=0}\right..\eeq
To first order in the interaction,\footnote{We work in Matsubara formalism, in which the interaction line comes with a minus sign.}  we have
\bea \Gamma^{\Omega}
_{\alpha,\beta;\gamma,\delta}
&=&\left(U_0
-\frac {U_{|\bp-\bpp|}}{ 2}\right)\delta_{\alpha\gamma}\delta_{\beta\delta}\notag\
-\frac{U_{|\bp-\bpp|}}{2} 
\bs_{\alpha\gamma}\cdot\bs_{\beta\delta},\label{gamma1}\\
\eea
which is the same as in the SU2S case.
 To see the non-SU2S terms, one needs to go to at least second order.
 
We start with a particle-particle diagram
\begin{figure}
\includegraphics[scale=0.55]{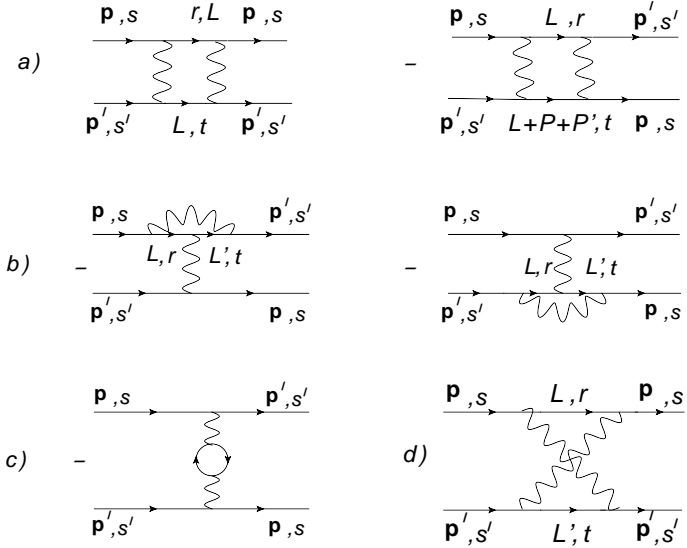}
\caption{All second-order diagrams for the $\Gamma^{\Omega}$ vertex  related to the Landau function. The rest of the second-order diagrams for $\Gamma$ vanishes in the $q/\Omega\to 0$ limit. The internal $(2+1)$ momentum $L'$ is  $L'=L+P'-P$. Exchange diagrams enter with a minus sign, shown explicitly in the figure. Both diagrams in $b)$ are the exchange ones. }\label{2ndorder}
\end{figure}
in Fig.~\ref{2ndorder}$a$,
 the direct (on the left) and exchange (on the right) parts of which are given by
 \bsu
\bea
 \Gamma^{\Omega,a1}_{\alpha,\beta;\gamma,\delta}(\bp,\bp')&=&-\int_L U^2_{|\bp-\bg{l}|}G^f_{\alpha\gamma}(L)G^f_{\beta\delta}(L')\notag\\
\label{gcd}\\
\Gamma^{\Omega,a2}_{\alpha,\beta;\gamma,\delta}(\bp,\bp')&=&\int_L U_{|\bp-\bg{l}|}U_{|\bpp-\bg{l}|}G^f_{\alpha\delta}(L)G^f_{\beta\gamma}(L'),\notag\\\label{gce}
\eea
\esu
correspondingly. Here, integration goes over
$L=(\bg{l},\omega)$, $L'=-L+K$ with $K=(\bk,0)=(\bp+\bpp,0)$, and the free Green's function in the spin basis reads 
\beq
\hat{G}^f(P)=\sum_{s}\frac{1}{2}\left(\mathds{1}+s\bs\times{\bf e}_{\bf p}\cdot\ez\right)g_{s}(P)
\eeq
where $g_s(P)$ is the same as in Eq.~(\ref{free_g}).

First, we focus on the direct part of the vertex, given by Eq.~(\ref{gcd}). The cross product of two unity matrices
merely renormalizes the $\mathds{1}\mathds{1}$ term in the Landau function 
of an SU2S FL, Eq.~(\ref{LFSU2})
Consider now the terms involving one unity matrix and one Pauli matrix: 
\bea \left(\Gamma^{\Omega,a1}_{\alpha,\beta;\gamma,\delta}\right)_g
&=& 
-\frac{
1
}
{4}
[\mathds{1}_{\alpha\gamma}\bs_{\beta\delta}\times\sum_{s,s'}s'{\bf A}'_{ss'}(\bom{{k}})\cdot {\bf e}_z\notag\\
&+&\mathds{1}_{\beta\delta}\bs_{\alpha\gamma}\times\sum_{s,s'}s{\bf A }_{ss'}(\bom{{k}})\cdot {\bf e}_z],
\eea
where
\begin{subequations}
\bea
{\bf A}_{ss'}(\bk)&=&\int_L \bg{l} U^2_{|\bp-\bg{l}|}
g^f_s(L)
g^f_{s'}(L'),\label{veca}\\
{\bf A}'_{ss'}(\bk)&=&\int_L \bg{l}'
U^2_{|\bp-\bg{l}|}
 g^f_s(L)
g^f_{s'}(L').\label{vecap}
\eea
\end{subequations}
Since the Green's function depends only on the magnitude of the electron momentum, the vectors in Eq.~(\ref{veca}) and (\ref{vecap})
are related to each other by
\beq {\bf A}_{ss'}(\bk)={\bf A}'_{s's}(\bk).\eeq
For the same reason, the directions of both vectors must coincide with that of $\bk$. Using these two properties, we obtain 
\beq \sum_{s,s'}s {\bf A}_{ss'}=\sum_{s,s'}s' {\bf A}'_{ss'}=\bk {\mathcal A}_{1}(k),\label{ida}\eeq
where ${\mathcal A}_1(k)$ is some scalar function of $k$, whose precise form depends on the choice of $U_q$.
Using the last result, we finally arrive at
\bea &&\left(\Gamma^{\Omega,a1}_{\alpha,\beta;\gamma,\delta}\right)_g=-\frac{1}{4}{\mathcal A}_{1}(|\bom{p+p'}|)\notag\\
&& 
\times\left[\mathds{1}_{\alpha\gamma}\bs_{\beta\delta}
+\mathds{1}_{\beta\delta}\bs_{\alpha\gamma}\right]\times(\bom{p+p'})\cdot {\bf e}_z\notag\\
&&\equiv \left[\mathds{1}_{\alpha\gamma}\bs_{\beta\delta}
+\mathds{1}_{\beta\delta}\bs_{\alpha\gamma}\right]\times {\bf C}_1\cdot\ez,
\label{cdir}
\eea
which coincides with the tensorial form of the $g^{\mathrm{pp}}$ term in Eq.~(\ref{LF}). 

Likewise, exchange diagram $a2$ gives
\bea &&\left(\Gamma^{\Omega,a2}_{\alpha,\beta;\gamma,\delta}\right)_g=\frac{1}{4}{\mathcal A}_{2}(|\bom{p+p'}|)\notag\\
&& 
\times\left[\mathds{1}_{\alpha\delta}\bs_{\beta\gamma}+\mathds{1}_{\beta\gamma}\bs_{\alpha\delta}\right]\times(\bom{p+p'})\cdot {\bf e}_z
\notag\\
&&\equiv \left[\mathds{1}_{\alpha\delta}\bs_{\beta\gamma}+\mathds{1}_{\beta\gamma}\bs_{\alpha\delta}\right]\times {\bf C}_2\cdot\ez,
\label{cex}
\eea
where ${\mathcal A}_2$ is defined by the same relations as (\ref{veca}) and (\ref{ida}), except for a replacement
\beq
U^2_{|\bp-\bg{l}|}\to  U_{|\bp-\bg{l}|}U_{|\bpp-\bg{l})|}.
\eeq
 Although the tensorial form of the exchange vertex looks differently
from that of the direct one, it is, in fact, the same. To see this, it is convenient to transform the vertex into the chiral basis 
using \bea
\Gamma^\Omega_{s,s';s,s'}(\bp,\bpp)&=&\sum_{\alpha\beta\gamma\delta}\Gamma^\Omega_{\alpha,\beta;\gamma,\delta}(\bp,\bpp)\la\alpha|s,\bp\ra\la\beta|s',\bpp\ra\notag\\
&&\times\la s,\bp|\gamma\ra\la s',\bpp|\delta\ra,
\label{rot}
\eea
where $|s,\bp\ra$ is the Rashba spinor defined by Eq.~(\ref{basis}). Applying (\ref{rot}) to Eqs.~(\ref{cdir}) and (\ref{cex}),
we obtain for the direct and exchange contributions, correspondingly
\bsu
\bea
\Gamma^{\Omega,a1}_{s,s';s,s'}&=&\left(\la s,\bp\vert\bs \vert s,\bp\ra+\la s',\bpp\vert\bs\vert s',\bpp\ra\right)\times{\bf C}_1\cdot\ez\notag\\
\label{cdirc}\\
\Gamma^{\Omega,a2}_{s,s';s,s'}&=&\frac 1 2\left[1+ss'e^{i\left(\theta_\bp-\theta_{\bpp}\right)}\right]\la s,\bp\vert\bs \vert s',\bpp\ra\times {\bf C}_2\cdot\ez\notag
\\&&+\frac 1 2\left[1+ss'e^{i\left(\theta_\bpp-\theta_{\bp}\right)}\right]\la s',\bpp\vert\bs \vert s,\bp\ra\times {\bf C}_2\cdot\ez.\notag\\
\label{cexc}
\eea
\esu
Calculating the matrix elements of the Pauli matrices, it can be readily shown that the vectors  to the left from ${\bf C}_1$ in  (\ref{cdirc}) and from ${\bf C_2}$ in (\ref{cexc}) are the same. Therefore, the tensorial forms of the direct and exchange vertices are the same also in the spin basis.
So far, we have reproduced the $g^{\mathrm{pp}}$ term of Eq.~(\ref{LF}).

Now, we consider the product of two Pauli matrices in Eq.~(\ref{gcd}):
\bea &&\left(\Gamma^{\Omega,a1}_{\alpha,\beta;\gamma,\delta}\right)_h=
-\frac 1 4
({\bf e}_z\times\bs_{\alpha\gamma})_i({\bf e}_z\times\bs_{\beta\delta})
_jS^{(1)}_{ij}(\bp,\bpp),\notag\\\eea
where $S^{(1)}_{ij}$ is a second-rank tensor
\beq S^{(1)}_{ij}(\bp,\bpp)=\sum_{s,s'}ss' \int_L \hat{\bg{l}}_i\hat{\bg{l}}_j'U^2_{|\bp-\bg{l}|}g^f_s(L)
g^f_{s'}(L').\label{tens}\eeq
This tensor depends only on $\bk=\bp+\bpp$, and can be formed only from the components of $\bk$ as
\beq
S^{(1)}_{ij}(\bk)=k_i k_j{\mathcal S}^{(1)}(k).\label{tens2}
\eeq
Hence
\bea &&\left(\Gamma^{\Omega,a1}_{\alpha,\beta;\gamma,\delta}\right)_h=-\frac 1 4 {\mathcal S}^{(1)} (|\bom{p+p'}|)\notag\\
&&\times\left[\bs_{\alpha\gamma}\times(\bp+\bpp)\cdot {\bf e}_z)\right]\left[\bs_{\beta\delta}\times(\bp+\bpp)\cdot{\bf e}_z\right],
\label{ch}
\eea
which reproduces the tensorial structure of the $h$ terms in Eq.~(\ref{LF}) (with
$h^{(1)}=h^{(2)}=h/2$ to this order of the perturbation theory).
As before, the exchange diagram produces same type of terms.

Since the tensorial structures of the direct and exchange particle-particle diagrams are the same, 
the non-SU2S 
part of the total particle-particle vertex $\Gamma^{\Omega,a}_{\alpha,\beta;\gamma,\delta}=\Gamma^{\Omega,a1}_{\alpha,\beta;\gamma,\delta}+\Gamma^{\Omega,a2}_{\alpha,\beta;\gamma,\delta}$ vanishes for a contact interaction, $U=\mathrm{const}$. 
Therefore, the $g^{\mathrm{pp}}$ term in the LF is absent in this case, whereas the $h$ terms come from only the particle-hole channel, considered below.

Next, we show that the crossed particle-hole diagram  (Fig.~\ref{2ndorder}$d$) produces both the $g^{\mathrm{ph}}$ and $h$ terms in Eq.~(\ref{LF}).
A product of the unity and Pauli matrices is processed in the same way
as the analogous term in the particle-particle diagram.
The only difference is that the vectors ${\bf A}_{ss'}$ and ${\bf A}'_{ss'}$ are now replaced by
\begin{subequations}
\bea
{\bf B}_{ss'}(\bk)&=&\int_L \bg{l} U^2_{|\bp-\bg{l}|}
g^f_s(L)
g^f_{s'}(L''),\label{vecd}\\
{\bf B}'_{ss'}(\bk)&=&\int_L \bg{l}'
U^2_{|\bp-\bg{l}|}
 g^f_s(L)
g^f_{s'}(L''),\label{vecdp}
\eea
\end{subequations}
where $L''=L+P'-P$, and ${\bf B}$ and ${\bf B}'$ are related by
\beq {\bf B}_{ss'}(\bk)=-{\bf B}'_{s's}(\bk).\eeq
Defining a scalar function ${\mathcal B}(k)$ via
\beq \sum_{s,s'}s {\bf B}_{ss'}=-\sum_{s,s'}s' {\bf B }'_{ss'}=\bk {\mathcal B}(k),\label{id}\eeq
 we obtain
\bea &&\left(\Gamma^{\Omega,
d}_{\alpha,\beta;\gamma,\delta}\right)_g=-\frac 1 4 {\mathcal B}(|\bom{p-p'}|)\notag\\
&& 
\times\left[\mathds{1}_{\alpha\gamma}\bs_{\beta\delta}\times(\bom{p'-p})\cdot {\bf e}_z
-\mathds{1}_{\beta\delta}\bs_{\alpha\gamma}\times(\bom{p'-p})\cdot {\bf e}_z\right],\notag\\\eea
which coincides with the tensorial form of the $g^{\mathrm {ph}}$ term in Eq.~(\ref{LF}).
The term involving two Pauli matrices is cast into a form similar to Eq.~(\ref{ch})
\bea
&&\left(\Gamma^{\Omega,
d}_{\alpha,\beta;\gamma,\delta}\right)_h=-\frac1 4 {\cal T}(|\bp-\bp'|)
\notag\\
&&\times\left[\bs_{\alpha\gamma}\times(\bp-\bpp)\cdot {\bf e}_z)\right]\left[\bs_{\beta\delta}\times(\bp-\bpp)\cdot{\bf e}_z\right],
\eea
where ${\mathcal T}(k)$ is defined similarly to Eqs.~(\ref{tens}) and (\ref{tens2}) except for $L'$ is replaced by $L''=L+P'-P$.
The last expression reproduces again the tensorial structure of the $h^{(1,2)}$ and $h$ terms, except for now $h^{(1)}=h^{(2)}=-h/2$.

Both diagrams $b$ produce either the non-SU2S terms, we have encountered before, or renormalize the SU2S terms.
Finally, diagram $c$ only renormalizes the SU2S terms. We have thus shown that the second-order perturbation theory
accounts for all terms in the phenomenological Landau function, Eq.~(\ref{LF}).. 
\subsection{\label{sec:results}
Mass renormalization}
In this section, we apply the formalism developed in previous sections  to a calculation of mass renormalization for Rashba fermions with the FL formalism.
We consider the case of a weak,  contact {\em ee} interaction ($U=\mathrm{const}$),
and assume also that the SO interaction is weak as well, i.e., $\alpha\ll v_F$.
(An assumption about the contact nature of the interaction is unessential; at the end of this section we comment on how the results are to be generalized for an arbitrary interaction.)

Our focus will be on the leading, 
$\mathcal{O}(U^2\alpha^2)$,  corrections to the effective masses of Rashba fermions.
Although such a correction seems to be perfectly analytic, in fact, it is not.  It will be shown that the correction to the effective mass
of the subband $s$ comes as $sU^2\alpha^2$. Since the eigenenergies contain $s$ and $\alpha$ only as a combination of $s\alpha$, the second-order term in the regular expansion in $s\alpha$  is the same for $s=\pm 1$. In this sense, the $sU^2\alpha^2$ correction is non-analytic, and will be shown to come from the Kohn anomalies of the various vertices. 

To simplify notations, we will suppress the superscript $\Omega$ in the vertices because all the vertices considered in this section are of the  $\Gamma^\Omega$ type. Also, we will suppress the superscript $f$ labeling the quantities pertinent to the interaction-free system: in a straightforward perturbation theory, considered here, the interaction enters only as the $U^2$ prefactor.

 A non-analytic part of the Landau function comes from particle-hole diagrams for $\Gamma^{\Omega}$ in Fig.~\ref{2ndorder}, i.e., from diagrams $b$-$d$. To maximize the effect of the Kohn anomaly, one needs to select such initial and final states that correspond to the minimal number of small matrix elements for intra-band backscattering.
 
Detailed calculations of the diagrams are presented in Appendix~\ref{app:LFsplitting}; here, we illustrate how a $\mathcal{O}(U^2\alpha^2)$ correction occurs using diagram $c$ as an example. Explicitly, this diagram reads
\beq
\Gamma^c_{ss'}(\bp_s,\bpp_s)=-\frac{U^2}{2} [1+ss'\cos(\theta_{\bf {p'}}-\theta_{\bf p})]\Pi(|\bom{p}_s-\bom{p'}_{s'}|),\eeq
where $\Pi(q)$ is the polarization bubble of non-interacting Rashba fermions. 
Without the SO interaction, the Kohn anomaly of $\Pi$ is located at $q=2p_F$, where $p_F$ is the Fermi momentum at $\alpha=0$.  In 2D, the polarization bubble is independent of $q$ for $q\leq 2p_F$ and exhibits a characteristic square-root anomaly for $q>2p_F$. SO coupling splits the spectrum into two Rashba subbands with Fermi momenta $p_{\pm}$ given by Eq.~(\ref{ppm}). 
In Ref.~\onlinecite{pletyukhov:2006}, it was shown that the static polarization bubble $\Pi(q)$ does not depend on $q$ for $q\leq 2p_+$  for an arbitrary value of $\alpha$ (but as long as both subbands are occupied). In the region $2p_+\leq q\leq 2p_-$, the polarization bubble exhibits a non-analytic dependence on $q$. At small $\alpha$, all three Fermi momenta are close to each other: $p_+\approx p_-\approx p_F$. In this case, the singular parts of  $\Pi$ can be written as (see Appendix~\ref{app:D})
\bse
\bea \Pi(q)&=&-\nu+\Pi_{++}(q)+\Pi_{+-}(q)\label{bubble1a}\\
\Pi_{++}(q)&=&-\frac{\nu}{
6}\Theta(q-2p_+)\left(\frac{q-2p_+}{p_F}\right)^{3/2}\label{bubble1b}\\
\Pi_{+-}(q)&=&\frac{\nu}{2}\Theta(q-2p_F)\left(\frac{q-2p_F}{p_F}\right)^{1/2}.
\label{bubble1}
\eea
\ese 
The Kohn anomalies of $\Pi$ affect the amplitudes of backscattering processes with $\bom{p'}\approx -\bom{p}$;
hence $\theta_{\bf{p'}}-\theta_{\bf{p}}=\pi-\theta$ with $|\theta|\ll 1$. Accordingly, the vertex in Eq.~(\ref{bubble1}) reduces to 
\beq
\Gamma^c_{ss'}=-\frac{U^2}{2} \left(1+ss'-ss'\frac 1 2\theta^2\right)\Pi(|\bom{p}_s-\bom{p'}_{s'}|),\label{vert1}\eeq
Furthermore, the Kohn anomaly of $\Pi_{++}$ affects scattering processes with momentum transfer in the interval $2p_{+}\leq q \leq 2p_{-}$. Since $\Pi_{++}$ already
contains an extra (compared to $\Pi_{+-}$) factor of $(q-2p_+)$, reflecting the smallness of the matrix element of intraband backscattering,
the non analytic contribution from $\Pi_{++}$ is maximal for interband processes ($s=-s'$), when the angular factor in (\ref{vert1}) is almost equal to $1$. The only vertex of this type is 
\beq
\Gamma^c_{+-}=-U^2\Pi_{++}(q
)=\frac{\nu U^2}{
6}\Theta(q-2p_+)\left(\frac{q-2p_+}{p_F}\right)^{3/2}.\label{Gcpm}
\eeq
The Kohn anomaly of $\Pi_{+-}$ gives an effect of the same order as in (\ref{Gcpm}) but for intraband processes:
\bea\Gamma^c_{--}=-U^2\frac{\theta^2}{2}\Pi_{+-}(q
).
\label{Gcmm}
\eea
For a backscattering process,  
\beq q=|\bom{p}_s-\bom{p'}_{s'}|\approx p_s+p_{s'}-p_F\theta^2/4, \label{q}\eeq
where the dependence of the prefactor of the $\theta^2$ term on $\alpha$ can be  and was neglected. 
For $\Gamma^c_{--}$, we have $q=2p_--p_F\theta^2/4$.
Expressing $\theta^2$ in terms of $q$, we arrive at
\bea\Gamma^c_{--}=-\nu U^2\Theta(q-2p_F)\frac{\left(q-2p_{-}\right)\left(q-2p_F\right)^{1/2}}{p_F^{3/2}}.
\notag\\
\label{Gcmmq}
\eea
Now it is obvious that $\Gamma^c_{+-}$ and $\Gamma^c_{--}$ are of the same order.

The remaining diagrams are evaluated in Appendix \ref{app:LFsplitting} with the following results
\bse
\bea
\Gamma^b_{--}&=&0\label{Gb--}\\
\Gamma^b_{+-}&=&-\Gamma^c_{+-}=U^2\Theta\paren{q-2p_F}\paren{\frac{q-2p_F}{p_F}}^{3/2}\label{Gb+-m}\\
\Gamma^d_{--}&=&\frac {U^2\nu}{12}\Theta(q-2p_F)\paren{\frac{q-2p_F}{p_F}}^{3/2}\label{Gd--}\\
\Gamma^d_{+-}&=&\frac {U^2\nu}{24}\Theta(q-2p_+)\paren{\frac{q-2p_+}{p_F}}^{3/2}
\label{Gd+-}\\
\Gamma^{b,c,d}_{++}&=&0.
\eea
\ese
One should substitute $q=|\bp_{-}-\bpp_{-}|$ in Eqs.~(\ref{Gcmmq}), (\ref{Gb--}) and (\ref{Gd--}), and $q=|\bp_{+}-\bpp_{-}|$ in
Eqs.~(\ref{Gcpm}), (\ref{Gb+-m}) and (\ref{Gd+-}).

Now we can calculate the angular harmonics of the vertices. Combining the vertices in the $(--)$ channel and transforming back from $q$ to $\theta$, we obtain for the $n^\mathrm{th}$ harmonic of the total vertex in this channel 
\bwt
\bea
\Gamma_{--}^{,n}&=&\Gamma^{c,n}_{--}+\Gamma^{d,n}_{--}=-(-1)^n\frac{U^2\nu}{4}\int_0^{2\sqrt{\frac{2m\alpha}{p_F}}}\frac{d\theta}{\pi}\theta^2\left(\frac{2m\alpha}{p_F}-\frac{\theta^2}4\right)^{1/2}
+(-1)^n\frac{U^2\nu}{12}\int_0^{2\sqrt{\frac{2m\alpha}{p_F}}}\frac{d\theta}{\pi}\left(\frac{2m\alpha}{p_F}-\frac{\theta^2}4\right)^{3/2}
\notag\\&=&-\frac 38(-1)^n U^2\nu\left(\frac{m\alpha}{p_F}\right)^2. \label{Gnmm}
\eea
\ewt
Equation (\ref{Gnmm}) is valid for $1\leq n\ll\sqrt{p_F/8m\alpha}$. Likewise, we find for the  $+-$ channel 
\bea
\Gamma_{+-}^{,n}=\Gamma^{d,n}_{+-}=\frac{1}{16}(-1)^n U^2\nu\left(\frac{m\alpha}{p_F}\right)^2.\label{Gnpm}
\eea
The harmonics of the Landau function are related to those of vertices via Eq.~(\ref{z7}), in which the proportionality coefficient $\nu_{s'}Z^2_{s'}/\nu_{s}$ is  taken to be equal to one to lowest order in $U$ and $\alpha$. 

As mentioned in Sec.~\ref{sec:m}, the formula for the effective mass, Eq.~(\ref{mstar1}), contains not only the components of the Landau function but also the ratio of the subband Fermi momenta which, in general, are renormalized by the interaction. However, combining previous results from Refs.~\onlinecite{chesi-theorem, zak:2012,aasen:2012}, one can show that this effect occurs only to higher orders in $\alpha$ and $U$. Indeed,  an analytic part of the ground state energy of an electron system with the Rashba SO interaction can be written as $E_{\mathrm{an}}=C\paren{\delta N/N-2\alpha/v_F}^2$, where $C=$const, $\delta N=N_{-}-N_{+}$ is the difference in the number of electrons occupying the two Rashba subbands, and $v_F$ is the {\em bare} Fermi velocity.\cite{chesi-theorem} This result is valid to {\em all} orders in the {\em ee} interaction of arbitrary type, and to lowest order in $\alpha$. Therefore, the minimum of $E_{\mathrm{an}}$ corresponds to the same value of $\delta N$ as for a non-interacting electron gas, which means that the Fermi momenta are not renormalized. Renormalization of $\delta N$  occurs only because of non-analytic terms in the ground state energy.\cite{zak:2012,aasen:2012} For a contact interaction, the first non-analytic correction occurs to fourth order in $U$: $E_{\mathrm{na}}\propto  U^4|\alpha|^3\ln^2|\alpha|$. (A cubic dependence on $|\alpha|$ is because  a non-analytic part of ground state the energy scales in 2D  as a cube of the parameter controlling non-analyticity which, in our case, is  $\alpha$; \cite{maslov:2006, *maslov:2009}
an additional factor of $\ln^2|\alpha|$ comes from renormalization of the interaction in the Cooper channel.) Then the minimum of $E_{\mathrm{an}}+ E_{\mathrm{na}}$ corresponds to a change in the Fermi momenta $p_--p_+\propto \delta 
N\propto U^4\alpha^2\ln^2|\alpha|$, which is beyond the order of the perturbation theory considered here. Therefore, the Fermi momenta entering  Eq.~(\ref{mstar1}) are to be considered as unrenormalized. In addition,
to lowest order in $\alpha$, we can take $p_+$ to be equal to $p_-$, so that the ratio of the Fermi momenta drops out from Eq.~(\ref{mstar1}).

Finally, substituting Eqs.~(\ref{Gnmm}) and (\ref{Gnpm}) into Eq.~(\ref{mstar1}), we obtain for the renormalized masses
\bse
\bea 
\frac{m^*_+-m_{+}}{m_+}
&=&
F^{,1}_{++}+F^{,1}_{+-}
=-\frac{1}{16}\left(\frac{\nu Um\alpha}{p_F}\right)^2\label{mplus}\\ 
\frac{m^*_--m_{-}}{m_-}
&=&
F^{,1}_{--}+F^{,1}_{-+}
=\frac{5}{16}\left(\frac{\nu U m\alpha}{p_F}\right)^2.\label{mminus}
\eea
\ese
We see that the masses of fermions with opposite chiralities are renormalized differently.  Since we have already established that the Fermi momenta are not renormalized to this order, there is no cancellation between renormalizations of masses and that of Fermi momenta. Therefore,
the degeneracy of the Fermi velocities of the Rashba subbands is lifted by the interaction: 
\beq v_{+}^*-v_{-}^*=\frac{p_{+}}{m^*_{+}}-\frac{p_{-}}{m^*_{-}}=\frac{3}{8}v_F\left(\frac{\nu U m\alpha}{p_F}\right)^2
\label{vstar}
\eeq
where $v_F=\sqrt{2\mu/m}$.

For a momentum-dependent interaction, $U_q$, one can simply replace $U$ in Eq.~(\ref{vstar}) by $U_{2p_F}$  because mass renormalization comes from backscattering processes. In particular, $U_{2p_F}=\nu^{-1}r_s/\sqrt{2}$ for the screened Coulomb potential, Eq.~(\ref{TF}), in the large-$r_s$ limit. In this case, Eq.~(\ref{vstar}) reduces to
\beq \frac{v_{+}^*-v_{-}^*}{v_F}=\frac{3}{16}r_s^2\left(\frac{\alpha}{v_F}\right)^2.\label{vtf}\eeq

 As mentioned in Sec.~\ref{sec:m}, degeneracy of the subbands' Fermi velocities survives to an arbitrary order in the {\em ee} interaction, provided that the SO interaction is treated to first order. \cite{chesi-theorem} Reference \onlinecite{aasen:2012} finds that, within the small-$q$ scattering approximation for the screened Coulomb potential,  velocity splitting occurs because of a non-analytic, $r_s|\alpha|^3\ln|\alpha|$, term in the electron self-energy. (In Appendix \ref{app:A}, we reproduce the result of Ref.~\onlinecite{aasen:2012}.) Our result, Eq.~(\ref{vstar}), differs from that of Ref.~\onlinecite{aasen:2012} because it comes from backscattering rather than small-$q$ scattering, and corresponds to order $r_s^2$ of the perturbation theory, one order higher than order $r_s$ considered in Ref.~\onlinecite{aasen:2012}.  In reality, both the $r_s|\alpha|^3\ln|\alpha|$ and $r_s^2\alpha^2$ terms are present, and the competition between the two is controlled by the ratio  $|\alpha|\ln|\alpha|/r_s$ of the two small parameters of the model, $\alpha$ and $r_s$. 
 
 At the same time, we do not reproduce a rather surprising result of Ref.~\onlinecite{yu:2013}, which finds velocity splitting already to order $U\alpha$ for $U$=const. While such a term contradicts the general results of Refs.~\onlinecite{saraga:2005} and \onlinecite{chesi-theorem},  we also found that, to first order in $U$,  the Landau function is the same as for an SU2S FL, see Eq.~(\ref{gamma1}).  For $U$=const, this Landau function reduces to a constant and thus cannot produce mass renormalization.
 \subsection{\label{app:B} 
Limiting forms of the spin susceptibility}
In this
section, we show that Eq.~(\ref{chiz}) reproduces the known results for $\chi_{zz}$ in the limiting cases of a weak {\em ee} interaction or weak SO coupling.
\subsubsection{Weak electron-electron interaction}
First, we examine the limit of a weak contact {\em ee} interaction, i.e., we work to first order in the interaction amplitude $U$
without making any assumption about the strength of SO coupling.)
\begin{figure}
\includegraphics[scale=0.5]{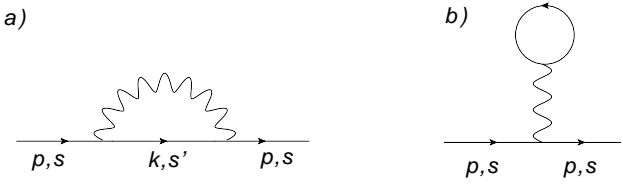}
\caption{Self-energy of chiral fermions to first order in the interaction.}
\label{fig:U}
\end{figure}
It follows from Eq.~(\ref{gamma1}) that $f^{\mathrm{a}\perp}=-U/2$ to first order in the interaction. We also need the self-energy of the quasiparticles to first order in the interaction given by (see Fig.~\ref{fig:U})
\beq \Sigma_s(P)=\Sigma^a_s(P)+\Sigma^b_s(P),\eeq
\bea \Sigma^a_s(P)&=&-\frac U 2\sum_{s'}\int_K [1+ss'\cos{(\theta_{\bk}-\theta_{\bp})}]g_s
(K)\notag=-\frac U 2 N,
\eea
and similarly $\Sigma^b_s(P)=UN$, where $N$ is the total number density of electrons. Therefore, the self-energies of both Rashba branches are constant and equal to each other.
Since the shift in the energy of quasiparticles with opposite chiralities is the same, we have $\varepsilon^{\bp}_+-\varepsilon^{\bp}_-=2\alpha p$ and $p_{-}-p_+=2m\alpha$, which are the same relations as for an interaction-free system. Substituting these results into Eq.~(\ref{chiz}), we find that to first order in $U$
\bea \chi_{zz}=\chi_0\left(1+\frac {mU}{2\pi}\right).
\label{chiz_1}\eea
As is to be expected, this result coincides with that of the first ladder diagram of the perturbation theory.
 \cite{*[{We use this opportunity to point out that the numerical coefficients in the denominators of Eqs.~(2.6) and (2.8) of} ]  [ {should be $1/2\pi$ instead of $1/\pi$.}] zak:2010}

 Notice that $\chi_{zz}$ in Eq.~(\ref{chiz_1}) does not depend on $\alpha$. This property survives to all orders in $U$ within the ladder approximation. Beyond the ladder approximation, the leading dependence on $\alpha$
occurs to order $U^2$ as a nonanalytic in $\alpha$ correction : $\delta\chi_{zz}=(2/3)\chi_0(mU/4\pi)^2 |\alpha|p_F/\epsilon_F$ (Refs.~\onlinecite{zak:2010,zak:2012}).
\subsubsection{Weak spin-orbit coupling}
Next, we consider the opposite limit of an arbitrary {\em ee} interactions but infinitesimally small SO coupling. In this limit,  one must recover the result of a SU2S FL. To obtain $\chi_{zz}$ in this limit, one needs to evaluate integrals of the type
\beq\int_{p_+}^{p_-} \frac {dp'p'}{2\pi}\frac{f^{\mathrm{a}\perp,0}(p,p')g^*(p')}{\varepsilon^{\bpp}_+-\varepsilon^{\bpp}_-},
\eeq
to zeroth order in $\alpha$. Infinitesimal SO coupling implies that the region of integration is infinitesimally small compared to the Fermi momentum, and it suffices to consider only the linear part of the dispersion near the FS
$\varepsilon_{s}^\bp=v_{s}(p-p_{s})$, where $v_{s}=\partial_{p}\varepsilon^{\bp}_{s}|_{p=p_s}$. Note that in the presence of the interaction, in general, $v_+\neq v_-$. However, to obtain $\chi_{zz}$ to zeroth order in $\alpha$, one can keep SO coupling only in the Fermi momenta of the two subbands 
and set $\alpha$ to zero everywhere else, including $f^{\mathrm{a}\perp,0}$ which reduces to $f^{\mathrm{a},0}$ of an SU2S FL.  As a result, $v_s$ can be set to equal to $v_F$, the Fermi velocity in the absence of SO coupling. Therefore,
\beq\int_{p_+}^{p_-}\frac {dp'p'}{2\pi}\frac{f^{\mathrm{a}\perp,0}(p,p')g^*(p')}{\varepsilon^{\bpp}_+-\varepsilon_-^{\bpp}}=\frac{p_F}{2\pi v_F}f^{\mathrm{a},0}(p_F,p_F)g^*(p_F),\label{int}\eeq
Substituting this result into Eq.~(\ref{g}) and solving for $g^*(p_F)$, we reproduce the result for an SU2S FL
\beq
g^*(p_F)=
\frac{g}{1+\nu f^{\mathrm{a},0}(p_F,p_F)},\label{gf}\eeq
where 
\beq \chi_{zz}\approx \frac{g^2\mu_B^2}2 \int_{p_+}^{p_-}\frac {dpp_F}{2\pi}\frac{g^*(p_F)}
{\varepsilon_+^\bp-\varepsilon_-^\bp}=\frac {g^2\mu_B^2} {4}\frac{\nu}{1+\nu f^{\mathrm{a},0}}.\label{chifl}\eeq
In both Eqs.~(\ref{gf}) and (\ref{chifl}), $\nu$ is the renormalized density of states.
Notice that SO coupling is eliminated from Eqs.~(\ref{int}) and (\ref{chifl}) as an {\em anomaly}, i.e., as a cancellation between a small denominator ($\varepsilon^{\bp}_+-\varepsilon^{\bp}_-$) and a narrow integration range ($p_--p_+$).

\section{\label{sec:Con}Conclusions}
We considered a 2D FL in the presence of the Rashba SO interaction, which breaks the SU(2) symmetry of the system. We constructed the phenomenological Landau function, Eq.~(\ref{LF}), satisfying all the symmetries ($C_{\infty v}$, permutation of particles, and time-reversal).  In Sec.\ref{sec:Pert.LF}, this form of the Landau function was also obtained by a second-order perturbation theory in the {\em ee} interaction. The key result of this paper is that  while the charge sector of a chiral FL can be fully described by the Landau function projected on the two spin-split FSs,
any quantity pertaining to the spin sector must involve the Landau function with momenta in between the two FSs. This feature is most explicitly demonstrated for the case of the static, out-of-plane spin susceptibility. Therefore, there is no conventional FL theory, i.e. a theory operating solely with free quasiparticles, for the spin sector of a chiral FL. This does not mean that we are dealing here with a non-Fermi liquid, because chiral quasiparticles are still well-defined near their respective FSs.  However, the spin sector of a chiral FL does not allow for a FL-type description. In other words, we a have a FL but without a full-fledged FL theory. In fact, this is true not only for a chiral but also for any FL with broken SU(2) symmetry, e.g. a 3D FL in the presence of a finite magnetic field. \cite{herring,meyerovich:1992a,meyerovich:1994,meyerovich:1994b}

The charge sector of a chiral FL is similar to any two-band FL. One interesting effect of the {\em ee} interaction in a system with either Rashba or Dresselhaus SO coupling is that it lifts the degeneracy of Fermi velocities of fermions with opposite chiralities, which is apparent already from the phenomenological formula for the effective masses,  Eq.~(\ref{mstar}) However, due to an exact property \cite{saraga:2005,chesi-theorem} no spin-splitting of the velocities occurs to first-order in SO coupling. We obtain an expression for splitting for the Coloumb-interaction case in the limit of strong SO coupling. 
In Appendix \ref{app:A}, we reproduce leading term of the analytic expansion of the effective masses in the SO coupling constant.  In addition, there are non-analytic corrections to the splitting, which we find by evaluating the non-analytic corrections to the Landau function, and show  velocity splitting occurs due to terms of order $sU^2\alpha^2$, where $s=\pm 1$ is the helicity.

All the results of this paper also hold for a system with the linear Dresselhaus rather than Rashba SO coupling. However, if both the Rashba and Dresselhaus couplings are present, the in-plane rotational symmetry of the FS is broken, and the FL becomes not only a chiral but also anisotropic.  A theory of such a FL can be constructed along the lines presented in this paper but we did not attempt such a construction here.

The formalism developed in this paper can also be readily generalized to chiral systems with a cubic-in-momentum SO coupling, e.g., to heavy holes in III-V  semiconductor heterostructures, surface state of SrTiO$_3$, and possibly  LaAlO$_3$/SrTiO$_3$ interfaces.~\cite{nakamura:2012,zhong:2013,kim:2013} In case of the $C_{\infty v}$ symmetry, the Rashba-type Hamiltonian for non-interacting particles with pseudospin $j_z=\pm 3/2$ reads~\cite{winkler:2000, *winkler:2002}
\beq \hat{H}_{j=3/2}=\frac{p^2}{2m}\mathds{1}+\frac{i\gamma}2\left(\hat\sigma_+p_-^3-\hat\sigma_-p_+^3\right),
\label{Hh}
\eeq
where $\hat\sigma_{\pm}=\sigma_x\pm i\sigma_y$, $p_{\pm}=p_x\pm ip_y$, and $\gamma$ is real. 
Selecting the invariants in the same way as for linear-in-momentum coupling, the Landau function corresponding to Hamiltonian (\ref{Hh}) is constructed as 
\bwt
\begin{eqnarray}\label{LFholes}
\hat{f}_{j=3/2}&=& f^{\mathrm{s}}\mathds{1}\mathds{1}^{\prime}+f^{\mathrm{a}\|}(\hat{\sigma}_x^{{}}\hat{\sigma}^{\prime}_x+\hat{\sigma}_y^{{}}\hat{\sigma}^{\prime}_y)
+f^{\mathrm{a}\perp}\hat{\sigma}_z^{{}}\hat{\sigma}^{\prime}_z
+ \frac i 2 g^{\mathrm{ph}}
\left[
\left(
\mathds{1}\hat\sigma'_+-\mathds{1'}\hat\sigma_+
\right)
\left(
p_-^{\prime 3}-p_-^3
\right)-
\left(
\mathds{1}\hat\sigma'_--\mathds{1'}\hat\sigma_-
\right)
\left(
p_+^{\prime 3}-p_+^3
\right)
\right]
\notag\\
&+& \frac i 2 g^{\mathrm{pp}}
\left[
\left(
\mathds{1}\hat\sigma'_++\mathds{1'}\hat\sigma_+
\right)
\left(
p_-^{\prime 3}+p_-^3
\right)-
\left(
\mathds{1}\hat\sigma'_-+\mathds{1'}\hat\sigma_-
\right)
\left(
p_+^{\prime 3}+p_+^3
\right)
\right]
+h^{(1)}\left(\hat\sigma_+p_-^3-\hat\sigma_-p_+^3\right)
\left(\hat\sigma'_+p_-^{\prime 3}-\hat\sigma'_-p_+^{\prime 3}\right)\notag\\
&+&h^{(2)} \left(\hat\sigma'_+p_-^3-\hat\sigma'_-p_+^3\right)
\left(\hat\sigma_+p_-^{\prime 3}-\hat\sigma_-p_+^{\prime 3}\right)
+\frac 1 2 h \left[\left(\hat\sigma_+p_-^3-\hat\sigma_-p_+^3\right)\left(\hat\sigma'_+p_-^3-\hat\sigma'_-p_+^3\right)\right.\notag\\
&+&\left.\left(\hat\sigma_+p_-^{\prime 3}-\hat\sigma_-p_+^{\prime 3}\right)\left(\hat\sigma'_+p_-^{\prime 3}-\hat\sigma'_-p_+^{\prime 3}\right)\right],
\label{LF3}
\end{eqnarray}  
\ewt
where all the scalar functions $f^{\mathrm{s}}\dots h$ are real.
The charge sector of the chiral FL with cubic-in-momentum SO coupling is qualitatively similar to that of a chiral FL with linear-in-momentum coupling, except for 
spin-splitting of the velocities now occurs already to the lowest order in the SO coupling, $\gamma$.~\cite{aasen:2012}
The major difference with the case of linear-in-momentum SO coupling is in how an in-plane magnetic field enters the Hamiltonian. To linear order in field,  the only form of coupling consistent with the $C_{\infty v}$ symmetry is $\hat\sigma_+p_-^2 H_-+\hat\sigma_-p_+^2H_+$ with $H_\pm=H_x\pm iH_y$. \cite{winkler:2000,winkler:2002,chesi:2007c} However, the structure of the theory remains qualitatively the same in that one still has to deal with off-diagonal components of the occupation number and  damped quasiparticles.
\acknowledgments
This work was supported by the NSF through the Materials World Network programs DMR-0908029 (A.A. and D.L.M) and DMR-0908070 (E.I.R.) We are grateful to S. Chesi, K. Ensslin,  A. Finkelstein, W.-M. Liu, D. Loss, C. Marcus, A. Meyerovich, S. Tarucha, X.-L.Yu, S.-S. Zhang, and D. Zumb{\"u}hl for stimulating discussions.
\appendix
\bwt
\section{\label{app:A} Mass renormalization via the self-energy} 
In Sec.~\ref{sec:m}, we analyzed mass renormalization of Rashba fermions to logarithmic accuracy in the parameter 
$p_{\mathrm{TF}}/p_{\pm}$ which controls the Random Phase Approximation for the Coulomb interaction, and showed that degeneracy of the subbands' Fermi velocities is lifted by the {\em ee} interaction, at least when SO coupling is so strong that $p_+\ll p_-$ (but still both subbands are occupied). 
In this Appendix, we derive the result for mass renormalization still to leading order in the {\em ee}  interaction but for an arbitrarily strong SO interaction.  For the time being, we assume that the interaction is described by a non-retarded and spin-independent, but otherwise arbitrary potential $U_q$. To obtain the result beyond the logarithmic accuracy, one needs to keep the matrix elements in Eq.~(\ref{selfenergy}) and relax the assumption of small momentum transfers. The only small parameter will now be the quasiparticle energy, 
$\varepsilon_{s}^{\bp,f}=\epsilon_s^{\bp,f}-\mu$.  From this point on, we suppress the superscript $f$ since all dispersions considered below are for non-interacting electrons, i.e, $\varepsilon_{s}^{\bp,f}\to \varepsilon_{s}^{\bf p}$.
The leading order renormalization is logarithmic, therefore the potential can be to taken as static since the dynamic part does not produce logarithmic integrals.
Subtracting the self-energy on the FS from Eq.~(\ref{selfenergy}), we obtain for the remainder
\begin{align}\Delta\Sigma_s =-\sum_{s^\prime }\sum_{q,\Omega }\frac {U_q}2 \left[(1+ss^\prime \cos\theta_{\bom{p+q}})g_{s^\prime }(\bom{p+q},\omega+\Omega) -(1+ss^\prime \cos\theta_{\bom{p}_s+\bom{q}})g_{s^\prime }(\bom{p}_s+\bom{q},\Omega)\right].
\end{align}
Because of the in-plane rotational symmetry, it is convenient to measure all angles from the direction of $\bp$ which, by definition, coincides with the direction of $\bp_s$.  
Using
\begin{align}
\epsilon_{s^\prime}^{\bom{p+q}}=\frac{|\bom{p+q}|^2}{2m}+s^\prime\alpha|\bom{p+q}|=\frac{p^2}{2m}
 +\frac {pq} m  \cos\theta+\frac{q^2}{2m}+s^\prime\alpha|\bom{p+q}|,
\end{align}
where $\theta$ is the angle between $\bp$ and $\bq$,
and 
\begin{align}
|\bom{ p+q}|=\left\vert\left(p_s+\frac {\varepsilon_{s}^{\bf p}}{v_0}\right)
\up+ \bom{q}\right\vert=| \bom{p}_s+\bom{q}|+\frac {\varepsilon_{s}^{\bp}}v \cos(\theta_{\bom{p}_s+\bom{q}}),
\end{align}
where $v_0$ is given by Eq.~(\ref{v0}), we get
\begin{align}
\varepsilon_{s^\prime}^{\bom{p+q}}
&=\varepsilon_s^{\bp}\left(1+\frac{q} {mv_0} \cos\theta -\frac {s\alpha}{v_0}+s^\prime\alpha\cos\theta_{\bom{p}_s+\bom{q}} \right)+\frac {p_sq} m  \cos\theta +\frac{q^2}{2m}+s^\prime\alpha|\bom{p}_s+\bom{q}|-s\alpha p_s,
\label{energy}
\end{align}
and
\begin{align}
\cos\theta_{\bom{p+q}}&=\cos\theta_{{\bp_s}+\bq}+\frac{m\varepsilon_{s}^{\bp}}
{p_s |\bom{p}_s+\bom{q}|}\sin^2\theta_{{\bp_s}+\bq}.
\end{align}
One can separate $\Delta\Sigma_s$ into two parts as $\Delta\Sigma=\Delta\Sigma^1_s+\Delta\Sigma^2_s$ with
\bse
\bea\Delta\Sigma_s^{1}&=&-\sum_{s^\prime }\sum_{q,\Omega }\frac {U_q}2 (1+ss^\prime \cos\theta_{\bom{p}_s+\bom{q}})\left[g_{s^\prime }(\bom{p+q},\omega+\Omega) -g_{s^\prime }(\bom{p}_s+\bom{q},\Omega)\right],\label{selfenergy1}\\
\Delta\Sigma_s^{2}&=&- \frac{m\varepsilon_{s}^{\bp}}{p_s}\sum_{s^\prime }ss'\sum_{q,\Omega }\frac {U_q}2
\frac {
\sin^2\theta_{\bom{p}_s+\bom{q}}} {
|\bom{p}_s+\bom{q}|}
g_{s^\prime }(\bom{p}_s+\bom{q},\omega+\Omega),\label{selfenergy2}
\eea
\ese
where the first part comes from only the immediate vicinity of the FS, while the second one 
comes from the entire band. Integrating over $\Omega$ in Eq.~(\ref{selfenergy1}), we obtain the difference of two Fermi functions
of the $s'$ subband. Expanding the first Fermi function with the help of Eq.~(\ref{energy}), we get
\begin{align}\Sigma^s_1 =
\varepsilon_{s}^{\bf p}\sum_{s^\prime }\sum_{q}\frac {U{(q)}}2 \left(1+ss^\prime \cos\theta_{\bom{p}_s+\bom{q}}\right)
\left(1+\frac{q} {mv_0} \cos\theta -\frac {s\alpha}v+s^\prime\alpha \cos\theta_{\bom{p}_s+\bom{q}}\right)\delta\left(\frac {p_sq} m  \cos\theta +\frac{q^2}{2m}+s^\prime\alpha|\bom{p}_s+\bom{q}|-s\alpha p_s\right).
\label{selfenergy1-1}
\end{align}
The $\delta$-function in (\ref{selfenergy1-1})
enforces the mass-shell condition
 $|\bom{p}_s+\bom{q}|=p_{s^\prime}$, from which one can find 
\begin{align}&\cos\theta= -\frac q{2p_s} +(s-s^\prime)\frac{m\alpha}q\left(1+s\frac{m\alpha}{p_s}\right)\\
&\cos\theta_{\bp_s+\bq}=1-\frac {q^2}{2p_s p_{s^\prime}}+|s-s'|\frac{(m\alpha)^2}{p_s p_{s^\prime}},
\end{align}
where we used that $\cos\theta_{\bp+\bq}=(p+q \cos\theta)/{|\bom{p+q}|}$ for arbitrary $\bp$ and $\bq$.
Integrating over $\theta$,  we arrive at
\begin{align}\Delta\Sigma_s^{1}=\frac{\varepsilon_{s}^{\bf p}}{2\pi^2v_0}&\left[\int^{2p_s}_0dq U_q\sqrt{1-\frac{q^2}{4p_s^2}}\left(1-\frac {q^2}{2p_s mv_0}-\frac{s\alpha}v\frac {q^2}{2p_s^2}\right)
\right.
\notag\\&+\left.
\int^{2mv_0}_{2m\alpha} dq U_q\frac{\left[\frac {q^2}{4p_F^2}-\left(\frac{m\alpha}{p_F}\right)^2\right]\left(1-\frac {q^2}{2p_s mv_0}+s\frac{\alpha}{v_0}\frac {q^2}{2p_F^2}+\frac{2m\alpha^2}{v_0p_s}-s\frac{2m^2\alpha^3}{v_0p_F^2}\right)}{\left[1-\frac{s\alpha}{v_0}\left(1+\frac{p_s}{p_{-s}}\right)\right]\sqrt{1-\left[\frac {q}{2p_s}-s\frac{2m\alpha}q\left(1+\frac{sm\alpha}{p_s}\right)\right]^2}}
\right],
\label{selfenergy1-2}
\end{align}
where the first (second) term is a result of intra-band (inter-band) transitions. Equation (\ref{selfenergy1-2}) is exact in $\alpha$.
If $U_q$ is replaced by the screened Coulomb potential from Eq.~(\ref{TF}), the first term in Eq.~(\ref{selfenergy1-2}) produces, in the small-$q$ limit,  the $r_s\ln r_s$ result of Eq.~(\ref{mass-splitting}):
\beq  \Delta\Sigma_s^{1}=\frac{\varepsilon_{s}^{\bf p}}{2\pi^2v_0}\int^{2p_s}_0dq U^{\mathrm{TF}}_q= \frac{\epsilon_{s}^\bp}{\pi v_F}e^2\ln\frac {p_s}{p_{\mathrm{TF}}}.\label{521a}\eeq
Note that, in contrast to Eq.~(\ref{mass-splitting}), where $p_{s}$ came as an upper cutoff a logarithmically divergent integral, the upper limit of the integral in Eq.~(\ref{521a}) is defined uniquely. 

Equation (\ref{selfenergy1-2}) also allows one to study the dependence of the self-energy on $\alpha$ at small $\alpha$, without assuming that this dependence is analytic. To first order in $\alpha$, the integrand can be expanded as 
\begin{align}\Delta\Sigma_s^{1}&=-\frac {\varepsilon_{s}^{\bf p}}{\pi^2v_F}\frac{s\alpha}{v_F}\int^{2p_F}_{0} dq U_q\frac{q^2}{2p_F^2}\sqrt{1-\frac{q^2}{4p_F^2}}.\label{sigma1}
\end{align}
where $p_F$ and $v_F$ are the Fermi momentum and Fermi velocity at $\alpha=0$, correspondingly.
On the other hand, integration over $\Omega$ in Eq.~(\ref{selfenergy2}) gives
\begin{align}\Delta\Sigma_s^{2}&=-\frac{m\varepsilon_{s}^{\bf p}}{p_s}\sum_{s^\prime }ss^\prime\sum_{q }\frac {U_q}2 
 \frac {\sin^2\theta_{\bom{p}_s+\bom{q}}} { |\bom{p}_s+\bom{q}|}
 \Theta\left({-\varepsilon_{s'}^{\bom{p}_s+\bom{q}}}\right)\notag\\
&=- \frac{m\varepsilon_{s}^{\bf p}}{p_s}\sum_{s^\prime }ss^\prime\sum_{q }\frac {U_q}2 
 \frac {\sin^2\theta_{\bom{p}_s+\bom{q}}} { |\bom{p}_s+\bom{q}|}\Theta\left(-\frac {p_sq} m  \cos\theta -\frac{q^2}{2m}-s^\prime\alpha|\bom{p}_s+\bom{q}|+s\alpha p_s\right).\label{sig2}
\end{align}
A term of order $\alpha$ can be obtained by expanding the $\Theta$-function with respect to $\alpha$
\begin{align}
\Delta\Sigma_s^{2}
&=\frac{m^2\varepsilon_{s}^{\bf p}}{p^2_s}\alpha s\sum_{s^\prime }\sum_{q }\frac {U_q}2 
\frac {
q^2\sin^2\theta
} {
 |\bom{p}_s+\bom{q}|^2}
\delta\left(\frac {p_Fq} m  \cos\theta +\frac{q^2}{2m}\right)\notag\\
&=\frac {\varepsilon_{s}^{\bf p}}{\pi^2v_F}\frac{s\alpha}{v_F}\int^{2p_F}_{0} dq U_q\frac{q^2}{2p_F^2}\sqrt{1-\frac{q^2}{4p_F^2}}.
\label{sigma2}\end{align}
Comparing Eqs.~(\ref{sigma1}) and (\ref{sigma2}), we see that  $\Delta\Sigma_s=\Delta\Sigma^{1}_s+\Delta\Sigma^{2}_s=0$ to order $\alpha$, which is an agreement with the previous results.\cite{saraga:2005,chesi-theorem} 

In the Coulomb case, Refs.~\onlinecite{saraga:2005,aasen:2012}  also find an $r_s\alpha^2\ln\alpha$ correction to the mass, which is the same for two Rashba subbands. This correction is produced by the second term in Eq.~(\ref{selfenergy1-2}). To see this, one can keep  only the logarithmic integrals in (\ref{selfenergy1-2}). Differentiating  (\ref{selfenergy1-2}) twice with respect to $\alpha$, we obtain
\beq \frac{ \partial^2\Delta\Sigma_s}{\partial\alpha^2}= -\frac{e^2\varepsilon_{s}^{\bf p}}{\pi v_F^3}
\int^{2mv_F}_{2m|\alpha|} \frac{dq}{q}=-
\frac{e^2\varepsilon_{s}^{\bf p}}{\pi v_F^3}
\ln
\frac{v_F}{\alpha},
\label{diff2}
\eeq
Integrating Eq.~(\ref{diff2}) back over $\alpha$, we reproduce the result of Ref.~\onlinecite{saraga:2005,aasen:2012}.
In addition, Ref.~\onlinecite{aasen:2012} obtains velocity splitting as 
$s r_s\alpha^3\ln \alpha$. Such a term is produced by both $\Sigma^1_s$ and $\Sigma_s^2$. Indeed, keeping again only logarithmic integrals and differentiating  (\ref{selfenergy1-2}) three times with respect to $\alpha$ gives
\beq \frac{ \partial^3\Delta\Sigma^1_s}{\partial\alpha^3}= -s\frac{3e^2\varepsilon_{s}^{\bf p}}{\pi v_F^4}
\int^{2mv_F}_{2m\alpha} \frac{dq}{q}=-s\frac{3e^2\varepsilon_{s}^{\bf p}}{\pi v_F^4}
\ln\frac{v_F}{\alpha}.
\label{d3_1}
\eeq
There is also another contribution of the same order coming from $\Delta\Sigma^2_s$. It follows from Eq.~ (\ref{sig2}) that
\begin{align}\Delta\Sigma_s^{2}&=-\frac {m\varepsilon_{s}^{\bf p}} {8\pi^2 p_s} \left [\int^{2p_s}_0 qdq U{(q)}\int^{-\frac q{2p_s}}_{-1}d(\cos\theta)\frac {q^2 \sin\theta}{(p_s^2+q^2+2qp_s\cos\theta)^{3/2}}\right.\notag\\
&\left. -\int^{2p_{-s}}_{4m|\alpha|} qdq U{(q)}\int^{-\frac q{2p_s}+\frac {4ms\alpha}q}_{-1}d(\cos\theta)\frac {q^2 \sin\theta}{(p_s^2+q^2+2qp_s \cos\theta)^{3/2}}
\right].\end{align}
Keeping again only logarithmic integrals, we find
\begin{align}\Delta\Sigma_s^{2}= \frac {m\varepsilon_{s}^{\bf p}} {16\pi^2 p_s}
\int^{2p_{-s}}_{4m|\alpha|} \frac{q^3}{p_s^3}dq U{(q)}\sqrt{1-
\left(\frac q{2p_s}-\frac {4ms\alpha}q\right)^2}\left(\frac {4ms\alpha}q-\frac q{2p_s}\right).\label{sigm2}\end{align}
Differentiating (\ref{sigm2}) three times with respect to $\alpha$ gives
\beq \frac{ \partial^3\Delta\Sigma^2_s}{\partial\alpha^3}= -s\frac{4e^2\varepsilon_{s}^{\bf p}}{\pi v_F^4}
\int^{2p_F}_{4m\alpha} \frac{dq}{q}=-s\frac{4e^2\varepsilon_{s}^{\bf p}}{\pi v_F^4}
\ln\frac{v_F}{\alpha}.
\label{d3_2}
\eeq
Combining Eqs.~(\ref{d3_1}) and (\ref{d3_2}),  we obtain
\beq \frac{ \partial^3\Delta\Sigma_s}{\partial\alpha^3}=-s\frac{7e^2\varepsilon_{s}^{\bf p}}{\pi v_F^4}
\ln\frac{v_F}{\alpha},
\eeq
which, after integration over $\alpha$, reproduces the main logarithmic term in the result of Ref.~\onlinecite{aasen:2012}.
[We believe that a constant inside the logarithm, also obtained in Ref.~\onlinecite{aasen:2012}, exceeds the overall accuracy of the calculation.]
Note that the velocity splitting from both contributions in Eqs.~(\ref{d3_1}) and (\ref{d3_2}) is a result of inter-band transitions only.

\section{\label{app:Bfree}Thermodynamic calculation of the out-of-plane spin susceptibility of non-interacting Rashba electrons}
In this Appendix, we show that the out-of-plane spin susceptibility, $\chi_{zz}$, of a non-interacting electron gas with Rashba SO coupling is determined by the states in between from the two spin-split FSs. Since we are dealing only with free states here, the index $f$ will be suppressed while quantities in the absence of the magnetic field will be denoted by the superscript $0$.  

In the presence of  a weak magnetic field $H$ in the $z$-direction, the electron spectrum changes to
\beq \epsilon_s(p)= \epsilon^{0}_s(p)+s \frac{\Delta^2}{2\alpha p}, \eeq 
where $\epsilon^{0}_s(p)$ coincides with Eq.~(\ref{e_free}) and $\Delta=g\mu_B H/2$.
The ground-state energy is given by
\beq E=\sum_s \int \frac{dpp}{2\pi}\epsilon_s(p) \Theta(p_s-p), \eeq
where the Fermi momenta of the subbands are found from 
\beq \epsilon_s(p_s)=\epsilon_F\label{appB1}.\eeq
It is easy to check that, at fixed number density, 
$\epsilon_F$ is not affected by the magnetic field.
One can thus replace $\epsilon_F$ by $\epsilon^0_s(p^0_s)$ in Eq.~(\ref{appB1}) which gives for the corrections to the Fermi momenta
\beq p_s=p_s^0+\delta p_s=p_s^0 - s\frac{\Delta^2}{2\alpha v_0 p_s^0}, \eeq
where 
is the Fermi velocity in each of the subbands, given by Eq.~(\ref{v0}). It is convenient to subtract the field-independent part from $E$, and split the remainder into two parts as
\beq E-E^0=E_{\mathrm{on}}+E_{\mathrm{off}}, \eeq
where $E_{\mathrm{on}}$ is the contribution from the states near the FSs
\beq
E_{\mathrm{on}}=\frac{1}{2\pi}\sum_s\int_{p^0_s}^{p^0_s+\delta p_s} dpp \epsilon_s^0(p)
\eeq
and $E_{\mathrm{off}}$ is the contribution from the states away from the FSs
\beq E_{\mathrm{off}}= \frac{1}{2\pi}\sum_s\int_{0}^{p^0_s} dpp \left(s\frac{\Delta^2}{2\alpha p}\right). \eeq
The on-the-Fermi-surface contribution vanishes to order $\Delta^2$
\beq E_{\mathrm{on}}= \frac{1}{2\pi}\sum_s p_s\delta p_s \epsilon_s^0(p_s)=0.
\eeq
 On the other hand, the off-FS contribution becomes
\beq E_{\mathrm{off}}= \frac{1}{2\pi}\int_{p^0_+}^{p^0_-} dpp \left(-\frac{\Delta^2}{2\alpha p}\right)=-\frac{m\Delta^2}{2\pi}, \eeq
which gives a correct result $\chi_{zz}^0=g^2\mu_B^2m/4\pi$ (Ref.~\onlinecite{zak:2010}). Therefore, the out-of-plane spin susceptibility of non-interacting Rashba fermions comes entirely from the states in between the two FSs.

\section{\label{app:D} Kohn anomaly 
in a free electron gas with Rashba spin-orbit coupling}
In this Appendix, we derive Eqs.~(\ref{bubble1a}-\ref{bubble1}) for the Kohn anomalies in
the polarization bubble of non-interacting Rashba fermions (Fig. \ref{bubblegraph}):
\begin{figure}
\includegraphics[scale=0.602]{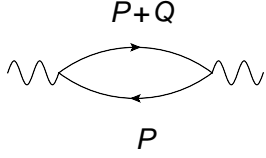}
\caption{\footnotesize Polarization bubble with $2+1$ momentum transfer $Q$, given by Eq. \ref{totalBubble}.}
\label{bubblegraph}
\end{figure}
\beq \Pi(Q)=\sum_{s,s'}\Pi_{ss'}(Q),\label{totalBubble}\eeq
where
\bea
\Pi_{ss'}(Q)&=&\frac{1}{2}\int_P [1+ss'\cos(\theta_\bom{p}-\theta_{\bom{p+q}})]g_s(P)g_{s'}(P+Q)=\frac{1}{2}\int\frac{d^2p}{(2\pi)^2} [1+ss'\cos(\theta_\bom{p}-\theta_{\bom{p+q}})]\Theta(-\epsilon_{s,\bf p})\notag\\
&&\times\left [\frac{1}{i\Omega-\epsilon_{s',\bf{p+q}}+\epsilon_{s,k}}-\frac{1}{i\Omega+\epsilon_{s,\bf{p-q}}-\epsilon_{s',\bf{p}}}\right].\nn\\
\label{bubble}
\eea
(As in Appendices~\ref{app:A} and \ref{app:Bfree}, we suppress the superscript $f$ denoting the properties of an interaction-free system.) 
For $q$ near $2p_F$,  we expand the difference of the quasiparticle energies as
\beq \epsilon_{s}^{\bf{p+q}}-\epsilon_{s}^{\bf p}=-2\epsilon_{s}^{\bf p}+ v_Fq'+v_Fp_F\theta^2,\label{disp} \eeq
where $q'=q-2p_s$ and the angle $\theta=\angle(-\bom{p},\bom{q})$ is taken to be small. To leading order in SO, the prefactors in the $q'$ and $\theta^2$ terms can be and were replaced by their values in the absence of SO. (For the second term in Eq.~(\ref{bubble}), the angle $\theta=\angle(\bom{p},\bom{q})$ is taken to be small.) Using Eq.~(\ref{disp}), replacing $[1+\cos(\theta_\bom{p}-\theta_{\bom{p+q}})]$ by $2\theta^2$, and integrating over $\epsilon_{\bp}$ in Eq.~(\ref{bubble}), we obtain for
 $\Omega=0$ and $s=s'$
\beq \Pi_{ss}(q)=\frac{\nu}{2\pi}\int_0^{\theta_c} \theta^2 \ln\left(\theta^2 +\frac{q'}{p_F}\right)d\theta, \label{t2}\eeq
where $\theta_c$ is a cutoff whose particular choice is not going to affect the singular part of the $q$-dependence. 
To extract the singular part, we differentiate $ \Pi_{ss}(q)$ twice with respect to $q$, which makes the integral to converge, and obtain
\beq \Pi_{ss}(q)=\Pi_{ss}(q=2p_s)-\frac{\nu}{6}\Theta(q-2p_s) \left(\frac{q-2p_s}{p_F}\right)^{3/2}. \label{pipp} \eeq
The $(q')^{3/2}$ anomaly in $\Pi_{ss}(q)$ is weaker than the square-root anomaly in the absence of SO. This is a consequence of the fact that backscattering within the same Rashba subband is forbidden, which is manifested by a small factor of $\theta^2$ in the integrand of Eq.~(\ref{t2}). A similar procedure is applied to $\Pi_{s,-s}$, which results from backscattering between different subbands and hence does not have  a small factor of $\theta^2$.  With $q'$ replaced now by $q-2p_F$, we obtain the usual square-root Kohn anomaly 
\beq \Pi_{s,-s}(q)=\Pi_{s,-s}(q=2p_F)+\frac{\nu}{2} \Theta(q-2p_F)\left(\frac{q-2p_F}{p_F}\right)^{1/2}. \label{pipm}\eeq

Note that when the bubble is inserted into the interaction vertex,  as in Fig.~\ref{2ndorder}, the maximum momentum entering the bubble is $2p_{-}$ (when both the incoming electrons are the $-$ subband). Therefore, the singularity in $\Pi_{--}$, present only for $q>2p_{-}$, is outside the range of allowed momenta.  With this in mind,  
the non-analytic part of $\Pi$, relevant for the calculation of the Landau function in Sec.~\ref{sec:results}, can be written as in  Eqs.~(\ref{bubble1a}-\ref{bubble1}) of the main text.
\section{\label{app:LFsplitting} Non-analytic contributions to the Landau function}
In this Appendix, we calculate non-analytic contributions from the second-order vertices $b$  and $d$ of Fig.~\ref{2ndorder} to the $s\alpha^2$ term in the Landau function, discussed in Sec.~\ref{sec:results}. A contribution from diagram $c$ is calculated in the main text of Sec.~\ref{sec:results}. 

Diagram $d$ reads
\bea
\Gamma^d_{ss'}&=&-\frac {U^2}4 \sum_{r,t,L}[1+sr\cos(\theta_{\bf p}-\theta_{\bf l})][1+s't\cos(\theta_{\bf p'}-\theta_{\bf l'})]g_r(L)g_t(L'),\label{Gd}
\eea
where $L'=(\bom{l+p'-p},\omega_l)$. The non-analytic contribution to $\Gamma^d_{ss'}$ comes from backscattering processes with $\bp\approx -\bpp$. 
Therefore, internal momenta must be chosen such that $\bom{l}\approx \bom{p}$ and $\bom{l'\approx p'\approx -p}$, and both cosine terms in Eq.~(\ref{Gd})  are near $+1$. (The magnitudes of the internal momenta may differ from external ones, but taking this effect into account is not necessary to lowest order in $\alpha$.) The $s=s'=+1$ channel does not contain a singularity, because the maximum momentum transfer in this channel is $2p_{+}$, while any convolution of two Green's functions does not depend on the difference of their momenta for $q\leq 2p_{+}$ (cf. Appendix \ref{app:D}).  We thus focus on the $s=s'=-1$ and $s=-s'$ channels.   In the $s=s'=-1$ channel, the angular-dependent factors are maximal for $r=t=-1$; however, the singularity in the convolution of two $g_{-}$ function is outside the allowed momentum transfer range. The next best choice $r=-t$, when one of angular-dependent factors is small but the second one is large. The choice $t=r=1$ makes both angular-dependent factors to be small and is discarded. With all angles measured from $\bp$  ($\theta_{\bp}=0$), we denote
$\theta_\bpp=\pi-\theta$, $\theta_{\bl}=\phi$, and $\theta_{\blp}=\pi-\phi'$ and take $\theta$, $\phi$, and $\phi'$ to small.   After some elementary geometry, we find that $\phi'=\phi+\theta$, which implies   that the angular-dependent factors in front of the $g_{+}g_{-}$ and $g_{-}g_{+}$ combinations are the same. Therefore,
\beq
\Gamma^d_{--}=-\frac {U^2}2\int_L \phi^2g_+(L)g_-(L'),
\label{Ga--}
\eeq
and similarly
\beq
\Gamma^d_{+-}=-\frac {U^2}4\int_L \phi^2g_+(L)g_+(L').
\label{Ga+-}
\eeq
The integrals in the equations above are the same as for the $\Pi_{++}$ component of the polarization bubble [cf. Eq.~(\ref{t2}], except for the momentum transfer in $\Gamma^d_{--}$ is measured from $2p_F$  rather than from $2p_+$. With this difference taken into account, we obtain the results for $\Gamma^d_{--}$ and 
$\Gamma^d_{+-}$ in Eqs.~(\ref{Gd--}) and (\ref{Gd+-}) of the main text.

The two \lq\lq wine-glass\rq\rq\/ diagrams $b$) are equal to each other, and their sum is given by 
\bea
\Gamma^b_{ss'}&=&\frac{U^2}{8} \sum_{r,t}\int_L[1-ss'e^{i(\theta_{\bf p'}-\theta_{\bf p})}]
[1+rte^{i(\theta_{\bf l}-\theta_{\bf l'})}]
[1+sre^{i(\theta_{\bf p}-\theta_{\bf l})}][1+s'te^{i(\theta_{\bf l'}-\theta_{\bf p'})}]g_r(L)g_t(L').\notag\\
\eea
Again, we have $\bom{l\approx p}$ and $\bom{l'\approx p'\sim -p}$, so that $r=-t$ for the $s=s'=-1$ channel.  
Using the definitions of small angles above, we obtain
\bea
\Gamma^b_{--}&=&\frac{U^2}{2}\paren{1-e^{-i\theta}}
\int_L\paren{1+e^{i(2\phi+\theta)}}\paren{1-e^{-2i\phi}}\bra{g_+(L)g_-(L')+g_-(L)g_+(L')}
\label{Gb--1}.
\eea
Expanding the internal angular-dependent factors to first order in $\phi$ and $\theta$ yields an integral which vanishes by parity. 
The second order term in $\phi$ is finite but it is still multiplied by a small factor of $\theta$ from the external matrix element. Therefore, $\Gamma^{b}_{--}$ does not contribute to order $\alpha^2$.

For the $s=-s'$ channel, only the combination $r=t=+1$ contributes. Following the same steps as for $\Gamma^b_{--}$, we arrive at  
\beq
\Gamma^b_{+-}=\frac{U^2}{8}\paren{1+e^{-i\theta}}\int_L\paren{1-e^{i(2\phi+\theta)}}\paren{1-e^{-2i\phi}} g_+(L)g_+(L').
\label{Gb+-1}
\eeq
Expanding the angular factors to lowest order, we obtain
\beq
\Gamma^b_{+-}=U^2\int_L \phi^2g_+(L)g_+(L'),
\label{Gb+-}
\eeq
which coincides with the integral for $\Pi_{ss}$ in Eq.~(\ref{t2}). Combining everything together, we obtain the vertices in Eqs.~(\ref{Gb--}) and (\ref{Gb+-m}) of the main text.
\ewt
\bibliography{SOFLI}

\begin{thebibliography}{103}%
\makeatletter
\providecommand \@ifxundefined [1]{%
 \@ifx{#1\undefined}
}%
\providecommand \@ifnum [1]{%
 \ifnum #1\expandafter \@firstoftwo
 \else \expandafter \@secondoftwo
 \fi
}%
\providecommand \@ifx [1]{%
 \ifx #1\expandafter \@firstoftwo
 \else \expandafter \@secondoftwo
 \fi
}%
\providecommand \natexlab [1]{#1}%
\providecommand \enquote  [1]{``#1''}%
\providecommand \bibnamefont  [1]{#1}%
\providecommand \bibfnamefont [1]{#1}%
\providecommand \citenamefont [1]{#1}%
\providecommand \href@noop [0]{\@secondoftwo}%
\providecommand \href [0]{\begingroup \@sanitize@url \@href}%
\providecommand \@href[1]{\@@startlink{#1}\@@href}%
\providecommand \@@href[1]{\endgroup#1\@@endlink}%
\providecommand \@sanitize@url [0]{\catcode `\\12\catcode `\$12\catcode
  `\&12\catcode `\#12\catcode `\^12\catcode `\_12\catcode `\%12\relax}%
\providecommand \@@startlink[1]{}%
\providecommand \@@endlink[0]{}%
\providecommand \url  [0]{\begingroup\@sanitize@url \@url }%
\providecommand \@url [1]{\endgroup\@href {#1}{\urlprefix }}%
\providecommand \urlprefix  [0]{URL }%
\providecommand \Eprint [0]{\href }%
\providecommand \doibase [0]{http://dx.doi.org/}%
\providecommand \selectlanguage [0]{\@gobble}%
\providecommand \bibinfo  [0]{\@secondoftwo}%
\providecommand \bibfield  [0]{\@secondoftwo}%
\providecommand \translation [1]{[#1]}%
\providecommand \BibitemOpen [0]{}%
\providecommand \bibitemStop [0]{}%
\providecommand \bibitemNoStop [0]{.\EOS\space}%
\providecommand \EOS [0]{\spacefactor3000\relax}%
\providecommand \BibitemShut  [1]{\csname bibitem#1\endcsname}%
\let\auto@bib@innerbib\@empty
\bibitem [{\citenamefont {\ifmmode \check{Z}\else
  \v{Z}\fi{}uti\ifmmode~\acute{c}\else \'{c}\fi{}}\ \emph
  {et~al.}(2004)\citenamefont {\ifmmode \check{Z}\else
  \v{Z}\fi{}uti\ifmmode~\acute{c}\else \'{c}\fi{}}, \citenamefont {Fabian},\
  and\ \citenamefont {Das~Sarma}}]{zutic:2004}%
  \BibitemOpen
  \bibfield  {author} {\bibinfo {author} {\bibfnamefont {I.}~\bibnamefont
  {\ifmmode \check{Z}\else \v{Z}\fi{}uti\ifmmode~\acute{c}\else \'{c}\fi{}}},
  \bibinfo {author} {\bibfnamefont {J.}~\bibnamefont {Fabian}}, \ and\ \bibinfo
  {author} {\bibfnamefont {S.}~\bibnamefont {Das~Sarma}},\ }\href {\doibase
  10.1103/RevModPhys.76.323} {\bibfield  {journal} {\bibinfo  {journal} {Rev.
  Mod. Phys.}\ }\textbf {\bibinfo {volume} {{\bf 76}}},\ \bibinfo {pages} {323}
  (\bibinfo {year} {2004})}\BibitemShut {NoStop}%
\bibitem [{\citenamefont {Winkler}(2003)}]{winkler:book}%
  \BibitemOpen
  \bibfield  {author} {\bibinfo {author} {\bibfnamefont {R.}~\bibnamefont
  {Winkler}},\ }\href@noop {} {\emph {\bibinfo {title} {Spin--Orbit Coupling
  Effects in Two-Dimensional Electron and Hole Systems}}}\ (\bibinfo
  {publisher} {Springer, Berlin/Heidelberg},\ \bibinfo {year}
  {2003})\BibitemShut {NoStop}%
\bibitem [{\citenamefont {{K.V. Samokhin}}(2009)}]{samokhin:2009}%
  \BibitemOpen
  \bibfield  {author} {\bibinfo {author} {\bibnamefont {{K.V. Samokhin}}},\
  }\href@noop {} {\bibfield  {journal} {\bibinfo  {journal} {Ann. Phys.}\
  }\textbf {\bibinfo {volume} {324}},\ \bibinfo {pages} {2385} (\bibinfo {year}
  {2009})}\BibitemShut {NoStop}%
\bibitem [{\citenamefont {Sigrist}\ and\ \citenamefont
  {Ueda}(1991)}]{sigrist:1991}%
  \BibitemOpen
  \bibfield  {author} {\bibinfo {author} {\bibfnamefont {M.}~\bibnamefont
  {Sigrist}}\ and\ \bibinfo {author} {\bibfnamefont {K.}~\bibnamefont {Ueda}},\
  }\href@noop {} {\bibfield  {journal} {\bibinfo  {journal} {Rev. Mod. Phys.}\
  }\textbf {\bibinfo {volume} {63}},\ \bibinfo {pages} {239} (\bibinfo {year}
  {1991})}\BibitemShut {NoStop}%
\bibitem [{\citenamefont {{V. P. Mineev and M.
  Sigrist}}(2012)}]{mineev_sigrist}%
  \BibitemOpen
  \bibfield  {author} {\bibinfo {author} {\bibnamefont {{V. P. Mineev and M.
  Sigrist}}},\ }in\ \href@noop {} {\emph {\bibinfo {booktitle}
  {Non-Centrosymmetric Superconductors}}},\ \bibinfo {series and number}
  {Lecture Notes in Physics},\ \bibinfo {editor} {edited by\ \bibinfo {editor}
  {\bibnamefont {{E. Bauer and M. Sigrist}}}}\ (\bibinfo  {publisher} {Springer
  Berlin / Heidelberg},\ \bibinfo {year} {2012})\ pp.\ \bibinfo {pages}
  {129--154}\BibitemShut {NoStop}%
\bibitem [{\citenamefont {Hasan}\ and\ \citenamefont
  {Kane}(2010)}]{hasan:2010}%
  \BibitemOpen
  \bibfield  {author} {\bibinfo {author} {\bibfnamefont {M.~Z.}\ \bibnamefont
  {Hasan}}\ and\ \bibinfo {author} {\bibfnamefont {C.~L.}\ \bibnamefont
  {Kane}},\ }\href {\doibase 10.1103/RevModPhys.82.3045} {\bibfield  {journal}
  {\bibinfo  {journal} {Rev. Mod. Phys.}\ }\textbf {\bibinfo {volume} {{\bf
  82}}},\ \bibinfo {pages} {3045} (\bibinfo {year} {2010})}\BibitemShut
  {NoStop}%
\bibitem [{\citenamefont {Hasan}\ and\ \citenamefont
  {Moore}(2011)}]{hasan:2011}%
  \BibitemOpen
  \bibfield  {author} {\bibinfo {author} {\bibfnamefont {M.~Z.}\ \bibnamefont
  {Hasan}}\ and\ \bibinfo {author} {\bibfnamefont {J.~E.}\ \bibnamefont
  {Moore}},\ }\href@noop {} {\bibfield  {journal} {\bibinfo  {journal} {Ann.
  Rev. Cond. Matt. Phys.}\ }\textbf {\bibinfo {volume} {{\bf 2}}},\ \bibinfo
  {pages} {55} (\bibinfo {year} {2011})}\BibitemShut {NoStop}%
\bibitem [{\citenamefont {Qi}\ and\ \citenamefont {Zhang}(2011)}]{qi:2011}%
  \BibitemOpen
  \bibfield  {author} {\bibinfo {author} {\bibfnamefont {X.-L.}\ \bibnamefont
  {Qi}}\ and\ \bibinfo {author} {\bibfnamefont {S.-C.}\ \bibnamefont {Zhang}},\
  }\href {\doibase 10.1103/RevModPhys.83.1057} {\bibfield  {journal} {\bibinfo
  {journal} {Rev. Mod. Phys.}\ }\textbf {\bibinfo {volume} {83}},\ \bibinfo
  {pages} {1057} (\bibinfo {year} {2011})}\BibitemShut {NoStop}%
\bibitem [{\citenamefont {Alicea}(2012)}]{alicea:2012}%
  \BibitemOpen
  \bibfield  {author} {\bibinfo {author} {\bibfnamefont {J.}~\bibnamefont
  {Alicea}},\ }\href@noop {} {\bibfield  {journal} {\bibinfo  {journal} {Rep.
  Prog. Phys.}\ }\textbf {\bibinfo {volume} {75}},\ \bibinfo {pages} {076501}
  (\bibinfo {year} {2012})}\BibitemShut {NoStop}%
\bibitem [{\citenamefont {Bardarson}\ and\ \citenamefont
  {Moore}(2013)}]{moore:2012}%
  \BibitemOpen
  \bibfield  {author} {\bibinfo {author} {\bibfnamefont {J.~H.}\ \bibnamefont
  {Bardarson}}\ and\ \bibinfo {author} {\bibfnamefont {J.~E.}\ \bibnamefont
  {Moore}},\ }\href@noop {} {\bibfield  {journal} {\bibinfo  {journal} {Rep.
  Prog. Phys.}\ }\textbf {\bibinfo {volume} {76}},\ \bibinfo {pages} {056501}
  (\bibinfo {year} {2013})}\BibitemShut {NoStop}%
\bibitem [{\citenamefont {{Cen}}\ \emph {et~al.}(2009)\citenamefont {{Cen}},
  \citenamefont {{Thiel}}, \citenamefont {{Mannhart}},\ and\ \citenamefont
  {{Levy}}}]{cen:2009}%
  \BibitemOpen
  \bibfield  {author} {\bibinfo {author} {\bibfnamefont {C.}~\bibnamefont
  {{Cen}}}, \bibinfo {author} {\bibfnamefont {S.}~\bibnamefont {{Thiel}}},
  \bibinfo {author} {\bibfnamefont {J.}~\bibnamefont {{Mannhart}}}, \ and\
  \bibinfo {author} {\bibfnamefont {J.}~\bibnamefont {{Levy}}},\ }\href
  {\doibase 10.1126/science.1168294} {\bibfield  {journal} {\bibinfo  {journal}
  {Science}\ }\textbf {\bibinfo {volume} {{\bf 323}}},\ \bibinfo {pages} {1026}
  (\bibinfo {year} {2009})}\BibitemShut {NoStop}%
\bibitem [{\citenamefont {Lin}\ \emph {et~al.}(2009)\citenamefont {Lin},
  \citenamefont {Compton}, \citenamefont {Perry}, \citenamefont {Phillips},
  \citenamefont {Porto},\ and\ \citenamefont {Spielman}}]{lin:2009}%
  \BibitemOpen
  \bibfield  {author} {\bibinfo {author} {\bibfnamefont {Y.-J.}\ \bibnamefont
  {Lin}}, \bibinfo {author} {\bibfnamefont {R.~L.}\ \bibnamefont {Compton}},
  \bibinfo {author} {\bibfnamefont {A.~R.}\ \bibnamefont {Perry}}, \bibinfo
  {author} {\bibfnamefont {W.~D.}\ \bibnamefont {Phillips}}, \bibinfo {author}
  {\bibfnamefont {J.~V.}\ \bibnamefont {Porto}}, \ and\ \bibinfo {author}
  {\bibfnamefont {I.~B.}\ \bibnamefont {Spielman}},\ }\href {\doibase
  10.1103/PhysRevLett.102.130401} {\bibfield  {journal} {\bibinfo  {journal}
  {Phys. Rev. Lett.}\ }\textbf {\bibinfo {volume} {102}},\ \bibinfo {pages}
  {130401} (\bibinfo {year} {2009})}\BibitemShut {NoStop}%
\bibitem [{\citenamefont {{Y.-J. Lin, K. Jim{\'e}nez-Garc{\'\i}a, and I. B.
  Spielman}}(2011)}]{cold_Rashba_Bose:2011}%
  \BibitemOpen
  \bibfield  {author} {\bibinfo {author} {\bibnamefont {{Y.-J. Lin, K.
  Jim{\'e}nez-Garc{\'\i}a, and I. B. Spielman}}},\ }\href@noop {} {\bibfield
  {journal} {\bibinfo  {journal} {Nature}\ }\textbf {\bibinfo {volume} {471}}
  (\bibinfo {year} {2011})}\BibitemShut {NoStop}%
\bibitem [{\citenamefont {Wang}\ \emph {et~al.}(2012)\citenamefont {Wang},
  \citenamefont {Yu}, \citenamefont {Fu}, \citenamefont {Miao}, \citenamefont
  {Huang}, \citenamefont {Chai}, \citenamefont {Zhai},\ and\ \citenamefont
  {Zhang}}]{wang:2012}%
  \BibitemOpen
  \bibfield  {author} {\bibinfo {author} {\bibfnamefont {P.}~\bibnamefont
  {Wang}}, \bibinfo {author} {\bibfnamefont {Z.-Q.}\ \bibnamefont {Yu}},
  \bibinfo {author} {\bibfnamefont {Z.}~\bibnamefont {Fu}}, \bibinfo {author}
  {\bibfnamefont {J.}~\bibnamefont {Miao}}, \bibinfo {author} {\bibfnamefont
  {L.}~\bibnamefont {Huang}}, \bibinfo {author} {\bibfnamefont
  {S.}~\bibnamefont {Chai}}, \bibinfo {author} {\bibfnamefont {H.}~\bibnamefont
  {Zhai}}, \ and\ \bibinfo {author} {\bibfnamefont {J.}~\bibnamefont {Zhang}},\
  }\href {\doibase 10.1103/PhysRevLett.109.095301} {\bibfield  {journal}
  {\bibinfo  {journal} {Phys. Rev. Lett.}\ }\textbf {\bibinfo {volume} {109}},\
  \bibinfo {pages} {095301} (\bibinfo {year} {2012})}\BibitemShut {NoStop}%
\bibitem [{\citenamefont {Cheuk}\ \emph {et~al.}(2012)\citenamefont {Cheuk},
  \citenamefont {Sommer}, \citenamefont {Hadzibabic}, \citenamefont {Yefsah},
  \citenamefont {Bakr},\ and\ \citenamefont {Zwierlein}}]{cheuk:2012}%
  \BibitemOpen
  \bibfield  {author} {\bibinfo {author} {\bibfnamefont {L.~W.}\ \bibnamefont
  {Cheuk}}, \bibinfo {author} {\bibfnamefont {A.~T.}\ \bibnamefont {Sommer}},
  \bibinfo {author} {\bibfnamefont {Z.}~\bibnamefont {Hadzibabic}}, \bibinfo
  {author} {\bibfnamefont {T.}~\bibnamefont {Yefsah}}, \bibinfo {author}
  {\bibfnamefont {W.~S.}\ \bibnamefont {Bakr}}, \ and\ \bibinfo {author}
  {\bibfnamefont {M.~W.}\ \bibnamefont {Zwierlein}},\ }\href {\doibase
  10.1103/PhysRevLett.109.095302} {\bibfield  {journal} {\bibinfo  {journal}
  {Phys. Rev. Lett.}\ }\textbf {\bibinfo {volume} {109}},\ \bibinfo {pages}
  {095302} (\bibinfo {year} {2012})}\BibitemShut {NoStop}%
\bibitem [{\citenamefont {Polini}\ \emph {et~al.}(2007)\citenamefont {Polini},
  \citenamefont {Asgari}, \citenamefont {Barlas}, \citenamefont
  {Pereg-Barnea},\ and\ \citenamefont {MacDonald}}]{polini:2007}%
  \BibitemOpen
  \bibfield  {author} {\bibinfo {author} {\bibfnamefont {M.}~\bibnamefont
  {Polini}}, \bibinfo {author} {\bibfnamefont {R.}~\bibnamefont {Asgari}},
  \bibinfo {author} {\bibfnamefont {Y.}~\bibnamefont {Barlas}}, \bibinfo
  {author} {\bibfnamefont {T.}~\bibnamefont {Pereg-Barnea}}, \ and\ \bibinfo
  {author} {\bibfnamefont {A.}~\bibnamefont {MacDonald}},\ }\href {\doibase
  10.1016/j.ssc.2007.04.035} {\bibfield  {journal} {\bibinfo  {journal} {Solid
  St. Comm.}\ }\textbf {\bibinfo {volume} {{\bf 143}}},\ \bibinfo {pages} {58}
  (\bibinfo {year} {2007})}\BibitemShut {NoStop}%
\bibitem [{\citenamefont {Quay}\ \emph {et~al.}(2010)\citenamefont {Quay},
  \citenamefont {Hughes}, \citenamefont {Sulpizio}, \citenamefont {Pfeiffer},
  \citenamefont {Baldwin}, \citenamefont {West}, \citenamefont
  {Goldhaber-Gordon},\ and\ \citenamefont {de~Picciotto}}]{quay:2010}%
  \BibitemOpen
  \bibfield  {author} {\bibinfo {author} {\bibfnamefont {C.~H.~L.}\
  \bibnamefont {Quay}}, \bibinfo {author} {\bibfnamefont {T.~L.}\ \bibnamefont
  {Hughes}}, \bibinfo {author} {\bibfnamefont {J.~A.}\ \bibnamefont
  {Sulpizio}}, \bibinfo {author} {\bibfnamefont {L.~N.}\ \bibnamefont
  {Pfeiffer}}, \bibinfo {author} {\bibfnamefont {K.~W.}\ \bibnamefont
  {Baldwin}}, \bibinfo {author} {\bibfnamefont {K.~W.}\ \bibnamefont {West}},
  \bibinfo {author} {\bibfnamefont {D.}~\bibnamefont {Goldhaber-Gordon}}, \
  and\ \bibinfo {author} {\bibfnamefont {R.}~\bibnamefont {de~Picciotto}},\
  }\href@noop {} {\bibfield  {journal} {\bibinfo  {journal} {{Nat. Phys.}}\
  }\textbf {\bibinfo {volume} {{\bf 6}}},\ \bibinfo {pages} {336} (\bibinfo
  {year} {2010})}\BibitemShut {NoStop}%
\bibitem [{\citenamefont {Mourik}\ \emph {et~al.}(2012)\citenamefont {Mourik},
  \citenamefont {Zuo}, \citenamefont {Frolov}, \citenamefont {Plissard},
  \citenamefont {Bakkers},\ and\ \citenamefont {Kouwenhoven}}]{mourik:2012}%
  \BibitemOpen
  \bibfield  {author} {\bibinfo {author} {\bibfnamefont {V.}~\bibnamefont
  {Mourik}}, \bibinfo {author} {\bibfnamefont {K.}~\bibnamefont {Zuo}},
  \bibinfo {author} {\bibfnamefont {S.~M.}\ \bibnamefont {Frolov}}, \bibinfo
  {author} {\bibfnamefont {S.~R.}\ \bibnamefont {Plissard}}, \bibinfo {author}
  {\bibfnamefont {E.~P. A.~M.}\ \bibnamefont {Bakkers}}, \ and\ \bibinfo
  {author} {\bibfnamefont {L.~P.}\ \bibnamefont {Kouwenhoven}},\ }\href
  {\doibase 10.1126/science.1222360} {\bibfield  {journal} {\bibinfo  {journal}
  {Science}\ }\textbf {\bibinfo {volume} {{\bf 336}}},\ \bibinfo {pages} {1003}
  (\bibinfo {year} {2012})}\BibitemShut {NoStop}%
\bibitem [{\citenamefont {Das}\ \emph {et~al.}(2012)\citenamefont {Das},
  \citenamefont {Ronen}, \citenamefont {Most}, \citenamefont {Oreg},
  \citenamefont {Heiblum},\ and\ \citenamefont {Shtrikman}}]{das:2012}%
  \BibitemOpen
  \bibfield  {author} {\bibinfo {author} {\bibfnamefont {A.}~\bibnamefont
  {Das}}, \bibinfo {author} {\bibfnamefont {Y.}~\bibnamefont {Ronen}}, \bibinfo
  {author} {\bibfnamefont {Y.}~\bibnamefont {Most}}, \bibinfo {author}
  {\bibfnamefont {Y.}~\bibnamefont {Oreg}}, \bibinfo {author} {\bibfnamefont
  {M.}~\bibnamefont {Heiblum}}, \ and\ \bibinfo {author} {\bibfnamefont
  {H.}~\bibnamefont {Shtrikman}},\ }\href@noop {} {\bibfield  {journal}
  {\bibinfo  {journal} {Nature Phys.}\ }\textbf {\bibinfo {volume} {8}},\
  \bibinfo {pages} {887} (\bibinfo {year} {2012})}\BibitemShut {NoStop}%
\bibitem [{\citenamefont {Varykhalov}\ \emph {et~al.}(2008)\citenamefont
  {Varykhalov}, \citenamefont {S\'anchez-Barriga}, \citenamefont {Shikin},
  \citenamefont {Biswas}, \citenamefont {Vescovo}, \citenamefont {Rybkin},
  \citenamefont {Marchenko},\ and\ \citenamefont {Rader}}]{varykhalov:2008}%
  \BibitemOpen
  \bibfield  {author} {\bibinfo {author} {\bibfnamefont {A.}~\bibnamefont
  {Varykhalov}}, \bibinfo {author} {\bibfnamefont {J.}~\bibnamefont
  {S\'anchez-Barriga}}, \bibinfo {author} {\bibfnamefont {A.~M.}\ \bibnamefont
  {Shikin}}, \bibinfo {author} {\bibfnamefont {C.}~\bibnamefont {Biswas}},
  \bibinfo {author} {\bibfnamefont {E.}~\bibnamefont {Vescovo}}, \bibinfo
  {author} {\bibfnamefont {A.}~\bibnamefont {Rybkin}}, \bibinfo {author}
  {\bibfnamefont {D.}~\bibnamefont {Marchenko}}, \ and\ \bibinfo {author}
  {\bibfnamefont {O.}~\bibnamefont {Rader}},\ }\href {\doibase
  10.1103/PhysRevLett.101.157601} {\bibfield  {journal} {\bibinfo  {journal}
  {Phys. Rev. Lett.}\ }\textbf {\bibinfo {volume} {{\bf 101}}},\ \bibinfo
  {pages} {157601} (\bibinfo {year} {2008})}\BibitemShut {NoStop}%
\bibitem [{\citenamefont {{D. Marchenko, A. Varykhalov, M.R. Scholz, G.
  Bihlmayer, E.I. Rashba, A. Rybkin, A.M. Shikin, and O.
  Rader}}(2012)}]{marchenko:2012}%
  \BibitemOpen
  \bibfield  {author} {\bibinfo {author} {\bibnamefont {{D. Marchenko, A.
  Varykhalov, M.R. Scholz, G. Bihlmayer, E.I. Rashba, A. Rybkin, A.M. Shikin,
  and O. Rader}}},\ }\href@noop {} {\bibfield  {journal} {\bibinfo  {journal}
  {Nature Commun.}\ }\textbf {\bibinfo {volume} {3}},\ \bibinfo {pages} {1232}
  (\bibinfo {year} {2012})}\BibitemShut {NoStop}%
\bibitem [{\citenamefont {Ast}\ \emph {et~al.}(2007)\citenamefont {Ast},
  \citenamefont {Henk}, \citenamefont {Ernst}, \citenamefont {Moreschini},
  \citenamefont {Falub}, \citenamefont {Pacil\'e}, \citenamefont {Bruno},
  \citenamefont {Kern},\ and\ \citenamefont {Grioni}}]{ast:2007}%
  \BibitemOpen
  \bibfield  {author} {\bibinfo {author} {\bibfnamefont {C.~R.}\ \bibnamefont
  {Ast}}, \bibinfo {author} {\bibfnamefont {J.}~\bibnamefont {Henk}}, \bibinfo
  {author} {\bibfnamefont {A.}~\bibnamefont {Ernst}}, \bibinfo {author}
  {\bibfnamefont {L.}~\bibnamefont {Moreschini}}, \bibinfo {author}
  {\bibfnamefont {M.~C.}\ \bibnamefont {Falub}}, \bibinfo {author}
  {\bibfnamefont {D.}~\bibnamefont {Pacil\'e}}, \bibinfo {author}
  {\bibfnamefont {P.}~\bibnamefont {Bruno}}, \bibinfo {author} {\bibfnamefont
  {K.}~\bibnamefont {Kern}}, \ and\ \bibinfo {author} {\bibfnamefont
  {M.}~\bibnamefont {Grioni}},\ }\href {\doibase 10.1103/PhysRevLett.98.186807}
  {\bibfield  {journal} {\bibinfo  {journal} {Phys. Rev. Lett.}\ }\textbf
  {\bibinfo {volume} {98}},\ \bibinfo {pages} {186807} (\bibinfo {year}
  {2007})}\BibitemShut {NoStop}%
\bibitem [{\citenamefont {Meier}\ \emph {et~al.}(2009)\citenamefont {Meier},
  \citenamefont {Petrov}, \citenamefont {Guerrero}, \citenamefont {Mudry},
  \citenamefont {Patthey}, \citenamefont {Osterwalder},\ and\ \citenamefont
  {Dil}}]{meier:2009}%
  \BibitemOpen
  \bibfield  {author} {\bibinfo {author} {\bibfnamefont {F.}~\bibnamefont
  {Meier}}, \bibinfo {author} {\bibfnamefont {V.}~\bibnamefont {Petrov}},
  \bibinfo {author} {\bibfnamefont {S.}~\bibnamefont {Guerrero}}, \bibinfo
  {author} {\bibfnamefont {C.}~\bibnamefont {Mudry}}, \bibinfo {author}
  {\bibfnamefont {L.}~\bibnamefont {Patthey}}, \bibinfo {author} {\bibfnamefont
  {J.}~\bibnamefont {Osterwalder}}, \ and\ \bibinfo {author} {\bibfnamefont
  {J.~H.}\ \bibnamefont {Dil}},\ }\href {\doibase 10.1103/PhysRevB.79.241408}
  {\bibfield  {journal} {\bibinfo  {journal} {Phys. Rev. B}\ }\textbf {\bibinfo
  {volume} {79}},\ \bibinfo {pages} {241408} (\bibinfo {year}
  {2009})}\BibitemShut {NoStop}%
\bibitem [{\citenamefont {{K. Ishizaka et al.}}(2011)}]{bitei:2011}%
  \BibitemOpen
  \bibfield  {author} {\bibinfo {author} {\bibnamefont {{K. Ishizaka et
  al.}}},\ }\href@noop {} {\bibfield  {journal} {\bibinfo  {journal} {Nature
  Mat.}\ }\textbf {\bibinfo {volume} {10}},\ \bibinfo {pages} {521} (\bibinfo
  {year} {2011})}\BibitemShut {NoStop}%
\bibitem [{\citenamefont {Eremeev}\ \emph {et~al.}(2012)\citenamefont
  {Eremeev}, \citenamefont {Nechaev}, \citenamefont {Koroteev}, \citenamefont
  {Echenique},\ and\ \citenamefont {Chulkov}}]{eremeev:2012}%
  \BibitemOpen
  \bibfield  {author} {\bibinfo {author} {\bibfnamefont {S.~V.}\ \bibnamefont
  {Eremeev}}, \bibinfo {author} {\bibfnamefont {I.~A.}\ \bibnamefont
  {Nechaev}}, \bibinfo {author} {\bibfnamefont {Y.~M.}\ \bibnamefont
  {Koroteev}}, \bibinfo {author} {\bibfnamefont {P.~M.}\ \bibnamefont
  {Echenique}}, \ and\ \bibinfo {author} {\bibfnamefont {E.~V.}\ \bibnamefont
  {Chulkov}},\ }\href {\doibase 10.1103/PhysRevLett.108.246802} {\bibfield
  {journal} {\bibinfo  {journal} {Phys. Rev. Lett.}\ }\textbf {\bibinfo
  {volume} {{\bf 108}}},\ \bibinfo {pages} {246802} (\bibinfo {year}
  {2012})}\BibitemShut {NoStop}%
\bibitem [{\citenamefont {Crepaldi}\ \emph {et~al.}(2012)\citenamefont
  {Crepaldi}, \citenamefont {Moreschini}, \citenamefont {Aut\`es},
  \citenamefont {Tournier-Colletta}, \citenamefont {Moser}, \citenamefont
  {Virk}, \citenamefont {Berger}, \citenamefont {Bugnon}, \citenamefont
  {Chang}, \citenamefont {Kern}, \citenamefont {Bostwick}, \citenamefont
  {Rotenberg}, \citenamefont {Yazyev},\ and\ \citenamefont
  {Grioni}}]{crepaldi:2012}%
  \BibitemOpen
  \bibfield  {author} {\bibinfo {author} {\bibfnamefont {A.}~\bibnamefont
  {Crepaldi}}, \bibinfo {author} {\bibfnamefont {L.}~\bibnamefont
  {Moreschini}}, \bibinfo {author} {\bibfnamefont {G.}~\bibnamefont {Aut\`es}},
  \bibinfo {author} {\bibfnamefont {C.}~\bibnamefont {Tournier-Colletta}},
  \bibinfo {author} {\bibfnamefont {S.}~\bibnamefont {Moser}}, \bibinfo
  {author} {\bibfnamefont {N.}~\bibnamefont {Virk}}, \bibinfo {author}
  {\bibfnamefont {H.}~\bibnamefont {Berger}}, \bibinfo {author} {\bibfnamefont
  {P.}~\bibnamefont {Bugnon}}, \bibinfo {author} {\bibfnamefont {Y.~J.}\
  \bibnamefont {Chang}}, \bibinfo {author} {\bibfnamefont {K.}~\bibnamefont
  {Kern}}, \bibinfo {author} {\bibfnamefont {A.}~\bibnamefont {Bostwick}},
  \bibinfo {author} {\bibfnamefont {E.}~\bibnamefont {Rotenberg}}, \bibinfo
  {author} {\bibfnamefont {O.~V.}\ \bibnamefont {Yazyev}}, \ and\ \bibinfo
  {author} {\bibfnamefont {M.}~\bibnamefont {Grioni}},\ }\href {\doibase
  10.1103/PhysRevLett.109.096803} {\bibfield  {journal} {\bibinfo  {journal}
  {Phys. Rev. Lett.}\ }\textbf {\bibinfo {volume} {{\bf 109}}},\ \bibinfo
  {pages} {096803} (\bibinfo {year} {2012})}\BibitemShut {NoStop}%
\bibitem [{\citenamefont {Chesi}\ \emph {et~al.}(2007)\citenamefont {Chesi},
  \citenamefont {Simion},\ and\ \citenamefont {Giuliani}}]{chesi:2007b}%
  \BibitemOpen
  \bibfield  {author} {\bibinfo {author} {\bibfnamefont {S.}~\bibnamefont
  {Chesi}}, \bibinfo {author} {\bibfnamefont {G.}~\bibnamefont {Simion}}, \
  and\ \bibinfo {author} {\bibfnamefont {G.~F.}\ \bibnamefont {Giuliani}},\
  }\href@noop {} {\bibfield  {journal} {\bibinfo  {journal}
  {arXiv:cond-mat/0702060}\ } (\bibinfo {year} {2007})}\BibitemShut {NoStop}%
\bibitem [{\citenamefont {Juri}\ and\ \citenamefont
  {Tamborenea}(2008)}]{juri:2008}%
  \BibitemOpen
  \bibfield  {author} {\bibinfo {author} {\bibfnamefont {L.~O.}\ \bibnamefont
  {Juri}}\ and\ \bibinfo {author} {\bibfnamefont {P.~I.}\ \bibnamefont
  {Tamborenea}},\ }\href {\doibase 10.1103/PhysRevB.77.233310} {\bibfield
  {journal} {\bibinfo  {journal} {Phys. Rev. B}\ }\textbf {\bibinfo {volume}
  {77}},\ \bibinfo {pages} {233310} (\bibinfo {year} {2008})}\BibitemShut
  {NoStop}%
\bibitem [{\citenamefont {Wu}\ and\ \citenamefont {Zhang}(2004)}]{wu:2004}%
  \BibitemOpen
  \bibfield  {author} {\bibinfo {author} {\bibfnamefont {C.}~\bibnamefont
  {Wu}}\ and\ \bibinfo {author} {\bibfnamefont {S.-C.}\ \bibnamefont {Zhang}},\
  }\href@noop {} {\bibfield  {journal} {\bibinfo  {journal} {Phys. Rev. Lett.}\
  }\textbf {\bibinfo {volume} {{\bf 93}}},\ \bibinfo {pages} {036403} (\bibinfo
  {year} {2004})}\BibitemShut {NoStop}%
\bibitem [{\citenamefont {Wu}\ \emph {et~al.}(2007)\citenamefont {Wu},
  \citenamefont {Sun}, \citenamefont {Fradkin},\ and\ \citenamefont
  {Zhang}}]{wu:2007}%
  \BibitemOpen
  \bibfield  {author} {\bibinfo {author} {\bibfnamefont {C.}~\bibnamefont
  {Wu}}, \bibinfo {author} {\bibfnamefont {K.}~\bibnamefont {Sun}}, \bibinfo
  {author} {\bibfnamefont {E.}~\bibnamefont {Fradkin}}, \ and\ \bibinfo
  {author} {\bibfnamefont {S.-C.}\ \bibnamefont {Zhang}},\ }\href {\doibase
  10.1103/PhysRevB.75.115103} {\bibfield  {journal} {\bibinfo  {journal} {Phys.
  Rev. B}\ }\textbf {\bibinfo {volume} {{\bf 75}}},\ \bibinfo {pages} {115103}
  (\bibinfo {year} {2007})}\BibitemShut {NoStop}%
\bibitem [{\citenamefont {Chubukov}\ and\ \citenamefont
  {Maslov}(2009)}]{chubukov:2009}%
  \BibitemOpen
  \bibfield  {author} {\bibinfo {author} {\bibfnamefont {A.~V.}\ \bibnamefont
  {Chubukov}}\ and\ \bibinfo {author} {\bibfnamefont {D.~L.}\ \bibnamefont
  {Maslov}},\ }\href@noop {} {\bibfield  {journal} {\bibinfo  {journal} {Phys.
  Rev. Lett.}\ }\textbf {\bibinfo {volume} {{\bf 103}}},\ \bibinfo {pages}
  {216401} (\bibinfo {year} {2009})}\BibitemShut {NoStop}%
\bibitem [{\citenamefont {Berg}\ \emph {et~al.}(2012)\citenamefont {Berg},
  \citenamefont {Rudner},\ and\ \citenamefont {Kivelson}}]{berg:2012}%
  \BibitemOpen
  \bibfield  {author} {\bibinfo {author} {\bibfnamefont {E.}~\bibnamefont
  {Berg}}, \bibinfo {author} {\bibfnamefont {M.~S.}\ \bibnamefont {Rudner}}, \
  and\ \bibinfo {author} {\bibfnamefont {S.~A.}\ \bibnamefont {Kivelson}},\
  }\href {\doibase 10.1103/PhysRevB.85.035116} {\bibfield  {journal} {\bibinfo
  {journal} {Phys. Rev. B}\ }\textbf {\bibinfo {volume} {85}},\ \bibinfo
  {pages} {035116} (\bibinfo {year} {2012})}\BibitemShut {NoStop}%
\bibitem [{\citenamefont {{A. A. Abrikosov, L. P. Gorkov, and I. E.
  Dzyaloshinski}}(1963)}]{agd:1963}%
  \BibitemOpen
  \bibfield  {author} {\bibinfo {author} {\bibnamefont {{A. A. Abrikosov, L. P.
  Gorkov, and I. E. Dzyaloshinski}}},\ }\href@noop {} {\emph {\bibinfo {title}
  {Methods of Quantum Field Theory in Statistical Physics}}}\ (\bibinfo
  {publisher} {Dover, New York},\ \bibinfo {year} {1963})\BibitemShut {NoStop}%
\bibitem [{\citenamefont {{P. Nozi{\`e}res and D.
  Pines}}(1966)}]{nozieres:1966}%
  \BibitemOpen
  \bibfield  {author} {\bibinfo {author} {\bibnamefont {{P. Nozi{\`e}res and D.
  Pines}}},\ }\href@noop {} {\emph {\bibinfo {title} {{The Theory of Quantum
  Liquids}}}},\ Vol.~\bibinfo {volume} {1}\ (\bibinfo  {publisher} {New York:
  Benjamin},\ \bibinfo {year} {1966})\BibitemShut {NoStop}%
\bibitem [{\citenamefont {{E. M. Lifshitz and L. P.
  Pitaevski}}(1980)}]{lifshitz:1980}%
  \BibitemOpen
  \bibfield  {author} {\bibinfo {author} {\bibnamefont {{E. M. Lifshitz and L.
  P. Pitaevski}}},\ }\href@noop {} {\emph {\bibinfo {title} {Statistical
  Physics}}}\ (\bibinfo  {publisher} {Pergamon Press, New York},\ \bibinfo
  {year} {1980})\BibitemShut {NoStop}%
\bibitem [{\citenamefont {Saraga}\ and\ \citenamefont
  {Loss}(2005)}]{saraga:2005}%
  \BibitemOpen
  \bibfield  {author} {\bibinfo {author} {\bibfnamefont {D.~S.}\ \bibnamefont
  {Saraga}}\ and\ \bibinfo {author} {\bibfnamefont {D.}~\bibnamefont {Loss}},\
  }\href {\doibase 10.1103/PhysRevB.72.195319} {\bibfield  {journal} {\bibinfo
  {journal} {Phys. Rev. B}\ }\textbf {\bibinfo {volume} {{\bf 72}}},\ \bibinfo
  {pages} {195319} (\bibinfo {year} {2005})}\BibitemShut {NoStop}%
\bibitem [{\citenamefont {Agarwal}\ \emph {et~al.}(2011)\citenamefont
  {Agarwal}, \citenamefont {Chesi}, \citenamefont {Jungwirth}, \citenamefont
  {Sinova}, \citenamefont {Vignale},\ and\ \citenamefont
  {Polini}}]{agarwal:2011}%
  \BibitemOpen
  \bibfield  {author} {\bibinfo {author} {\bibfnamefont {A.}~\bibnamefont
  {Agarwal}}, \bibinfo {author} {\bibfnamefont {S.}~\bibnamefont {Chesi}},
  \bibinfo {author} {\bibfnamefont {T.}~\bibnamefont {Jungwirth}}, \bibinfo
  {author} {\bibfnamefont {J.}~\bibnamefont {Sinova}}, \bibinfo {author}
  {\bibfnamefont {G.}~\bibnamefont {Vignale}}, \ and\ \bibinfo {author}
  {\bibfnamefont {M.}~\bibnamefont {Polini}},\ }\href {\doibase
  10.1103/PhysRevB.83.115135} {\bibfield  {journal} {\bibinfo  {journal} {Phys.
  Rev. B}\ }\textbf {\bibinfo {volume} {{\bf 83}}},\ \bibinfo {pages} {115135}
  (\bibinfo {year} {2011})}\BibitemShut {NoStop}%
\bibitem [{\citenamefont {Aasen}\ \emph {et~al.}(2012)\citenamefont {Aasen},
  \citenamefont {Chesi},\ and\ \citenamefont {Coish}}]{aasen:2012}%
  \BibitemOpen
  \bibfield  {author} {\bibinfo {author} {\bibfnamefont {D.}~\bibnamefont
  {Aasen}}, \bibinfo {author} {\bibfnamefont {S.}~\bibnamefont {Chesi}}, \ and\
  \bibinfo {author} {\bibfnamefont {W.~A.}\ \bibnamefont {Coish}},\ }\href
  {\doibase 10.1103/PhysRevB.85.075321} {\bibfield  {journal} {\bibinfo
  {journal} {Phys. Rev. B}\ }\textbf {\bibinfo {volume} {85}},\ \bibinfo
  {pages} {075321} (\bibinfo {year} {2012})}\BibitemShut {NoStop}%
\bibitem [{\citenamefont {{Yu}}\ \emph {et~al.}(2013)\citenamefont {{Yu}},
  \citenamefont {{Zhang}},\ and\ \citenamefont {{Liu}}}]{yu:2013}%
  \BibitemOpen
  \bibfield  {author} {\bibinfo {author} {\bibfnamefont {X.-L.}\ \bibnamefont
  {{Yu}}}, \bibinfo {author} {\bibfnamefont {S.-S.}\ \bibnamefont {{Zhang}}}, \
  and\ \bibinfo {author} {\bibfnamefont {W.-M.}\ \bibnamefont {{Liu}}},\ }\href
  {\doibase 10.1103/PhysRevA.87.043633} {\bibfield  {journal} {\bibinfo
  {journal} {\pra}\ }\textbf {\bibinfo {volume} {87}},\ \bibinfo {eid} {043633}
  (\bibinfo {year} {2013})}\BibitemShut {NoStop}%
\bibitem [{\citenamefont {Pletyukhov}\ and\ \citenamefont
  {Gritsev}(2006)}]{pletyukhov:2006}%
  \BibitemOpen
  \bibfield  {author} {\bibinfo {author} {\bibfnamefont {M.}~\bibnamefont
  {Pletyukhov}}\ and\ \bibinfo {author} {\bibfnamefont {V.}~\bibnamefont
  {Gritsev}},\ }\href {\doibase 10.1103/PhysRevB.74.045307} {\bibfield
  {journal} {\bibinfo  {journal} {Phys. Rev. B}\ }\textbf {\bibinfo {volume}
  {{\bf 74}}},\ \bibinfo {pages} {045307} (\bibinfo {year} {2006})}\BibitemShut
  {NoStop}%
\bibitem [{\citenamefont {Pletyukhov}\ and\ \citenamefont
  {Konschuh}(2007)}]{pletyukhov:2007}%
  \BibitemOpen
  \bibfield  {author} {\bibinfo {author} {\bibfnamefont {M.}~\bibnamefont
  {Pletyukhov}}\ and\ \bibinfo {author} {\bibfnamefont {S.}~\bibnamefont
  {Konschuh}},\ }\href@noop {} {\bibfield  {journal} {\bibinfo  {journal} {Eur.
  Phys. Journal B}\ }\textbf {\bibinfo {volume} {{\bf 60}}},\ \bibinfo {pages}
  {29} (\bibinfo {year} {2007})}\BibitemShut {NoStop}%
\bibitem [{\citenamefont {Barlas}\ \emph {et~al.}(2007)\citenamefont {Barlas},
  \citenamefont {Pereg-Barnea}, \citenamefont {Polini}, \citenamefont
  {Asgari},\ and\ \citenamefont {MacDonald}}]{barlas:2007}%
  \BibitemOpen
  \bibfield  {author} {\bibinfo {author} {\bibfnamefont {Y.}~\bibnamefont
  {Barlas}}, \bibinfo {author} {\bibfnamefont {T.}~\bibnamefont
  {Pereg-Barnea}}, \bibinfo {author} {\bibfnamefont {M.}~\bibnamefont
  {Polini}}, \bibinfo {author} {\bibfnamefont {R.}~\bibnamefont {Asgari}}, \
  and\ \bibinfo {author} {\bibfnamefont {A.~H.}\ \bibnamefont {MacDonald}},\
  }\href {\doibase 10.1103/PhysRevLett.98.236601} {\bibfield  {journal}
  {\bibinfo  {journal} {Phys. Rev. Lett.}\ }\textbf {\bibinfo {volume} {{\bf
  98}}},\ \bibinfo {pages} {236601} (\bibinfo {year} {2007})}\BibitemShut
  {NoStop}%
\bibitem [{\citenamefont {Chesi}\ and\ \citenamefont
  {Giuliani}(2007)}]{chesi:2007c}%
  \BibitemOpen
  \bibfield  {author} {\bibinfo {author} {\bibfnamefont {S.}~\bibnamefont
  {Chesi}}\ and\ \bibinfo {author} {\bibfnamefont {G.~F.}\ \bibnamefont
  {Giuliani}},\ }\href {\doibase 10.1103/PhysRevB.75.155305} {\bibfield
  {journal} {\bibinfo  {journal} {Phys. Rev. B}\ }\textbf {\bibinfo {volume}
  {{\bf 75}}},\ \bibinfo {pages} {155305} (\bibinfo {year} {2007})}\BibitemShut
  {NoStop}%
\bibitem [{\citenamefont {Chesi}(2007)}]{chesi:thesis}%
  \BibitemOpen
  \bibfield  {author} {\bibinfo {author} {\bibfnamefont {S.}~\bibnamefont
  {Chesi}},\ }\emph {\bibinfo {title} {Effects of structural spin-orbit
  coupling in two-dimensional electron and hole liquids}},\ \href@noop {}
  {Ph.D. thesis},\ \bibinfo  {school} {Purdue University} (\bibinfo {year}
  {2007})\BibitemShut {NoStop}%
\bibitem [{\citenamefont {Chesi}\ \emph {et~al.}(2009)\citenamefont {Chesi},
  \citenamefont {\ifmmode~\dot{Z}\else \.{Z}\fi{}ak}, \citenamefont {Simon},\
  and\ \citenamefont {Loss}}]{chesi:2009}%
  \BibitemOpen
  \bibfield  {author} {\bibinfo {author} {\bibfnamefont {S.}~\bibnamefont
  {Chesi}}, \bibinfo {author} {\bibfnamefont {R.~A.}\ \bibnamefont
  {\ifmmode~\dot{Z}\else \.{Z}\fi{}ak}}, \bibinfo {author} {\bibfnamefont
  {P.}~\bibnamefont {Simon}}, \ and\ \bibinfo {author} {\bibfnamefont
  {D.}~\bibnamefont {Loss}},\ }\href {\doibase 10.1103/PhysRevB.79.115445}
  {\bibfield  {journal} {\bibinfo  {journal} {Phys. Rev. B}\ }\textbf {\bibinfo
  {volume} {{\bf 79}}},\ \bibinfo {pages} {115445} (\bibinfo {year}
  {2009})}\BibitemShut {NoStop}%
\bibitem [{\citenamefont {Zak}\ \emph {et~al.}(2010)\citenamefont {Zak},
  \citenamefont {Maslov},\ and\ \citenamefont {Loss}}]{zak:2010}%
  \BibitemOpen
  \bibfield  {author} {\bibinfo {author} {\bibfnamefont {R.~A.}\ \bibnamefont
  {Zak}}, \bibinfo {author} {\bibfnamefont {D.~L.}\ \bibnamefont {Maslov}}, \
  and\ \bibinfo {author} {\bibfnamefont {D.}~\bibnamefont {Loss}},\ }\href
  {\doibase 10.1103/PhysRevB.82.115415} {\bibfield  {journal} {\bibinfo
  {journal} {Phys. Rev. B}\ }\textbf {\bibinfo {volume} {{\bf 82}}},\ \bibinfo
  {pages} {115415} (\bibinfo {year} {2010})}\BibitemShut {NoStop}%
\bibitem [{\citenamefont {Culcer}(2011)}]{culcer:2011}%
  \BibitemOpen
  \bibfield  {author} {\bibinfo {author} {\bibfnamefont {D.}~\bibnamefont
  {Culcer}},\ }\href@noop {} {\bibfield  {journal} {\bibinfo  {journal} {Phys.
  Rev. B}\ }\textbf {\bibinfo {volume} {{\bf 84}}},\ \bibinfo {pages} {235411}
  (\bibinfo {year} {2011})}\BibitemShut {NoStop}%
\bibitem [{\citenamefont {Zak}\ \emph {et~al.}(2012)\citenamefont {Zak},
  \citenamefont {Maslov},\ and\ \citenamefont {Loss}}]{zak:2012}%
  \BibitemOpen
  \bibfield  {author} {\bibinfo {author} {\bibfnamefont {R.~A.}\ \bibnamefont
  {Zak}}, \bibinfo {author} {\bibfnamefont {D.~L.}\ \bibnamefont {Maslov}}, \
  and\ \bibinfo {author} {\bibfnamefont {D.}~\bibnamefont {Loss}},\ }\href
  {\doibase {10.1103/PhysRevB.85.115424}} {\bibfield  {journal} {\bibinfo
  {journal} {{Phys. Rev. B}}\ }\textbf {\bibinfo {volume} {{\bf 85}}},\
  \bibinfo {pages} {15424} (\bibinfo {year} {2012})}\BibitemShut {NoStop}%
\bibitem [{\citenamefont {Shekhter}\ \emph {et~al.}(2005)\citenamefont
  {Shekhter}, \citenamefont {Khodas},\ and\ \citenamefont
  {Finkel'stein}}]{shekhter:2005}%
  \BibitemOpen
  \bibfield  {author} {\bibinfo {author} {\bibfnamefont {A.}~\bibnamefont
  {Shekhter}}, \bibinfo {author} {\bibfnamefont {M.}~\bibnamefont {Khodas}}, \
  and\ \bibinfo {author} {\bibfnamefont {A.~M.}\ \bibnamefont {Finkel'stein}},\
  }\href {\doibase 10.1103/PhysRevB.71.165329} {\bibfield  {journal} {\bibinfo
  {journal} {Phys. Rev. B}\ }\textbf {\bibinfo {volume} {{\bf 71}}},\ \bibinfo
  {pages} {165329} (\bibinfo {year} {2005})}\BibitemShut {NoStop}%
\bibitem [{\citenamefont {Hankiewicz}\ and\ \citenamefont
  {Vignale}(2006)}]{vignale:2006}%
  \BibitemOpen
  \bibfield  {author} {\bibinfo {author} {\bibfnamefont {E.~M.}\ \bibnamefont
  {Hankiewicz}}\ and\ \bibinfo {author} {\bibfnamefont {G.}~\bibnamefont
  {Vignale}},\ }\href {\doibase 10.1103/PhysRevB.73.115339} {\bibfield
  {journal} {\bibinfo  {journal} {Phys. Rev. B}\ }\textbf {\bibinfo {volume}
  {73}},\ \bibinfo {pages} {115339} (\bibinfo {year} {2006})}\BibitemShut
  {NoStop}%
\bibitem [{\citenamefont {Tse}\ and\ \citenamefont
  {Das~Sarma}(2007)}]{tse:2007}%
  \BibitemOpen
  \bibfield  {author} {\bibinfo {author} {\bibfnamefont {W.-K.}\ \bibnamefont
  {Tse}}\ and\ \bibinfo {author} {\bibfnamefont {S.}~\bibnamefont
  {Das~Sarma}},\ }\href {\doibase 10.1103/PhysRevB.75.045333} {\bibfield
  {journal} {\bibinfo  {journal} {Phys. Rev. B}\ }\textbf {\bibinfo {volume}
  {{\bf 75}}},\ \bibinfo {pages} {045333} (\bibinfo {year} {2007})}\BibitemShut
  {NoStop}%
\bibitem [{\citenamefont {D'Amico}\ and\ \citenamefont
  {Vignale}(2002)}]{vignale:2002}%
  \BibitemOpen
  \bibfield  {author} {\bibinfo {author} {\bibfnamefont {I.}~\bibnamefont
  {D'Amico}}\ and\ \bibinfo {author} {\bibfnamefont {G.}~\bibnamefont
  {Vignale}},\ }\href {\doibase 10.1103/PhysRevB.65.085109} {\bibfield
  {journal} {\bibinfo  {journal} {Phys. Rev. B}\ }\textbf {\bibinfo {volume}
  {{\bf 65}}},\ \bibinfo {pages} {085109} (\bibinfo {year} {2002})}\BibitemShut
  {NoStop}%
\bibitem [{\citenamefont {Badalyan}\ \emph {et~al.}(2010)\citenamefont
  {Badalyan}, \citenamefont {Matos-Abiague}, \citenamefont {Vignale},\ and\
  \citenamefont {Fabian}}]{badalyan:2010}%
  \BibitemOpen
  \bibfield  {author} {\bibinfo {author} {\bibfnamefont {S.~M.}\ \bibnamefont
  {Badalyan}}, \bibinfo {author} {\bibfnamefont {A.}~\bibnamefont
  {Matos-Abiague}}, \bibinfo {author} {\bibfnamefont {G.}~\bibnamefont
  {Vignale}}, \ and\ \bibinfo {author} {\bibfnamefont {J.}~\bibnamefont
  {Fabian}},\ }\href {\doibase 10.1103/PhysRevB.81.205314} {\bibfield
  {journal} {\bibinfo  {journal} {Phys. Rev. B}\ }\textbf {\bibinfo {volume}
  {{\bf 81}}},\ \bibinfo {pages} {205314} (\bibinfo {year} {2010})}\BibitemShut
  {NoStop}%
\bibitem [{\citenamefont {Baym}\ and\ \citenamefont {Chin}(1976)}]{baym:1976}%
  \BibitemOpen
  \bibfield  {author} {\bibinfo {author} {\bibfnamefont {G.}~\bibnamefont
  {Baym}}\ and\ \bibinfo {author} {\bibfnamefont {S.~A.}\ \bibnamefont
  {Chin}},\ }\href {\doibase 10.1016/0375-9474(76)90513-3} {\bibfield
  {journal} {\bibinfo  {journal} {Nucl. Phys. A}\ }\textbf {\bibinfo {volume}
  {{\bf 262}}},\ \bibinfo {pages} {527} (\bibinfo {year} {1976})}\BibitemShut
  {NoStop}%
\bibitem [{\citenamefont {Song}(2001)}]{song:2001}%
  \BibitemOpen
  \bibfield  {author} {\bibinfo {author} {\bibfnamefont {C.}~\bibnamefont
  {Song}},\ }\href {\doibase 10.1016/S0370-1573(00)00108-3} {\bibfield
  {journal} {\bibinfo  {journal} {Phys. Rep.}\ }\textbf {\bibinfo {volume}
  {{\bf 347}}},\ \bibinfo {pages} {289} (\bibinfo {year} {2001})}\BibitemShut
  {NoStop}%
\bibitem [{\citenamefont {Quintanilla}\ \emph {et~al.}(2009)\citenamefont
  {Quintanilla}, \citenamefont {Carr},\ and\ \citenamefont
  {Betouras}}]{betouras:2009}%
  \BibitemOpen
  \bibfield  {author} {\bibinfo {author} {\bibfnamefont {J.}~\bibnamefont
  {Quintanilla}}, \bibinfo {author} {\bibfnamefont {S.~T.}\ \bibnamefont
  {Carr}}, \ and\ \bibinfo {author} {\bibfnamefont {J.~J.}\ \bibnamefont
  {Betouras}},\ }\href {\doibase 10.1103/PhysRevA.79.031601} {\bibfield
  {journal} {\bibinfo  {journal} {Phys. Rev. A}\ }\textbf {\bibinfo {volume}
  {79}},\ \bibinfo {pages} {031601} (\bibinfo {year} {2009})}\BibitemShut
  {NoStop}%
\bibitem [{\citenamefont {Fregoso}\ and\ \citenamefont
  {Fradkin}(2009)}]{fregoso:2009}%
  \BibitemOpen
  \bibfield  {author} {\bibinfo {author} {\bibfnamefont {B.~M.}\ \bibnamefont
  {Fregoso}}\ and\ \bibinfo {author} {\bibfnamefont {E.}~\bibnamefont
  {Fradkin}},\ }\href {\doibase 10.1103/PhysRevLett.103.205301} {\bibfield
  {journal} {\bibinfo  {journal} {Phys. Rev. Lett.}\ }\textbf {\bibinfo
  {volume} {103}},\ \bibinfo {pages} {205301} (\bibinfo {year}
  {2009})}\BibitemShut {NoStop}%
\bibitem [{\citenamefont {Fregoso}\ \emph {et~al.}(2009)\citenamefont
  {Fregoso}, \citenamefont {Sun}, \citenamefont {Fradkin},\ and\ \citenamefont
  {Lev}}]{fregoso:2009a}%
  \BibitemOpen
  \bibfield  {author} {\bibinfo {author} {\bibfnamefont {B.~M.}\ \bibnamefont
  {Fregoso}}, \bibinfo {author} {\bibfnamefont {K.}~\bibnamefont {Sun}},
  \bibinfo {author} {\bibfnamefont {E.}~\bibnamefont {Fradkin}}, \ and\
  \bibinfo {author} {\bibfnamefont {B.~L.}\ \bibnamefont {Lev}},\ }\href@noop
  {} {\bibfield  {journal} {\bibinfo  {journal} {New J. Phys.}\ }\textbf
  {\bibinfo {volume} {11}},\ \bibinfo {pages} {103003} (\bibinfo {year}
  {2009})}\BibitemShut {NoStop}%
\bibitem [{\citenamefont {{Li}}\ and\ \citenamefont {{Wu}}(2012)}]{li:2012}%
  \BibitemOpen
  \bibfield  {author} {\bibinfo {author} {\bibfnamefont {Y.}~\bibnamefont
  {{Li}}}\ and\ \bibinfo {author} {\bibfnamefont {C.}~\bibnamefont {{Wu}}},\
  }\href {\doibase 10.1103/PhysRevB.85.205126} {\bibfield  {journal} {\bibinfo
  {journal} {\prb}\ }\textbf {\bibinfo {volume} {\bf 85}},\ \bibinfo {eid}
  {205126} (\bibinfo {year} {2012})}\BibitemShut {NoStop}%
\bibitem [{\citenamefont {Abrikosov}\ and\ \citenamefont
  {Dzyaloshinskii}(1958)}]{abrikosov:1958}%
  \BibitemOpen
  \bibfield  {author} {\bibinfo {author} {\bibfnamefont {A.~A.}\ \bibnamefont
  {Abrikosov}}\ and\ \bibinfo {author} {\bibfnamefont {I.~E.}\ \bibnamefont
  {Dzyaloshinskii}},\ }\href@noop {} {\bibfield  {journal} {\bibinfo  {journal}
  {Sov. Phys. JETP}\ }\textbf {\bibinfo {volume} {{\bf 8}}},\ \bibinfo {pages}
  {535} (\bibinfo {year} {1958})}\BibitemShut {NoStop}%
\bibitem [{\citenamefont {{Silin}}(1958)}]{silin:1958}%
  \BibitemOpen
  \bibfield  {author} {\bibinfo {author} {\bibfnamefont {V.~P.}\ \bibnamefont
  {{Silin}}},\ }\href@noop {} {\bibfield  {journal} {\bibinfo  {journal} {Sov.
  Phys. JETP}\ }\textbf {\bibinfo {volume} {{\bf 6}}},\ \bibinfo {pages} {945}
  (\bibinfo {year} {1958})}\BibitemShut {NoStop}%
\bibitem [{\citenamefont {Silin}(1959)}]{silin:1959}%
  \BibitemOpen
  \bibfield  {author} {\bibinfo {author} {\bibfnamefont {V.~P.}\ \bibnamefont
  {Silin}},\ }\href@noop {} {\bibfield  {journal} {\bibinfo  {journal} {Sov.
  Phys. JETP}\ }\textbf {\bibinfo {volume} {{\bf 8}}},\ \bibinfo {pages} {870}
  (\bibinfo {year} {1959})}\BibitemShut {NoStop}%
\bibitem [{\citenamefont {Kondratenko}(1965)}]{kondratenko:1965}%
  \BibitemOpen
  \bibfield  {author} {\bibinfo {author} {\bibfnamefont {P.}~\bibnamefont
  {Kondratenko}},\ }\href@noop {} {\bibfield  {journal} {\bibinfo  {journal}
  {Sov. Phys. JETP}\ }\textbf {\bibinfo {volume} {{\bf 20}}},\ \bibinfo {pages}
  {1023} (\bibinfo {year} {{(1965).}})}\BibitemShut {NoStop}%
\bibitem [{\citenamefont {Platzman}\ and\ \citenamefont
  {Walsh}(1967)}]{platzman:1967}%
  \BibitemOpen
  \bibfield  {author} {\bibinfo {author} {\bibfnamefont {P.~M.}\ \bibnamefont
  {Platzman}}\ and\ \bibinfo {author} {\bibfnamefont {W.~M.}\ \bibnamefont
  {Walsh}},\ }\href {\doibase 10.1103/PhysRevLett.19.514} {\bibfield  {journal}
  {\bibinfo  {journal} {Phys. Rev. Lett.}\ }\textbf {\bibinfo {volume} {{\bf
  19}}},\ \bibinfo {pages} {514} (\bibinfo {year} {1967})}\BibitemShut
  {NoStop}%
\bibitem [{\citenamefont {Ma}\ \emph {et~al.}(1968)\citenamefont {Ma},
  \citenamefont {B\'eal-Monod},\ and\ \citenamefont {Fredkin}}]{ma:1968}%
  \BibitemOpen
  \bibfield  {author} {\bibinfo {author} {\bibfnamefont {S.-k.}\ \bibnamefont
  {Ma}}, \bibinfo {author} {\bibfnamefont {M.~T.}\ \bibnamefont
  {B\'eal-Monod}}, \ and\ \bibinfo {author} {\bibfnamefont {D.~R.}\
  \bibnamefont {Fredkin}},\ }\href {\doibase 10.1103/PhysRev.174.227}
  {\bibfield  {journal} {\bibinfo  {journal} {Phys. Rev.}\ }\textbf {\bibinfo
  {volume} {{\bf 174}}},\ \bibinfo {pages} {227} (\bibinfo {year}
  {1968})}\BibitemShut {NoStop}%
\bibitem [{\citenamefont {Leggett}(1970)}]{leggett:1970}%
  \BibitemOpen
  \bibfield  {author} {\bibinfo {author} {\bibfnamefont {A.~J.}\ \bibnamefont
  {Leggett}},\ }\href {http://stacks.iop.org/0022-3719/3/i=2/a=027} {\bibfield
  {journal} {\bibinfo  {journal} {J. Phys. C: Solid State Phys.}\ }\textbf
  {\bibinfo {volume} {{\bf 3}}},\ \bibinfo {pages} {448} (\bibinfo {year}
  {1970})}\BibitemShut {NoStop}%
\bibitem [{\citenamefont {{I. E. Dzyaloshinskii and P. S.
  Kondratenko}}(1976)}]{dzyalosh:1976}%
  \BibitemOpen
  \bibfield  {author} {\bibinfo {author} {\bibnamefont {{I. E. Dzyaloshinskii
  and P. S. Kondratenko}}},\ }\href@noop {} {\bibfield  {journal} {\bibinfo
  {journal} {Sov. Phys. JETP}\ }\textbf {\bibinfo {volume} {{43}}},\ \bibinfo
  {pages} {1036} (\bibinfo {year} {{1976}})}\BibitemShut {NoStop}%
\bibitem [{\citenamefont {{E. P. Bashkin and A. E.
  Meyerovich}}(1981)}]{bashkin:1981}%
  \BibitemOpen
  \bibfield  {author} {\bibinfo {author} {\bibnamefont {{E. P. Bashkin and A.
  E. Meyerovich}}},\ }\href@noop {} {\bibfield  {journal} {\bibinfo  {journal}
  {Adv. Phys.}\ }\textbf {\bibinfo {volume} {{\bf 30}}},\ \bibinfo {pages} {1}
  (\bibinfo {year} {1981})}\BibitemShut {NoStop}%
\bibitem [{\citenamefont {Meyerovich}(1983)}]{meyerovich:1983}%
  \BibitemOpen
  \bibfield  {author} {\bibinfo {author} {\bibfnamefont {A.~E.}\ \bibnamefont
  {Meyerovich}},\ }\href@noop {} {\bibfield  {journal} {\bibinfo  {journal} {J.
  Low Temp. Phys.}\ }\textbf {\bibinfo {volume} {{\bf 53}}},\ \bibinfo {pages}
  {487} (\bibinfo {year} {1983})}\BibitemShut {NoStop}%
\bibitem [{\citenamefont {Meyerovich}\ and\ \citenamefont
  {Musaelian}(1992)}]{meyerovich:1992a}%
  \BibitemOpen
  \bibfield  {author} {\bibinfo {author} {\bibfnamefont {A.~E.}\ \bibnamefont
  {Meyerovich}}\ and\ \bibinfo {author} {\bibfnamefont {K.~A.}\ \bibnamefont
  {Musaelian}},\ }\href@noop {} {\bibfield  {journal} {\bibinfo  {journal} {J.
  Low Temp. Phys.}\ }\textbf {\bibinfo {volume} {{\bf 89}}},\ \bibinfo {pages}
  {781} (\bibinfo {year} {1992})}\BibitemShut {NoStop}%
\bibitem [{\citenamefont {Meyerovich}\ and\ \citenamefont
  {Musaelian}(1994{\natexlab{a}})}]{meyerovich:1994}%
  \BibitemOpen
  \bibfield  {author} {\bibinfo {author} {\bibfnamefont {A.~E.}\ \bibnamefont
  {Meyerovich}}\ and\ \bibinfo {author} {\bibfnamefont {K.~A.}\ \bibnamefont
  {Musaelian}},\ }\href@noop {} {\bibfield  {journal} {\bibinfo  {journal} {J.
  Low Temp. Phys.}\ }\textbf {\bibinfo {volume} {{\bf 94}}},\ \bibinfo {pages}
  {249} (\bibinfo {year} {1994}{\natexlab{a}})}\BibitemShut {NoStop}%
\bibitem [{\citenamefont {Meyerovich}\ and\ \citenamefont
  {Musaelian}(1994{\natexlab{b}})}]{meyerovich:1994b}%
  \BibitemOpen
  \bibfield  {author} {\bibinfo {author} {\bibfnamefont {A.~E.}\ \bibnamefont
  {Meyerovich}}\ and\ \bibinfo {author} {\bibfnamefont {K.~A.}\ \bibnamefont
  {Musaelian}},\ }\href@noop {} {\bibfield  {journal} {\bibinfo  {journal} {J.
  Low Temp. Phys.}\ }\textbf {\bibinfo {volume} {{\bf 95}}},\ \bibinfo {pages}
  {789} (\bibinfo {year} {1994}{\natexlab{b}})}\BibitemShut {NoStop}%
\bibitem [{\citenamefont {Mineev}(2004)}]{mineev:2004}%
  \BibitemOpen
  \bibfield  {author} {\bibinfo {author} {\bibfnamefont {V.~P.}\ \bibnamefont
  {Mineev}},\ }\href {\doibase 10.1103/PhysRevB.69.144429} {\bibfield
  {journal} {\bibinfo  {journal} {Phys. Rev. B}\ }\textbf {\bibinfo {volume}
  {69}},\ \bibinfo {pages} {144429} (\bibinfo {year} {2004})}\BibitemShut
  {NoStop}%
\bibitem [{\citenamefont {Mineev}(2005)}]{mineev:2005}%
  \BibitemOpen
  \bibfield  {author} {\bibinfo {author} {\bibfnamefont {V.~P.}\ \bibnamefont
  {Mineev}},\ }\href {\doibase 10.1103/PhysRevB.72.144418} {\bibfield
  {journal} {\bibinfo  {journal} {Phys. Rev. B}\ }\textbf {\bibinfo {volume}
  {72}},\ \bibinfo {pages} {144418} (\bibinfo {year} {2005})}\BibitemShut
  {NoStop}%
\bibitem [{\citenamefont {Mineev}()}]{mineev:2011}%
  \BibitemOpen
  \bibfield  {author} {\bibinfo {author} {\bibfnamefont {V.~P.}\ \bibnamefont
  {Mineev}},\ }\href@noop {} {\bibinfo  {journal} {arXiv:1111.3208}\
  }\BibitemShut {NoStop}%
\bibitem [{\citenamefont {Oliva}\ and\ \citenamefont
  {Ashcroft}(1981)}]{ashcroft:1981}%
  \BibitemOpen
\bibfield  {journal} {  }\bibfield  {author} {\bibinfo {author} {\bibfnamefont
  {J.}~\bibnamefont {Oliva}}\ and\ \bibinfo {author} {\bibfnamefont {N.~W.}\
  \bibnamefont {Ashcroft}},\ }\href {\doibase 10.1103/PhysRevB.23.6399}
  {\bibfield  {journal} {\bibinfo  {journal} {Phys. Rev. B}\ }\textbf {\bibinfo
  {volume} {23}},\ \bibinfo {pages} {6399} (\bibinfo {year}
  {1981})}\BibitemShut {NoStop}%
\bibitem [{\citenamefont {Sanchez-Castro}\ and\ \citenamefont
  {Bedell}(1991)}]{bedell:1991}%
  \BibitemOpen
  \bibfield  {author} {\bibinfo {author} {\bibfnamefont {C.}~\bibnamefont
  {Sanchez-Castro}}\ and\ \bibinfo {author} {\bibfnamefont {K.~S.}\
  \bibnamefont {Bedell}},\ }\href {\doibase 10.1103/PhysRevB.43.12874}
  {\bibfield  {journal} {\bibinfo  {journal} {Phys. Rev. B}\ }\textbf {\bibinfo
  {volume} {43}},\ \bibinfo {pages} {12874} (\bibinfo {year}
  {1991})}\BibitemShut {NoStop}%
\bibitem [{\citenamefont {Chen}\ and\ \citenamefont
  {Raikh}(1999)}]{raikh:1999}%
  \BibitemOpen
  \bibfield  {author} {\bibinfo {author} {\bibfnamefont {G.-H.}\ \bibnamefont
  {Chen}}\ and\ \bibinfo {author} {\bibfnamefont {M.~E.}\ \bibnamefont
  {Raikh}},\ }\href {\doibase 10.1103/PhysRevB.60.4826} {\bibfield  {journal}
  {\bibinfo  {journal} {Phys. Rev. B}\ }\textbf {\bibinfo {volume} {{\bf
  60}}},\ \bibinfo {pages} {4826} (\bibinfo {year} {1999})}\BibitemShut
  {NoStop}%
\bibitem [{\citenamefont {Fujita}\ and\ \citenamefont
  {Quader}(1987)}]{fujita:1987}%
  \BibitemOpen
  \bibfield  {author} {\bibinfo {author} {\bibfnamefont {T.}~\bibnamefont
  {Fujita}}\ and\ \bibinfo {author} {\bibfnamefont {K.~F.}\ \bibnamefont
  {Quader}},\ }\href {\doibase 10.1103/PhysRevB.36.5152} {\bibfield  {journal}
  {\bibinfo  {journal} {Phys. Rev. B}\ }\textbf {\bibinfo {volume} {{\bf
  36}}},\ \bibinfo {pages} {5152} (\bibinfo {year} {1987})}\BibitemShut
  {NoStop}%
\bibitem [{\citenamefont {Winkler}(2000)}]{winkler:2000}%
  \BibitemOpen
  \bibfield  {author} {\bibinfo {author} {\bibfnamefont {R.}~\bibnamefont
  {Winkler}},\ }\href {\doibase 10.1103/PhysRevB.62.4245} {\bibfield  {journal}
  {\bibinfo  {journal} {Phys. Rev. B}\ }\textbf {\bibinfo {volume} {62}},\
  \bibinfo {pages} {4245} (\bibinfo {year} {2000})}\BibitemShut {NoStop}%
\bibitem [{\citenamefont {Winkler}\ \emph {et~al.}(2002)\citenamefont
  {Winkler}, \citenamefont {Noh}, \citenamefont {Tutuc},\ and\ \citenamefont
  {Shayegan}}]{winkler:2002}%
  \BibitemOpen
  \bibfield  {author} {\bibinfo {author} {\bibfnamefont {R.}~\bibnamefont
  {Winkler}}, \bibinfo {author} {\bibfnamefont {H.}~\bibnamefont {Noh}},
  \bibinfo {author} {\bibfnamefont {E.}~\bibnamefont {Tutuc}}, \ and\ \bibinfo
  {author} {\bibfnamefont {M.}~\bibnamefont {Shayegan}},\ }\href@noop {}
  {\bibfield  {journal} {\bibinfo  {journal} {Phys. Rev. B}\ }\textbf {\bibinfo
  {volume} {65}},\ \bibinfo {pages} {155303} (\bibinfo {year}
  {2002})}\BibitemShut {NoStop}%
\bibitem [{\citenamefont {Nakamura}\ \emph {et~al.}(2012)\citenamefont
  {Nakamura}, \citenamefont {Koga},\ and\ \citenamefont
  {Kimura}}]{nakamura:2012}%
  \BibitemOpen
  \bibfield  {author} {\bibinfo {author} {\bibfnamefont {H.}~\bibnamefont
  {Nakamura}}, \bibinfo {author} {\bibfnamefont {T.}~\bibnamefont {Koga}}, \
  and\ \bibinfo {author} {\bibfnamefont {T.}~\bibnamefont {Kimura}},\ }\href
  {\doibase 10.1103/PhysRevLett.108.206601} {\bibfield  {journal} {\bibinfo
  {journal} {Phys. Rev. Lett.}\ }\textbf {\bibinfo {volume} {{\bf 108}}},\
  \bibinfo {pages} {206601} (\bibinfo {year} {2012})}\BibitemShut {NoStop}%
\bibitem [{\citenamefont {Zhong}\ \emph {et~al.}(2013)\citenamefont {Zhong},
  \citenamefont {T\'oth},\ and\ \citenamefont {Held}}]{zhong:2013}%
  \BibitemOpen
  \bibfield  {author} {\bibinfo {author} {\bibfnamefont {Z.}~\bibnamefont
  {Zhong}}, \bibinfo {author} {\bibfnamefont {A.}~\bibnamefont {T\'oth}}, \
  and\ \bibinfo {author} {\bibfnamefont {K.}~\bibnamefont {Held}},\ }\href
  {\doibase 10.1103/PhysRevB.87.161102} {\bibfield  {journal} {\bibinfo
  {journal} {Phys. Rev. B}\ }\textbf {\bibinfo {volume} {87}},\ \bibinfo
  {pages} {161102} (\bibinfo {year} {2013})}\BibitemShut {NoStop}%
\bibitem [{Note1()}]{Note1}%
  \BibitemOpen
  \bibinfo {note} {Although orbital and spin degrees of freedoms cannot be,
  strictly speaking, separated in the presence of SO coupling, we will still
  loosely refer to ``charge-''\ and ``spin-sector''\/ quantities, in the sense
  specified above.}\BibitemShut {Stop}%
\bibitem [{\citenamefont {Herring}(1966)}]{herring}%
  \BibitemOpen
  \bibfield  {author} {\bibinfo {author} {\bibfnamefont {C.}~\bibnamefont
  {Herring}},\ }\href@noop {} {\emph {\bibinfo {title} {Exchange Interactions
  among Itinerant Electrons}}},\ edited by\ \bibinfo {editor} {\bibfnamefont
  {G.}~\bibnamefont {Rado}}\ and\ \bibinfo {editor} {\bibfnamefont
  {H.}~\bibnamefont {Suhl}},\ \bibinfo {series} {Magnetism}, Vol.~\bibinfo
  {volume} {IV}\ (\bibinfo  {publisher} {Academic Press},\ \bibinfo {year}
  {1966})\BibitemShut {NoStop}%
\bibitem [{\citenamefont {Raghu}\ \emph {et~al.}(2010)\citenamefont {Raghu},
  \citenamefont {Chung}, \citenamefont {Qi},\ and\ \citenamefont
  {Zhang}}]{raghu:2010}%
  \BibitemOpen
  \bibfield  {author} {\bibinfo {author} {\bibfnamefont {S.}~\bibnamefont
  {Raghu}}, \bibinfo {author} {\bibfnamefont {S.~B.}\ \bibnamefont {Chung}},
  \bibinfo {author} {\bibfnamefont {X.-L.}\ \bibnamefont {Qi}}, \ and\ \bibinfo
  {author} {\bibfnamefont {S.-C.}\ \bibnamefont {Zhang}},\ }\href {\doibase
  10.1103/PhysRevLett.104.116401} {\bibfield  {journal} {\bibinfo  {journal}
  {Phys. Rev. Lett.}\ }\textbf {\bibinfo {volume} {{\bf 104}}},\ \bibinfo
  {pages} {116401} (\bibinfo {year} {{(2010).}})}\BibitemShut {NoStop}%
\bibitem [{\citenamefont {{Ashrafi}}\ and\ \citenamefont
  {{Maslov}}(2012)}]{ashrafi:2012}%
  \BibitemOpen
  \bibfield  {author} {\bibinfo {author} {\bibfnamefont {A.}~\bibnamefont
  {{Ashrafi}}}\ and\ \bibinfo {author} {\bibfnamefont {D.~L.}\ \bibnamefont
  {{Maslov}}},\ }\href@noop {} {\bibfield  {journal} {\bibinfo  {journal}
  {\prl}\ }\textbf {\bibinfo {volume} {109}},\ \bibinfo {pages} {227201}
  (\bibinfo {year} {2012})}\BibitemShut {NoStop}%
\bibitem [{\citenamefont {{Zhang}}\ \emph {et~al.}()\citenamefont {{Zhang}},
  \citenamefont {{Yu}}, \citenamefont {{Ye}},\ and\ \citenamefont
  {{Liu}}}]{zhang:2012a}%
  \BibitemOpen
  \bibfield  {author} {\bibinfo {author} {\bibfnamefont {S.-S.}\ \bibnamefont
  {{Zhang}}}, \bibinfo {author} {\bibfnamefont {X.-L.}\ \bibnamefont {{Yu}}},
  \bibinfo {author} {\bibfnamefont {J.}~\bibnamefont {{Ye}}}, \ and\ \bibinfo
  {author} {\bibfnamefont {W.-M.}\ \bibnamefont {{Liu}}},\ }\href@noop {}
  {\bibinfo  {journal} {arXiv:1212.0424}\ }\BibitemShut {NoStop}%
\bibitem [{\citenamefont {Ashrafi}\ and\ \citenamefont
  {Maslov}()}]{ashrafi:partII}%
  \BibitemOpen
\bibfield  {journal} {  }\bibfield  {author} {\bibinfo {author} {\bibfnamefont
  {A.}~\bibnamefont {Ashrafi}}\ and\ \bibinfo {author} {\bibfnamefont {D.~L.}\
  \bibnamefont {Maslov}},\ }\href@noop {} {\bibinfo  {journal} {unpublished}\
  }\BibitemShut {NoStop}%
\bibitem [{\citenamefont {Overhauser}(1971)}]{overhauser:1971}%
  \BibitemOpen
\bibfield  {journal} {  }\bibfield  {author} {\bibinfo {author} {\bibfnamefont
  {A.~W.}\ \bibnamefont {Overhauser}},\ }\href {\doibase
  10.1103/PhysRevB.4.3318} {\bibfield  {journal} {\bibinfo  {journal} {Phys.
  Rev. B}\ }\textbf {\bibinfo {volume} {{\bf 4}}},\ \bibinfo {pages} {3318}
  (\bibinfo {year} {1971})}\BibitemShut {NoStop}%
\bibitem [{\citenamefont {Gangadharaiah}\ and\ \citenamefont
  {Maslov}(2005)}]{suhas:2005}%
  \BibitemOpen
  \bibfield  {author} {\bibinfo {author} {\bibfnamefont {S.}~\bibnamefont
  {Gangadharaiah}}\ and\ \bibinfo {author} {\bibfnamefont {D.~L.}\ \bibnamefont
  {Maslov}},\ }\href {\doibase 10.1103/PhysRevLett.95.186801} {\bibfield
  {journal} {\bibinfo  {journal} {Phys. Rev. Lett.}\ }\textbf {\bibinfo
  {volume} {{\bf 95}}},\ \bibinfo {pages} {186801} (\bibinfo {year}
  {2005})}\BibitemShut {NoStop}%
\bibitem [{\citenamefont {{L. M. Wei et al.}}(2011)}]{wei:2011}%
  \BibitemOpen
  \bibfield  {author} {\bibinfo {author} {\bibnamefont {{L. M. Wei et al.}}},\
  }\href {\doibase 10.1063/1.3633509} {\bibfield  {journal} {\bibinfo
  {journal} {J. Appl. Phys.}\ }\textbf {\bibinfo {volume} {{\bf 110}}},\
  \bibinfo {eid} {063707} (\bibinfo {year} {2011})}\BibitemShut {NoStop}%
\bibitem [{\citenamefont {Bychkov}\ and\ \citenamefont
  {Rashba}(1984)}]{bychkov:1984}%
  \BibitemOpen
  \bibfield  {author} {\bibinfo {author} {\bibfnamefont {Y.~A.}\ \bibnamefont
  {Bychkov}}\ and\ \bibinfo {author} {\bibfnamefont {E.~I.}\ \bibnamefont
  {Rashba}},\ }\href@noop {} {\bibfield  {journal} {\bibinfo  {journal} {Sov.
  Phys. JETP Lett.}\ }\textbf {\bibinfo {volume} {{\bf 39}}},\ \bibinfo {pages}
  {78} (\bibinfo {year} {1984})}\BibitemShut {NoStop}%
\bibitem [{\citenamefont {Luttinger}(1960)}]{luttinger:1961b}%
  \BibitemOpen
  \bibfield  {author} {\bibinfo {author} {\bibfnamefont {J.~M.}\ \bibnamefont
  {Luttinger}},\ }\href {\doibase 10.1103/PhysRev.119.1153} {\bibfield
  {journal} {\bibinfo  {journal} {Phys. Rev.}\ }\textbf {\bibinfo {volume}
  {119}},\ \bibinfo {pages} {1153} (\bibinfo {year} {1960})}\BibitemShut
  {NoStop}%
\bibitem [{\citenamefont {Rashba}(2005)}]{rashba:2005}%
  \BibitemOpen
  \bibfield  {author} {\bibinfo {author} {\bibfnamefont {E.}~\bibnamefont
  {Rashba}},\ }\href@noop {} {\bibfield  {journal} {\bibinfo  {journal} {J.
  Supercond.}\ }\textbf {\bibinfo {volume} {{\bf 18}}},\ \bibinfo {pages} {137}
  (\bibinfo {year} {2005})}\BibitemShut {NoStop}%
\bibitem [{\citenamefont {Chesi}\ and\ \citenamefont
  {Giuliani}(2011)}]{chesi-theorem}%
  \BibitemOpen
  \bibfield  {author} {\bibinfo {author} {\bibfnamefont {S.}~\bibnamefont
  {Chesi}}\ and\ \bibinfo {author} {\bibfnamefont {G.~F.}\ \bibnamefont
  {Giuliani}},\ }\href {\doibase 10.1103/PhysRevB.83.235308} {\bibfield
  {journal} {\bibinfo  {journal} {Phys. Rev. B}\ }\textbf {\bibinfo {volume}
  {{\bf 83}}},\ \bibinfo {pages} {235308} (\bibinfo {year} {2011})}\BibitemShut
  {NoStop}%
\bibitem [{\citenamefont {Pomeranchuk}(1958)}]{pomeranchuk:1958}%
  \BibitemOpen
  \bibfield  {author} {\bibinfo {author} {\bibfnamefont {I.~J.}\ \bibnamefont
  {Pomeranchuk}},\ }\href@noop {} {\bibfield  {journal} {\bibinfo  {journal}
  {Sov. Phys. JETP}\ }\textbf {\bibinfo {volume} {{\bf 8}}},\ \bibinfo {pages}
  {361} (\bibinfo {year} {1958})}\BibitemShut {NoStop}%
\bibitem [{\citenamefont {Skinner}\ and\ \citenamefont
  {Shklovskii}(2013)}]{skinner:2013}%
  \BibitemOpen
  \bibfield  {author} {\bibinfo {author} {\bibfnamefont {B.}~\bibnamefont
  {Skinner}}\ and\ \bibinfo {author} {\bibfnamefont {B.~I.}\ \bibnamefont
  {Shklovskii}},\ }\href@noop {} {\bibfield  {journal} {\bibinfo  {journal}
  {Phys. Rev. B}\ }\textbf {\bibinfo {volume} {87}},\ \bibinfo {pages} {035409}
  (\bibinfo {year} {2013})}\BibitemShut {NoStop}%
\bibitem [{Note2()}]{Note2}%
  \BibitemOpen
  \bibinfo {note} {In the SU2S case, the choice of the spin quantization axis
  is arbitrary, and the Zeeman energy of a quasiparticle can always be written
  in the diagonal form as $\sigma _z H$. In the chiral case, the choice of the
  spin quantization axis is unique, and the Zeeman energy cannot always be
  reduced to the diagonal form.}\BibitemShut {Stop}%
\bibitem [{Note3()}]{Note3}%
  \BibitemOpen
  \bibinfo {note} {We work in Matsubara formalism, in which the interaction
  line comes with a minus sign.}\BibitemShut {Stop}%
\bibitem [{\citenamefont {Maslov}\ \emph {et~al.}(2006)\citenamefont {Maslov},
  \citenamefont {Chubukov},\ and\ \citenamefont {Saha}}]{maslov:2006}%
  \BibitemOpen
  \bibfield  {author} {\bibinfo {author} {\bibfnamefont {D.~L.}\ \bibnamefont
  {Maslov}}, \bibinfo {author} {\bibfnamefont {A.~V.}\ \bibnamefont
  {Chubukov}}, \ and\ \bibinfo {author} {\bibfnamefont {R.}~\bibnamefont
  {Saha}},\ }\href {\doibase 10.1103/PhysRevB.74.220402} {\bibfield  {journal}
  {\bibinfo  {journal} {Phys. Rev. B}\ }\textbf {\bibinfo {volume} {{\bf
  74}}},\ \bibinfo {pages} {220402} (\bibinfo {year} {2006})}\BibitemShut
  {NoStop}%
\bibitem [{\citenamefont {Maslov}\ and\ \citenamefont
  {Chubukov}(2009)}]{maslov:2009}%
  \BibitemOpen
  \bibfield  {author} {\bibinfo {author} {\bibfnamefont {D.~L.}\ \bibnamefont
  {Maslov}}\ and\ \bibinfo {author} {\bibfnamefont {A.~V.}\ \bibnamefont
  {Chubukov}},\ }\href {\doibase 10.1103/PhysRevB.79.075112} {\bibfield
  {journal} {\bibinfo  {journal} {Phys. Rev. B}\ }\textbf {\bibinfo {volume}
  {{\bf 79}}},\ \bibinfo {pages} {075112} (\bibinfo {year} {2009})}\BibitemShut
  {NoStop}%
\bibitem [{\citenamefont {{Kim}}\ \emph {et~al.}(2013)\citenamefont {{Kim}},
  \citenamefont {{Lutchyn}},\ and\ \citenamefont {{Nayak}}}]{kim:2013}%
  \BibitemOpen
  \bibfield  {author} {\bibinfo {author} {\bibfnamefont {Y.}~\bibnamefont
  {{Kim}}}, \bibinfo {author} {\bibfnamefont {R.~M.}\ \bibnamefont
  {{Lutchyn}}}, \ and\ \bibinfo {author} {\bibfnamefont {C.}~\bibnamefont
  {{Nayak}}},\ }\href@noop {} {\bibfield  {journal} {\bibinfo  {journal}
  {arXiv:1304.0464}\ } (\bibinfo {year} {2013})}\BibitemShut {NoStop}%
\end{thebibliography}%
\end{document}